\documentclass [a4paper,12pt]{article}
\pdfoutput=1
\usepackage[english]{babel}
\usepackage{graphicx,amsmath,amsfonts,amssymb,slashed,upgreek,color,caption}

\newcommand{\be}{\begin{equation}}
\newcommand{\ee}{\end{equation}}
\newcommand{\bea}{\begin{eqnarray}}
\newcommand{\eea}{\end{eqnarray}}

\newcommand{\HH}{{\cal H}}

\def\ie{{\em i.e.}}

\def\d{\delta}
\def\tt{\Uptheta}
\def\kv{\vec k}
\def\qv{\vec q}
\def\xv{\vec x}
\def\sv{\vec s}

\def\Q{{\cal Q}}

\def\cGpc{\, h^{-3} \, {\rm Gpc}^3}
\def\kMpc{\, h \, {\rm Mpc}^{-1}}

\begin{document}
\def\thefootnote{\fnsymbol{footnote}}

\begin{center}
\Large{\textbf{Cosmological structure formation with clustering quintessence}} \\[0.4cm]
 
\large{Emiliano Sefusatti\footnote{email: \texttt{emiliano.sefusatti@cea.fr}} and Filippo Vernizzi\footnote{email: \texttt{filippo.vernizzi@cea.fr}}}
\\[0.4cm]

\small{
\textit{Institut de Physique Th\'eorique\\ CEA, IPhT, 91191 Gif-sur-Yvette c\'edex, France\\ CNRS, URA-2306, 91191 Gif-sur-Yvette c\'edex, France}}

\end{center}

\vspace{.5cm}

\hrule \vspace{0.3cm}
\noindent \small{\textbf{Abstract}
\vspace{0.2cm}\\
\noindent We study large-scale structure formation in the presence of a quintessence component with zero speed of sound in the framework of Eulerian Perturbation Theory. Due to the absence of pressure gradients, quintessence and dark matter are comoving and can be studied as a unique fluid in terms of  the total energy density contrast and the common velocity.
In this description the clustering of quintessence enhances the linear term proportional to the velocity divergence in the continuity equation by a factor $(1+w) \Omega_Q/\Omega_m$. This is responsible for a rapid evolution of the growth rate at low redshifts, and modifies the standard relation between the velocity divergence and the growth factor. For the total fluid, the solutions for the linear growth function and growth rate can be written in integral forms and admit simple fitting formulae, as in the $\Lambda$CDM case. At second order in perturbation theory, we derive an explicit expression for the kernels $F_2$ and $G_2$. They receive modifications of the order of the ratio between quintessence and total energy density perturbations, which affect the corresponding tree-level bispectra.  We finally compute the cumulative signal-to-noise  in the power spectrum, bispectrum and reduced bispectrum, expected for departures from a $\Lambda$CDM cosmology both in the clustering and smooth quintessence scenarios. The {\em reduced} bispectrum, in particular, receives sensible modifications only in the clustering case and can potentially be used to detect or rule out the model.\\

\hrule
\def\thefootnote{\arabic{footnote}}
\setcounter{footnote}{0}

\tableofcontents


\section{Introduction}

The simplest  and most popular model of dynamical dark energy is quintessence, a single scalar field whose vacuum energy dominates the Universe driving its acceleration. Quintessence energy density varies with time and a way to distinguish it against a cosmological constant is to observe the effect of the different expansion history on dark matter structure formation \cite{Wang:1998gt}. In its standard version, quintessence is described by a minimally-coupled canonical field \cite{Zlatev:1998tr}. In this case scalar fluctuations propagate at the speed of light maintaining quintessence homogeneous even in the presence of dark matter clumps \cite{Ferreira:1997au}. Quintessence can cluster only on scales larger than the horizon, where fluctuations have no time to propagate. However, observations on such large scales are strongly limited by cosmic variance and this effect is difficult to observe.

A model of quintessence that can cluster on all observable scales has been recently proposed in \cite{Creminelli:2008wc,Creminelli:2009mu}. It is based on a single scalar degree of freedom with fluctuations characterized by a practically zero speed of sound. As explained in \cite{Creminelli:2008wc}, there are several theoretical motivations to consider this case. In the limit of zero sound speed one recovers the Ghost Condensate theory \cite{ArkaniHamed:2003uy}, which is invariant under shift symmetry. Thus, there is no fine tuning in assuming that the speed of sound is very small: quintessence models with vanishing speed of sound should be thought of as deformations of this particular limit where shift symmetry is recovered \cite{Creminelli:2006xe,Senatore:2004rj}. Moreover, using the tools developed in \cite{Creminelli:2006xe,Cheung:2007st}, formulated in the context of an effective field theory, it has been shown that quintessence with an equation of state $w<-1$ can be free from ghosts and gradient instabilities only if the speed of sound is very tiny, $|c_s| \lesssim 10^{-15}$ \cite{Creminelli:2008wc}. Stability can be guaranteed by the presence of higher derivative operators \cite{Creminelli:2006xe,ArkaniHamed:2003uy}, although their effect is absent on cosmologically relevant scales \cite{Creminelli:2008wc}.

Apart from these theoretical considerations, a very important motivation to consider this model is that a series of galaxy and cosmic shear surveys are currently planned with the aim of understanding the nature of dark energy through its role in the structure formation. In this context the clustering scenario represents a phenomenologically interesting counterpart to the case of a smooth quintessence component. Indeed, quintessence with vanishing speed of sound actively participates to the formation of structures together with the dark matter and gives distinct modifications to the standard picture that can be strongly constrained by future data. 

In the past, several articles have investigated the observational consequences of a clustering quintessence in the linear regime,
in particular, on the cosmic microwave background \cite{DeDeo:2003te,Weller:2003hw,BeanDore,Hannestad:2005ak,Sapone:2009mb}, galaxy
redshift surveys \cite{Takada:2006xs}, large neutral hydrogen surveys \cite{TorresRodriguez:2007mk}, the cross-correlation of the integrated Sachs-Wolfe effect in the cosmic microwave background with the large-scale structures \cite{Hu:2004yd,Corasaniti:2005pq}, or on weak lensing \cite{Sapone:2010uy}. 

Theoretical investigations of the effect of dark energy on the nonlinear evolution of structures are particularly crucial. First of all, from linear theory alone it is difficult to distinguish the effects of quintessence on structure formation through its modification of the expansion history from those genuinely due to its perturbations. As we will see, the nonlinear evolution breaks this degeneracy. Furthermore, numerical simulations taking into account the gravitationally coupled evolution of dark matter particles and a clustering scalar field are still under construction and for the clustering scenario considered here they are totally missing. On the other hand, future redshift and weak lensing surveys will require very accurate predictions, both for the dark matter density and galaxy correlators, particularly on nonlinear scales where the signal is larger. Finally, to conclude this series of motivations we remind that the study of nonlinearities is receiving a lot of attention in the context of primordial non-Gaussianities \cite{Komatsu:2009kd}. It is pertinent to ask whether a second clustering component could mimic the effect of primordial non-Gaussianities on the nonlinear evolution.

A first description of clustering quintessence in the nonlinear regime was given in \cite{Creminelli:2009mu}. There it was shown that in the limit of zero sound speed  pressure gradients are negligible and, as long as the fluid approximation is valid, quintessence follows geodesics remaining comoving with the dark matter (see also \cite{Lim:2010yk} for a more recent model with identical phenomenology). 
In particular, reference \cite{Creminelli:2009mu} studied the effect of  quintessence with vanishing sound speed on the structure formation in the nonlinear regime, in the context of the spherical collapse model (see \cite{Bjaelde:2010qp} for a study of the spherical collapse when $c_s^2$ of quintessence is small but finite). Due to the absence of pressure gradients, comoving regions behave as closed FRW universes and the spherical collapse can be solved exactly. The modifications to the critical threshold of collapse are small and the effects on the dark matter mass function  are dominated by the modification on the linear dark matter growth function, which are also small. Today they are of the order of few per cent for realistic values of $w$.
A larger effect occurs when one considers the {\em total} mass function, which includes the contribution of quintessence overdensities to the virialized halos. Indeed, quintessence contributes to the total halo mass by a fraction which increases at lower redshifts and is proportional to the ratio between quintessence and dark matter energy densities, {\em i.e.}~$\sim (1 + w)\,\Omega_Q/ \Omega_m$.

In this paper we study the nonlinear regime of clustering quintessence using Eulerian Perturbation Theory (EPT). In particular, we extend the standard EPT approach for dark matter \cite{Peebles,Fry:1983cj,Bernardeau:2001qr} to the presence of a second fluid, a clustering quintessence, comoving and coupled only gravitationally to dark matter. This is the first natural step to the study of nonlinear perturbations beyond the spherical approximation. In contrast to the spherical collapse model, this approach is perturbative and solutions can be found order by order. Notice that on small scales the EPT perturbative expansion for density correlators is not well defined, because it presents large cancellations between contributions of the same order. However, in the standard case it has been shown that classes of higher-order corrections can be resummed, leading to a well established perturbative scheme known, in its first formulation, as Renormalized Perturbation Theory (RPT) \cite{Crocce:2005xy,Crocce:2005xz,Crocce:2007dt,Bernardeau:2008fa,Bernardeau:2010md}. Complementary approaches can be found in \cite{Matarrese:2007wc, Matarrese:2007aj,Pietroni:2008jx, Anselmi:2010fs, Saracco:2009df}.

In this work, we begin by considering the continuity equations for the dark matter and quintessence density contrasts and the Euler equation for their common velocity. Since gravitational observables are sensitive only to the sum of dark matter and quintessence fluctuations, we derive the continuity equation for the {\em total density contrast}. As both fluids are comoving, this equation and the Euler equation (together with the Poisson equation relating the total density to the gravitational potential) form a closed system in Fourier space, which can be solved perturbatively, as in the standard EPT approach. 

As in the standard case, the nonlinear couplings in the continuity and Euler equations are at most quadratic and the vertices are the same as those for a single dark matter fluid. The only  difference is that the  velocity divergence in the continuity equation is enhanced by the factor $(1 + w)\, \Omega_Q/ \Omega_m$. At linear order, this term is responsible for a rapid evolution of the growth rate at low redshifts, and changes the standard relation between the velocity divergence and the growth factor. Due to the absence of pressure gradients, the solutions for the linear growth and the linear growth rate of the total fluid can be written in integral form, as in the  standard $\Lambda$CDM case. Using these solutions we are able to find simple fitting functions for these quantities, which generalize those currently employed in $\Lambda$CDM cosmologies \cite{Lahav:1991wc,Carroll:1991mt}. 

At higher order in the perturbative expansion  clustering dark energy is responsible for an additional time-dependence of the kernels $F_n$ and $G_n$ defining the $n$-th order nonlinear corrections. The effect on $F_2$ and $G_2$ is of the order of the ratio between quintessence and total density perturbations, $\sim {\delta \rho_Q}/({\delta \rho_m +\delta \rho_Q})$, and gives distinctive signatures in the higher-order correlation functions such as the bispectrum. In particular, the {\em reduced} bispectrum, whose expression at leading order in EPT is independent of the linear power spectrum normalization, presents corrections {\em only} in the clustering case. Analogous corrections have been found in the halo mass function from the contribution of the quintessence mass to collapsed objects \cite{Creminelli:2009mu}.

It is not the first time that EPT is generalized to the presence of several components. For instance, in \cite{Somogyi:2009mh} EPT has been applied to the problem of following the nonlinear evolution of baryon and cold dark matter perturbations evolving from distinct initial conditions and in \cite{Wong:2008ws,Shoji:2009gg,Saito:2008bp,Saito:2009ah,Lesgourgues:2009am,Brouzakis:2010md} to the study of nonlinear perturbations in the presence of massive neutrinos. For modified gravity models it has been used in \cite{Koyama:2009me,Scoccimarro:2009eu, Chan:2009ew} to calculate the nonlinear power spectrum and, in particular, in \cite{Scoccimarro:2009eu, Chan:2009ew, BernardeauBrax2010} to compute the matter bispectrum. Higher-order observables, such as the normalized skewness $S_3 \equiv \langle \delta^3 \rangle / \langle \delta^2 \rangle^2$, have been also studied in \cite{Multamaki:2003vs,Lue:2003ky,Amendola:2004wa} in the context of modified gravity models, where variations up to $\sim 10\%$ have been found.

This paper is organized as follows. In section~\ref{sec:eom} we present the equations of motion describing the coupled evolution of matter and quintessence perturbations. In section~\ref{sec:linear} we solve the linearized equations for the density growth factor and the density growth rate and we discuss the solutions and fitting formulae. In section~\ref{sec:nonlinear} we discuss the perturbative solutions in EPT. In particular, we derive the second-order solutions for the density and velocity fields, while in section~\ref{sec:bispectrum} we derive the lowest-order observables: the density tree-level power spectrum and bispectrum. As a practical illustration of these results, in section~\ref{sec:StoN} we compare the signal-to-noise expected for the effect of clustering and smooth quintessence with respect to the $\Lambda$CDM case in ideal measurements of the density large-scale power spectrum, bispectrum and reduced bispectrum in a box of $1\cGpc$ at redshift $z=0.5$. Finally, we present our conclusions in section~\ref{sec:conclusions}.

In addition, we present in appendix \ref{app:scalar} a discussion on the analogy between the scalar field and the perfect fluid, with a derivation of the continuity, Euler and Poisson equations in the regime considered in this paper. In appendix  \ref{app:vertices} we derive evolution equations for the vertices in the spherical collapse approximation at all orders and in appendix \ref{app:zdist} we discuss the redshift-space distortion effects in the clustering quintessence case.


\section{Equations of motion}
\label{sec:eom}

We consider a flat FRW background universe with metric $ds^2 = a^2(\tau) (- d \tau^2 + d \vec x^2)$, where $\tau$ is the conformal time, and a generic perfect fluid $\alpha$ with energy density $\rho_\alpha$, pressure $p_\alpha$, and peculiar velocity with respect to the Hubble flow $\vec v_\alpha$.
The continuity and Euler equations in an expanding background read 
\begin{align}
&\frac{\partial \rho_\alpha}{\partial \tau} + 3 \HH (\rho_\alpha +p_\alpha) + \vec \nabla \cdot \left[ (\rho_\alpha+p_\alpha) \vec v_\alpha \right] = 0 \label{continuity}\,,\\
&\frac{\partial {\vec{ v}_\alpha}}{\partial \tau}  + \HH \vec v_\alpha+ (\vec v_\alpha \cdot \vec \nabla) \vec v_\alpha =- \frac{1}{\rho_\alpha +p_\alpha} \left(\vec \nabla p_\alpha + \vec v_\alpha \,\frac{\partial p_\alpha}{\partial \tau} \right)-\vec \nabla \Phi\;, \label{euler}
\end{align}
where $\HH \equiv d \ln a  /d \tau $ is the conformal Hubble time and $\Phi$ is the gravitational potential satisfying the Poisson equation,
\be
\nabla^2 \Phi = 4 \pi G a^2 \sum_\alpha \left(\delta \rho_\alpha + 3 \delta p_\alpha \right)\;. \label{poisson}
\ee
 As explained in appendix~\ref{app:scalar}, these equations are valid only on scales much smaller than the Hubble radius. Furthermore, they assume non-relativistic fluid velocities, $v \ll c$. For small density and pressure perturbations velocities remain small, independently of the speed of sound of the fluid. For the particular case of a fluid with zero speed of sound, such as dust or clustering quintessence, pressure gradients are suppressed and fluid velocities remain small even in the nonlinear regime.

For small velocities one can neglect the time derivative of the pressure in front of the pressure gradient on the right hand side of eq.~\eqref{euler}, $\vec v \; \partial_\tau p \ll \vec \nabla p$. In this regime the speed of sound is simply the ratio between pressure and energy density fluctuations, 
\be
c_{\alpha,s}^2 \equiv  \delta p_\alpha/{\delta \rho_\alpha}\;,\label{deltapvsdeltarho_rf}
\ee
and eq.~\eqref{euler} reduces to
\be
\frac{\partial {\vec{ v}_\alpha}}{\partial \tau}  + \HH \vec v_\alpha+ (\vec v_\alpha \cdot \vec \nabla) \vec v_\alpha =- \frac{c_s^2 \vec \nabla \rho_\alpha}{\rho_\alpha +p_\alpha}   -\vec \nabla \Phi\;. \label{euler_2}
\ee
For a  fluid with vanishing speed of sound the first term on the right hand side of the Euler equation vanishes and the fluid follows geodesics. 
This is the case for both dark matter and clustering quintessence. Thus, in their growing solution dark matter and quintessence are comoving \cite{Creminelli:2009mu} and we can take their velocities to be the same, \ie~$\vec v_m = \vec v_Q \equiv \vec v$.\footnote{{During matter dominance quintessence energy density is negligible and the dynamics is dominated by the gravitational potential wells of the dark matter. Assuming that quintessence velocity has not unreasonably ``extreme'' initial conditions, it will be rapidly driven to the dark matter velocity values by the force term in eq.~\eqref{euler_2}, while velocity differences decay with the expansion. This situation is thus different from that described in \cite{Somogyi:2009mh} to study the gravitationally coupled evolution of baryons and dark matter, where the baryon fraction is always non-negligible with respect to the dark matter.}} 

Let us define the density contrast $\delta_\alpha$ as $\delta_\alpha \equiv \delta \rho_\alpha / \bar \rho_\alpha$, where $\bar \rho_\alpha$ is the background value of the energy density. In terms of this quantity, the continuity equation (\ref{continuity}) for dark matter and quintessence becomes, respectively,
\begin{align}
&\frac{\partial \delta_m}{\partial \tau}  + \vec \nabla \cdot \big[ (1+\delta_m) \vec v \big] = 0 \label{continuity_m}\,,\\
&\frac{\partial \delta_Q}{\partial \tau}  -3 w \HH \delta_Q+ \vec \nabla \cdot \big[ (1+w+\delta_Q) \vec v\big] = 0 \label{continuity_Q}\,,
\end{align}
where $w \equiv \bar p_Q/\bar \rho_Q$ is the equation of state of quintessence. In the limit $c_{Q,s} =0$, the Euler equation for the common dark matter and quintessence velocity is the same, \ie
\be
\frac{\partial {\vec{ v}}}{\partial \tau}  + \HH \vec v+ (\vec v \cdot \vec \nabla) \vec v =-\vec \nabla \Phi\;. \label{euler_common}
\ee
These are the equations describing dark matter and quintessence in the regime $c_{Q,s} =0$. Note that the nonlinear couplings are only quadratic, as in the standard, pure dark matter case.

In EPT it is useful to define the velocity divergence $\theta\equiv\vec \nabla \cdot \vec v$. In fact, assuming $\vec v$ to be irrotational, which is a good approximation up to shell-crossing, the peculiar velocity $\vec v$ is completely described by its divergence $\theta$ \cite{Bernardeau:2001qr}. In Fourier space, the continuity equations for dark matter and quintessence become, respectively, 
\begin{align}
&\frac{\partial \delta_{m,\kv}}{\partial \tau}  + \theta_{\kv} = -\alpha(\qv_1, \qv_2)\, \theta_{\vec q_1} \delta_{m,\vec q_2} \label{continuity_m_FS}\,,\\
&\frac{\partial \delta_{Q,\vec k}}{\partial \tau}  -3 w \HH \delta_{Q,\vec k}+ (1+w) \theta_{\vec k} = -\alpha(\vec q_1, \vec q_2)\, \theta_{\vec q_1} \delta_{Q,\vec q_2} \label{continuity_Q_FS}\,,
\end{align}
where
\be
\alpha(\vec q_1, \vec q_2)  \equiv  1 + \frac{\vec q_1 \cdot \vec q_2}{q_1^2}  \;,  \label{alpha_def}
\ee
with $q_i \equiv | \vec q_i|$. Here and in the following, an integral $\int d^3 q_1 d^3 q_2 \delta_D (\vec k - \vec q_1 - \vec q_2)$ is implied over quadratic terms with repeated wavenumbers, such as on the right hand side of eqs.~(\ref{continuity_m_FS}) and (\ref{continuity_Q_FS}). There are few terms in eq.~\eqref{continuity_Q_FS} that differ from eq.~\eqref{continuity_m_FS}. The second term on the left hand side of eq.~(\ref{continuity_Q_FS}), absent when $w=0$, comes from the fact that the energy of quintessence does not scale as the volume, while the factor $1+w$ in front  of the third term comes from the fact that the 3-momentum density of quintessence sourcing the variation of the density contrast is proportional to $\rho_Q+p_Q$. The nonlinear couplings on the right hand side of eqs.~(\ref{continuity_m_FS}) and (\ref{continuity_Q_FS}) are both expressed in terms of $\alpha(\vec q_1, \vec q_2)$ and are the same for dark matter and quintessence. This is because pressure gradients are absent in both fluids.

We can now use the Friedmann equation $3 \HH^2 = 8 \pi G (\bar \rho_m+\bar \rho_Q)$ and $\delta p_m = \delta p_Q=0$ to rewrite the Poisson equation (\ref{poisson}) as
\be
\nabla^2 \Phi = \frac{3}{2}\, \HH^2\, \Omega_m \left( \delta_m +  \delta_Q \frac{\Omega_Q}{\Omega_m} \right),
\ee
where $\Omega_{\alpha} =\Omega_{\alpha} (\tau) \equiv \bar \rho_{\alpha}/(\bar \rho_m+\bar \rho_Q)$ for both matter and quintessence. Using this equation to replace the gravitational potential on the right hand side of eq.~\eqref{euler_common}, we can rewrite the Euler equation in Fourier space as
\be
\frac{\partial \theta_{\kv}}{\partial \tau} + \HH\, \theta_{\kv} + \frac32\, \Omega_m \HH^2 \left( \delta_{m,\kv} + \delta_{Q,\kv}\, \frac{\Omega_Q}{\Omega_m} \right) = - \,\beta(\vec q_1, \vec q_2)\, \theta_{\vec q_1} \,\theta_{\vec q_2}\;, \label{euler_sum}
\ee
where
\be
\beta(\vec q_1, \vec q_2)\equiv \frac{(\vec q_1 + \vec q_2)^2 \,\vec q_1 \cdot \vec q_2 }{2\, q_1^2\, q_2^2}\;.\label{beta_def}
\ee

To an observer measuring cosmological clustering through its gravitational effect, there is no distinction between $\delta \rho_m$ and $\delta \rho_Q$: the observer will be only sensitive to the gravitational potential sourced by the {\em total} density perturbation $\delta \rho = \delta \rho_m+\delta \rho_Q$, via the Poisson equation \eqref{poisson}. We will thus concentrate on the description of the total density perturbation $\delta \rho$. A convenient way of describing the evolution of perturbations is to define the {\em total density contrast}
\be
\delta \equiv \frac{\delta \rho}{\bar \rho_m} = \delta_m +  \delta_Q \frac{\Omega_Q}{\Omega_m}\;, \label{tdc}
\ee
which takes into account both dark matter and dark energy perturbations. With this definition the Poisson equation reads 
\be 
\nabla^2 \Phi = 4 \pi\, G\, a^2 \bar \rho_m\, \delta \;. \label{poisson_2}
\ee
An inadvertent observer could interpret $\delta$ as due to dark matter perturbations only. However, as we will see below the time evolution of $\delta$ when quintessence clusters is very different from the standard smooth case.
Note that we could have defined $\delta$ by dividing the total density perturbation $\delta \rho$, not by the background matter energy density, but by the total background energy density $\bar \rho = \bar \rho_m + \bar \rho_Q$. This would have had little consequence. Indeed, only the combination $\bar \rho_m \delta $ is an observable {\em via} the Poisson equation so that the only effect, apart from a trivial rescaling in the evolution equations, would have been to change eq.~\eqref{poisson_2} into $\nabla^2 \Phi = 4 \pi G \bar \rho \delta$. Our definition has the advantage of recovering the usual expressions when quintessence perturbations are set to zero.\footnote{While the matter density contrast has, by definition, the lower bound $\d_m>-1$, for the total density contrast we have instead $\d>-[1+\Omega_Q(\tau)/\Omega_m(\tau)]$.}$^,$\footnote{To capture the change in the gravitational potential induced by dark energy perturbations, Ref.~\cite{Amendola:2007rr} defines a parameter $Q$ such that the Poisson equation reads $\nabla^2 \Phi = 4 \pi GQ\bar \rho_m  \delta_m$. In our notation $Q \equiv \delta /\delta_m$.}
 
By noticing that ${\partial_\tau} (\Omega_Q/\Omega_m) = -3\, w\, \HH\, \Omega_Q/\Omega_m$, we can combine eqs.~(\ref{continuity_m_FS}) and (\ref{continuity_Q_FS}) into a single equation for $\delta$ and write a closed set of equations involving just the total density contrast $\delta$ and the velocity divergence $\theta$, 
\begin{align}
&\frac{\partial \delta_{\vec k}}{\partial \tau}  + C  \theta_{\vec k} = -\alpha(\vec q_1, \vec q_2) \theta_{\vec q_1} \delta_{\vec q_2} \label{continuity_tot}\,,\\
&\frac{\partial \theta_{\vec k}}{\partial \tau} + \HH \theta_{\vec k} + \frac{3}{2} \Omega_m \HH^2 \delta_{\vec k} = - \beta(\vec q_1, \vec q_2) \theta_{\vec q_1} \theta_{\vec q_2}\;, \label{euler_tot}
\end{align}
where we introduced
\be\label{eq:C}
C(\tau) \equiv  1+ (1+w)\frac{\Omega_Q(\tau)}{\Omega_m(\tau)}\;.
\ee
As expected, the effect of quintessence perturbations on the growth of the total density contrast is proportional to $(1+w) {\Omega_Q}/{\Omega_m}$, so that it vanishes at early times, when ${\Omega_Q}/{\Omega_m} \to 0$ or for $w=-1$. Indeed, from eq.~(\ref{euler_sum}) the effect of $\delta_Q$ on the common velocity divergence is proportional to $\Omega_Q/\Omega_m$. Furthermore, from eq.~(\ref{continuity_Q_FS}) the feedback of $\theta$ on the density contrast is proportional to $1+w$.
Note that the sign of the effect depends on the sign of $1+w$ \cite{Weller:2003hw,Creminelli:2008wc}. As inside an overdensity $\theta < 0$, the total density contrast increases faster when $1+w>0$, while increasing slower in the opposite case. 
For $C=1$ we recover the standard case of canonical scalar field quintessence with $c_s^2= 1$, for which dark energy perturbations propagate as acoustic waves at the speed of light and quintessence remains smooth within a Hubble patch \cite{Wang:1998gt}. In this case there is no distinction between $\delta$ and $ \delta_m$.

\begin{figure}[t]
\begin{center}
{\includegraphics[width=0.49\textwidth]{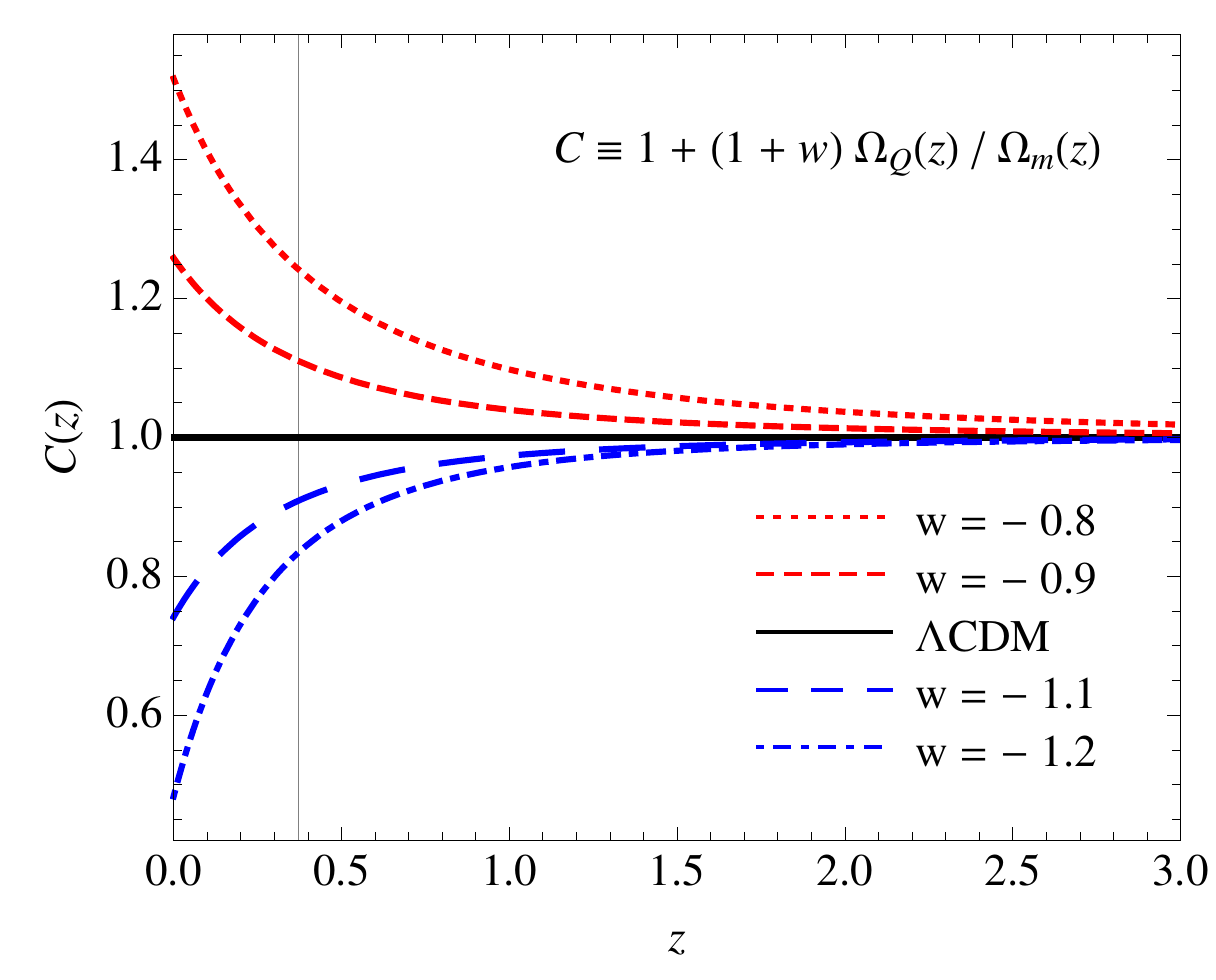}}
\caption{The quantity $C(z)$ as defined in eq.~(\ref{eq:C}), as a function of redshift for several values of $w$. Unless otherwise specified, here and afterward we will denote $w=-0.8$, $-0.9$, $-1.1$ and $-1.2$ respectively by red dotted, red short-dashed, blue long-dashed and blue dotted-dashed lines. Continuos black lines will denote $\Lambda$CDM. The vertical line indicates the redshift of equality between matter and quintessence for a $\Lambda$CDM cosmology, \ie~$z_{eq}=0.37$ for the assumed $\Omega_{m,0}=0.279$.}
\label{fig:C}
\end{center}
\end{figure}
Since the function $C$ captures all the modifications to the equations of motion in the clustering case, it is useful to plot such quantity. In Fig.~\ref{fig:C} we show $C(z)$ for several values of $w$ as a function of redshift. The vertical line indicates the redshift of equality between matter and  the cosmological constant in a $\Lambda$CDM cosmology, that is $z_{eq}\simeq 0.37$ for the assumed $\Omega_{m,0}=0.279$, where the index ``$0$" denotes quantities evaluated today.


\section{Linear theory}
\label{sec:linear}

We will now study the linear solutions of eqs.~(\ref{continuity_tot}) and (\ref{euler_tot}) by neglecting the quadratic terms on the right hand sides of these equations. In Fourier space, we can write the linear solutions as
\begin{align}
\delta^{\rm lin}_{\vec k} (\tau) &\equiv D(\tau) \delta^{\rm in}_{\vec k}\;,\label{delta_lin}\\
\theta^{\rm lin}_{\vec k}(\tau) &\equiv - \frac{\HH(\tau) f(\tau)}{C(\tau)} D(\tau) \delta^{\rm in}_{\vec k}\;, \label{theta_D}
\end{align}
where $D$ is the linear {\em growth function}. To derive the second equation we have used the linearization of eq.~(\ref{continuity_tot}) and introduced the linear {\em growth rate} $f$ as
\be
f\equiv \frac{d \ln D}{d \ln a}\;, \label{f_def}
\ee
a useful quantity to relate the velocity to the density perturbation in linear theory. Note that on the right hand side of eq.~\eqref{theta_D}, unlike the usual case, the function $C(z)$ is present at the denominator. 
Indeed, this comes from the factor in front of $\theta$ in the continuity equation for the total density contrast, eq.~\eqref{continuity_tot}.

We will now study the evolution of $D$ and $f$ in the clustering quintessence case and we will derive fitting formulae for these two quantities, generalizing to the case of clustering quintessence the expressions employed in $\Lambda$CDM cosmology. For simplicity we will assume a constant $w$. See appendix~\ref{app:zdist} for an extension of the linear theory to redshift space.


\subsection{Linear growth function}

By combining eq.~(\ref{continuity_tot}) and the time derivative of eq.~(\ref{euler_tot}) at linear order one obtains the evolution equation of the linear growth function $D$. As a function of the scale factor $a$ this reads
\be
\frac{d^2 D}{d\ln a^2} + \left[ \frac12 ( 1 - 3\, w\, \Omega_Q)  -  \frac{d \ln C}{d \ln a} \right] \frac{d D}{d\ln a} - \frac32 \Omega_m  C D =0\;, \label{growth_function_evol}
\ee
where the time derivative of $C$ is given by
\be
\frac{d \ln C}{d \ln a} = - \frac{3\, w\, (C-1)}{C}\;.
\label{C_evol}
\ee
For $C=1$ we recover the evolution equation of the growth function in the case of a smooth quintessence, derived in \cite{Wang:1998gt}. 

Assuming that the scale factor today is $a_0=1$, we can use the Friedmann equation to write the Hubble rate $H \equiv d \ln a /dt$, where $t$ is the cosmic time, as
\be
H(a) = H_0 \left[\Omega_{m,0}\,a^{-3} + \Omega_{Q,0}\,a^{-3(1+w)} \right]^{1/2}\;. \label{Hubble}
\ee
Then, deriving this equation with respect to the scale factor one obtains
\be
\frac{d \ln H}{d \ln a}= - \frac32 \,\Omega_m\, C\;,  \label{Ha}
\ee
where $\Omega_m = {\Omega_{m,0}}/({H^2 a^3})$. 
Using the relation
\be
\frac12 (1 - 3 w \Omega_Q) = 2 - \frac32 \Omega_m C\;,
\ee
valid only for $\Lambda$CDM {\em and} for clustering quintessence,
we can rewrite eq.~\eqref{growth_function_evol} as 
\be
\frac{d^2 D}{d\ln a^2} + \left[ 2 +  \frac{d \ln H}{d \ln a} -  \frac{d \ln C}{d \ln a} \right] \frac{d D}{d\ln a} + \frac{d \ln H}{d \ln a} D =0\;. \label{growth_function_evol_2}
\ee
We stress that this equation and the results that follow are valid {\em only} in the case of quintessence with zero speed of sound (or, in the limit where $C=1$, for $\Lambda$CDM). Written in this form, it is easy to check that the evolution equation of $D$ has two solutions. One is a decaying mode, $D_- \propto H$. The growing mode can be written in integral form as
\be
D_+(a) = \frac52\, H_0^2\, \Omega_{m,0}\, H(a)\! \int_0^a\!\! \frac{C(\tilde a)}{[\tilde a\, H(\tilde a)]^3} d \,\tilde a\;, \label{int_sol}
\ee
where 
we have normalized $D_+$ in such a way that at early time, during matter domination, $D_+$ is equal to the scale factor $a$. This is the first of the main results of this work. Although derived for a constant $w$, one can check that eqs.~\eqref{growth_function_evol_2} and \eqref{int_sol} hold also when $w$ depends on time.
Alternatively, an explicit solution can be found in terms of hypergeometric functions by means of eq.~(3.194) of \cite{GR}. This yields
\be
\frac{D_+(a)}{a} = \frac{1}{\sqrt{1+x}}\! \left[{}_2F_1\! \left(\frac32, - \frac5{6w}, 1 - \frac{5}{6w} , -x \right)
 + \,x  \frac{5 (1+w)}{5-6w} {}_2F_1\! \left(\frac32, 1 - \frac5{6w}, 2 - \frac5{6w} , -x \right) \right]\,, \label{D_sol}
\ee
where
\be
x (a) \equiv \frac{\Omega_Q (a) }{ \Omega_m (a)}= \frac{1 -\Omega_{m,0}}{\Omega_{m,0}} a^{-3w}\;.
\ee

As already remarked, the solution \eqref{int_sol} {\em does not} describe the case of a smooth dark energy component with $w\neq -1$. Indeed, such an integral solution relies on the absence of pressure gradients and on the fact that dark matter and quintessence move together along geodesics. In this case---and {\em only} in this case---each comoving region evolves as an independent, unperturbed FRW universe. 
We can take advantage of this fact to re-derive eq.~\eqref{int_sol} in an alternative way, similarly to what is commonly done in the $\Lambda$CDM case \cite{Heath:1977} (see also \cite{Peebles,Peebles:1994xt,Lyth:2009zz}).
Indeed, a spherical overdensity of dark matter and clustering quintessence of radius $R$ can be described by the Friedmann equation for a closed universe \cite{Creminelli:2009mu},
\be
\bigg( \frac{\dot R}{R} \bigg)^2 = \frac{8 \pi G}{3} (\rho_m + \rho_Q) - \frac{K}{R^2}\;, \label{FRW_closed}
\ee
where $K$ is the curvature constant and the dot denotes the derivative with respect to the cosmic time.
Let us describe deviations of the radius $R$ from the scale factor of the background universe $a$ by
\be
\alpha \equiv 1 - R/a\;.
\ee
Linearizing eq.~\eqref{FRW_closed} we obtain
\be
2 H \dot \alpha = - H^2 \Omega_m \delta + \frac{K}{a^2}\;;
\ee
using the relation $\dot H = -(3/2) H^2 \Omega_m C$, which can be derived by taking the time derivative of eq.~\eqref{Hubble}, we can rewrite this equation as
\be
 (3\, \dot\alpha\, C) \; H = \dot H \delta + \frac32 \frac{K}{a^2} C\;. \label{dot_alpha}
\ee
We can relate $\dot \alpha$ to $\dot \delta$ by using the continuity equation inside the overdense region, {\em i.e.}~\cite{Creminelli:2009mu}
\be
\dot \rho + 3 \frac{\dot R}{R} (\rho+ \bar p_Q) =0\;,
\ee
where $\rho = \rho_m + \rho_Q$ and the quintessence pressure  $\bar p_Q$ is unperturbed due to the absence of pressure gradients. Linearizing this equation one obtains $\dot \delta = 3 \,\dot\alpha\, C$, which can be used to replace the parenthesis on the left hand side of eq.~\eqref{dot_alpha}. With this replacement eq.~\eqref{dot_alpha} can be easily integrated to yield
\be
\delta (a) = \frac32 H (a) K \! \int_0^a\!\! \frac{C(\tilde a)}{[\tilde a\, H(\tilde a)]^3} d \,\tilde a\;,
\ee
which presents the same time evolution as eq.~\eqref{int_sol}. In this way we have consistently recovered the linear growing solution by linearizing the spherical collapse model, assuming that spherical overdensities behave as closed Friedmann universes.

Now we turn our attention to the individual matter and quintessence perturbations. We introduce the linear growth functions of dark matter $D_m$ and quintessence $D_Q$ as
\begin{align}
\delta^{\rm lin}_{m,\,\vec k} (\tau) &\equiv D_m (\tau) \delta^{\rm in}_{\vec k}\;, \label{deltam_lin} \\
\delta^{\rm lin}_{Q,\,\vec k} (\tau) &\equiv D_Q (\tau) \delta^{\rm in}_{\vec k}\;.
\end{align}
These components do not evolve independently. However, once the solution for $D$ is given, their evolution can be found by linearizing eqs.~(\ref{continuity_m}) and (\ref{continuity_Q}), to obtain
\begin{align}
\frac{d D_m}{d\ln a} & = \frac1C \frac{d D}{d\ln a}\,,\label{eq:Dm}\\
\frac{d D_Q}{d\ln a} - 3 w D_Q & =\frac{1+w}{C} \frac{d D}{d\ln a}\,.\label{eq:DQ}
\end{align}
Notice that, from eqs.~(\ref{eq:Dm}), (\ref{eq:DQ}) and \eqref{int_sol}, after few manipulations we can write $D_m(a)$ and $D_Q(a)$ in integral form as a function of $D(a)$ as
\begin{align}
D_m(a)&=\int_0^a \left[\frac52-\frac32\frac{D(\tilde{a})}{\tilde{a}}\right]\Omega_m(\tilde{a})\,d\tilde{a} \;, \\
D_Q(a)&= (1+w) \frac{\Omega_m(a)}{\Omega_Q(a)} \int_0^a \left[\frac52-\frac32\frac{D(\tilde{a})}{\tilde{a}}\right]\Omega_Q(\tilde{a}) \,d\tilde{a} \;.
\end{align}

In {\em matter dominance}, when $D_+ = D_{m,+} \propto a$, the growing mode for quintessence simply reads \cite{Creminelli:2008wc} 
\be
D_{Q,+} = \frac{1+w}{1-3w} D_{m,+} \qquad {\rm (matter \ dom.)}\;,
\ee
so that
\be
\frac{D_+}{a} = \left(1 + \frac{1+w}{1-3w} \frac{\Omega_Q}{\Omega_m} \right) \qquad {\rm (matter \ dom.)} \;.
\ee
When quintessence dominates this is a poor approximation. Below, we will present a much better approximation, valid during both matter and dark energy domination.

\begin{figure}[t]
\begin{center}
{\includegraphics[width=0.49\textwidth]{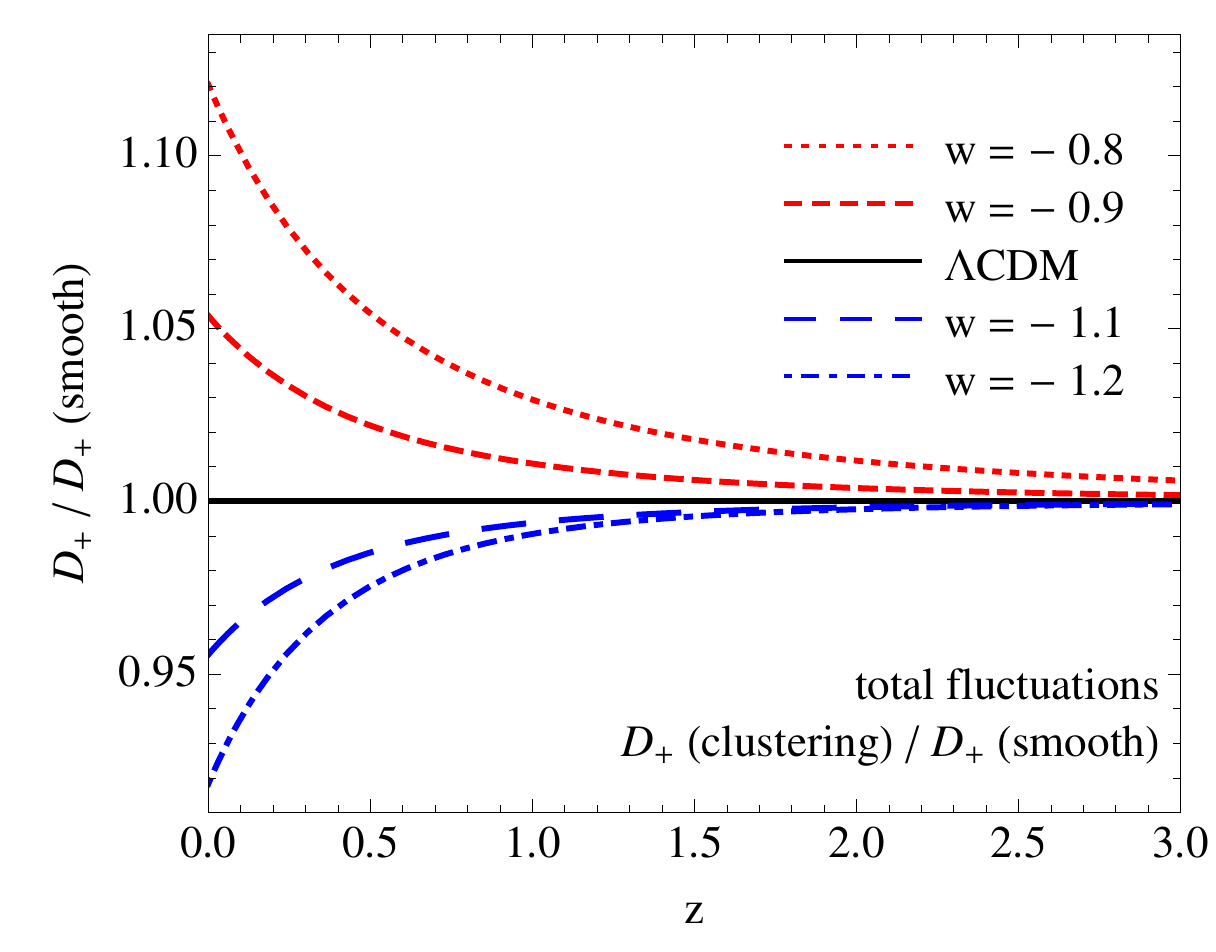}}
{\includegraphics[width=0.49\textwidth]{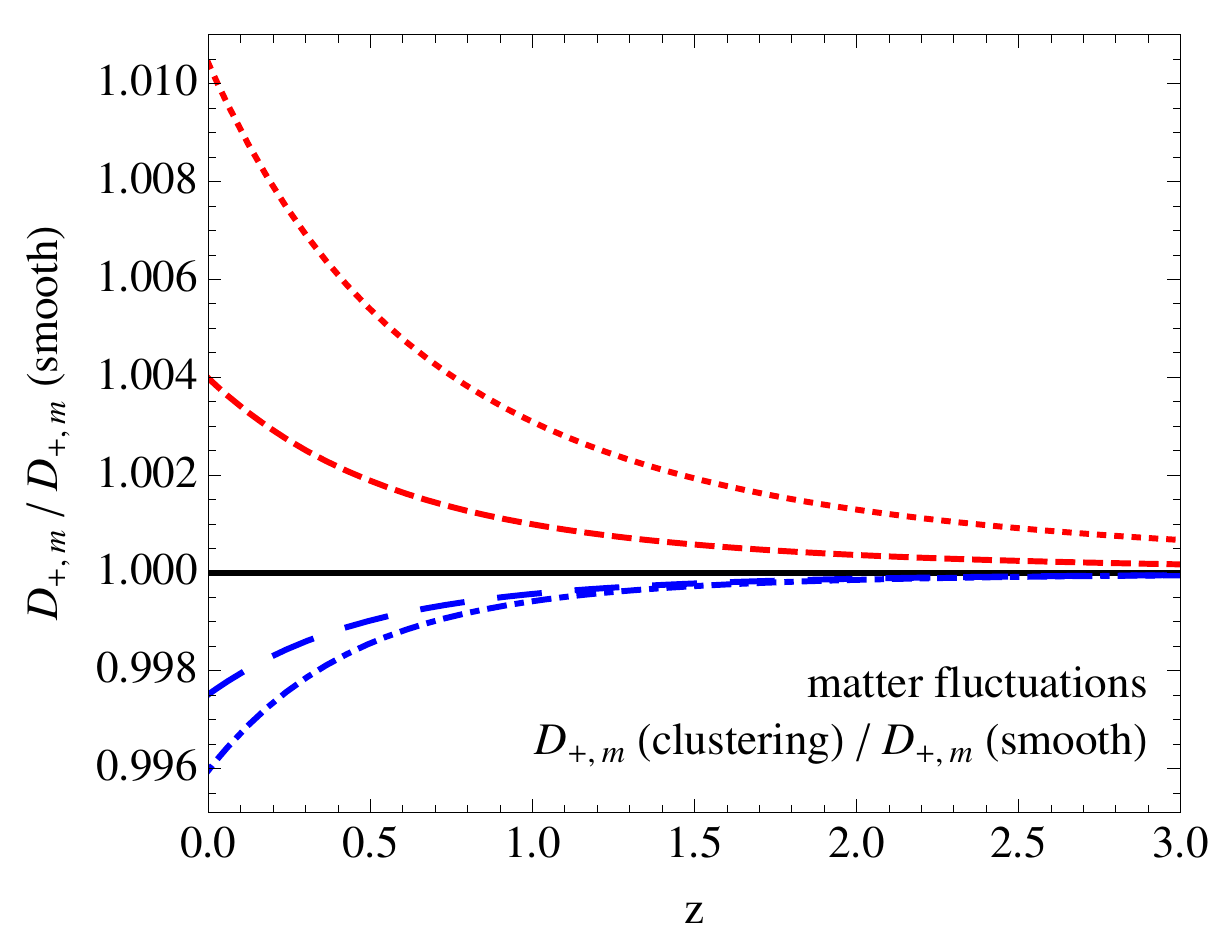}}
{\includegraphics[width=0.49\textwidth]{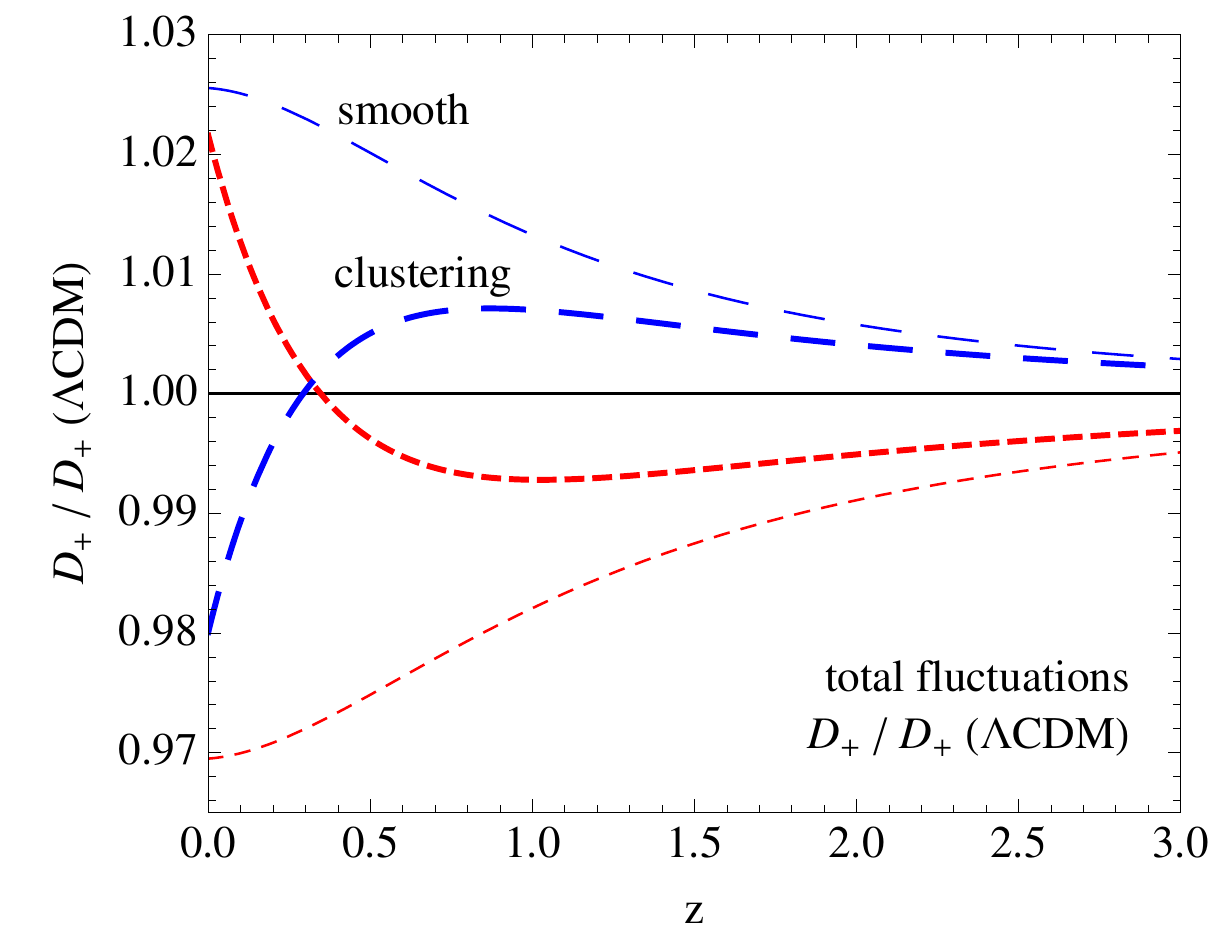}}
{\includegraphics[width=0.49\textwidth]{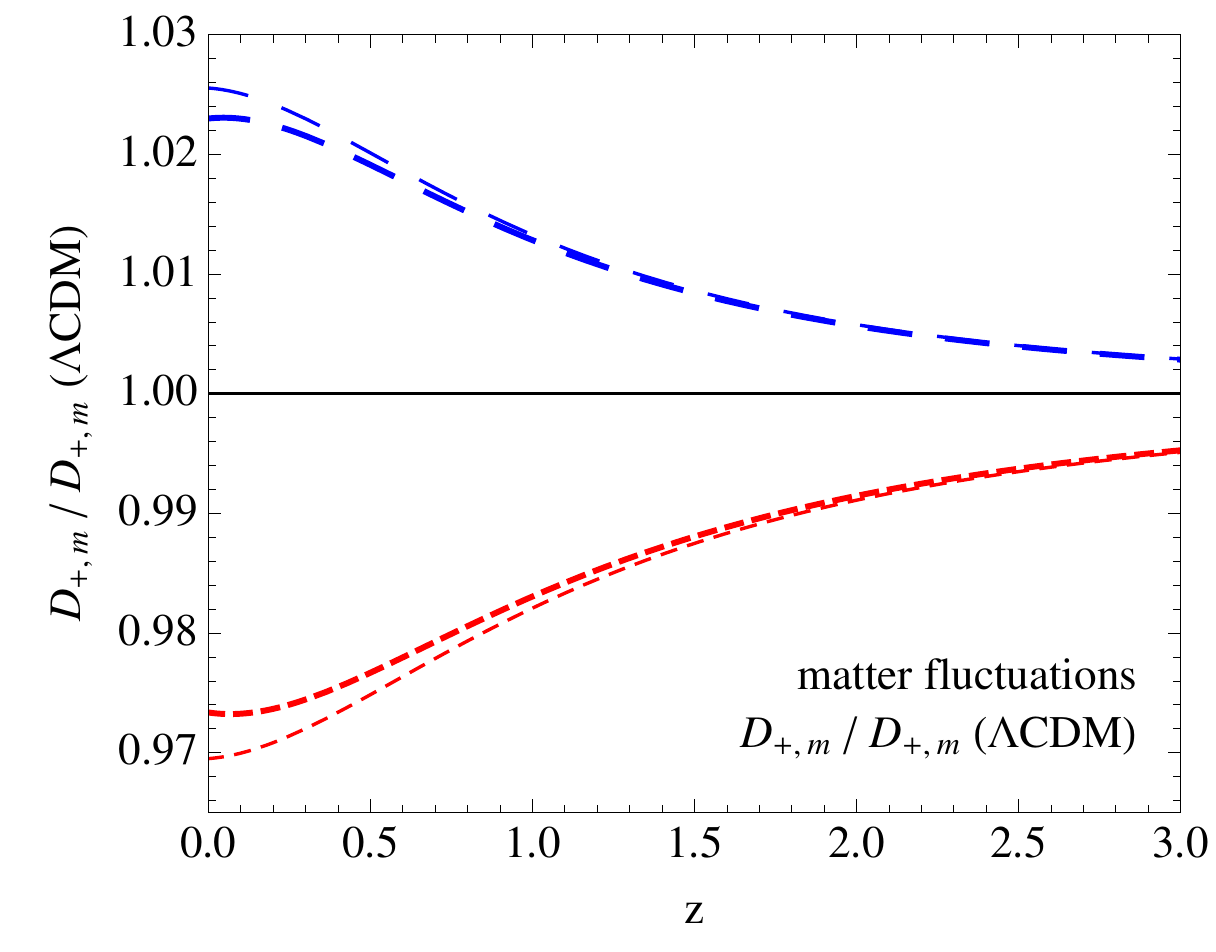}}
\caption{{\em Upper left panel}: ratio of the linear growth factor of the total perturbation,  $D_+$, in the clustering case to the smooth quintessence case. {\em Upper right panel}: the same as in the left panel for the matter growth factor, $D_{m,+}$. {\em Lower left panel}: ratio of the linear growth factor of the total perturbation  $D_+$ in the clustering (thick line) and smooth (thin line) cases to the $\Lambda$CDM case, for fixed $\Omega_{m,0}$ today. {\em Lower right panel}: the same as in the left panel for $D_{m,+}$.}
\label{fig:D}
\end{center}
\end{figure}
The effect of a clustering quintessence on the evolution of the total linear growing solution $D_+$ and of the linear dark matter growth function $D_{m,+}$ is shown in Fig.~\ref{fig:D}, in the left and right panels, respectively.
In the upper left panel we plot the ratio between $D_+$ in the clustering and smooth quintessence cases for several values of $w$. In the upper right panel we do the same for $D_{m,+}$. As expected, quintessence with zero sound speed enhances the clustering when $w>-1$, while clustering is hindered for $w<-1$. As the effect is proportional to $(1+w) \Omega_Q/\Omega_m$, it increases in time and vanishes for $w=-1$.
While the effect on the dark matter fluctuations is below the percent level for $-1.2\le w \le -0.8$ (and for the assumed value $\Omega_{m,0}=0.279$), the growth factor for the total perturbations $\d$ is affected by modifications as large as 10\% at redshift zero for models with $|1+w| \simeq 0.2 $.

In the lower left panel we plot the ratio of $D_+$ in the clustering (thick line) and smooth (thin line) cases to the $\Lambda$CDM case, for two values of $w$. In the lower right panel we do the same for $D_{m,+}$. In this case two opposite effects are into play, as one can clearly see from the bottom left panel of Fig.~\ref{fig:D}. On one hand, when $w>-1$ quintessence dominates the Universe earlier than a cosmological constant, anticipating the accelerated expansion phase. This has the initial effect of suppressing the growth of fluctuations. However, since clustering quintessence contributes to the total perturbation $\delta$, its effect on the linear growth function becomes important at low redshift, winning over the one of the accelerated expansion. 
As a consequence, the evolution of the gravitational potential is modified with respect to the smooth case and this distinctive signature can be used to constrain the dark energy parameter using the CMB  \cite{DeDeo:2003te,Weller:2003hw,BeanDore,Hannestad:2005ak} or by cross-correlating the integrated Sachs-Wolfe effect with the large-scale structures \cite{Hu:2004yd,Corasaniti:2005pq}. The effect of clustering quintessence on the {\em matter} linear growth is, on the other hand, much weaker, as shown in the bottom right panel.


\subsection{Linear growth rate}

From eq.~(\ref{growth_function_evol}) we can derive an evolution equation for the growth rate $f$, defined in eq.~\eqref{f_def}, given by 
\be
\frac{d\, f}{d \ln a} + f^2 + \left[ \frac12 ( 1 - 3\, w\, \Omega_Q) - \frac{d \ln C}{d \ln a}\right] f = \frac32 \Omega_m C \;. \label{f_evol}
\ee
We remind the reader that the linear growth rate relates the linear velocity divergence $\theta^{\rm lin}$ to the linear density contrast $\delta^{\rm lin}$ as
\be
\theta^{\rm lin} = - {\cal H} \, \frac{ f}{C} \, \delta^{\rm lin}\;. \label{v_delta}
\ee
Thus, the significant quantity is given by the ratio $f/C$ rather than $f$ alone.
Its explicit growing solution can be found using the definition \eqref{f_def} together with the integral solution for $D_+$, eq.~\eqref{int_sol}. This yields
\be
f_+ (a)=    \left[ \frac52\, ( {D_+}/{a} )^{-1}- \frac32 \right] \Omega_m C \;, \label{f_D}
\ee
where the ratio $D_+/a$ is given by eq.~\eqref{D_sol}.
The decaying solution is obtained from eq.~\eqref{f_def} with $D_- \propto H$ and from eq.~\eqref{Ha},
\be
f_- (a)= - \frac{3}{2} \Omega_m C\;. \label{f-}
\ee

\begin{figure}[t]
\begin{center}
{\includegraphics[width=0.49\textwidth]{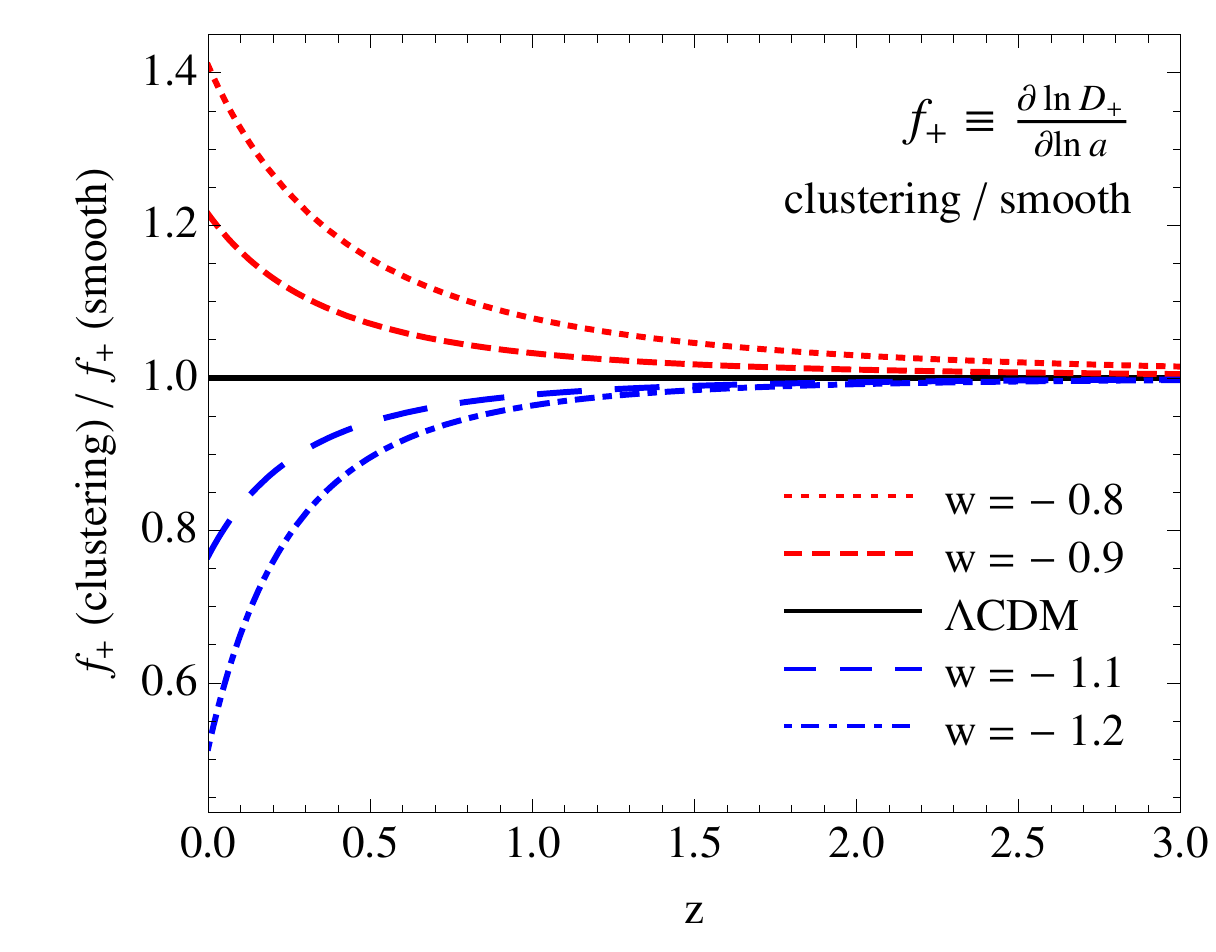}}
{\includegraphics[width=0.49\textwidth]{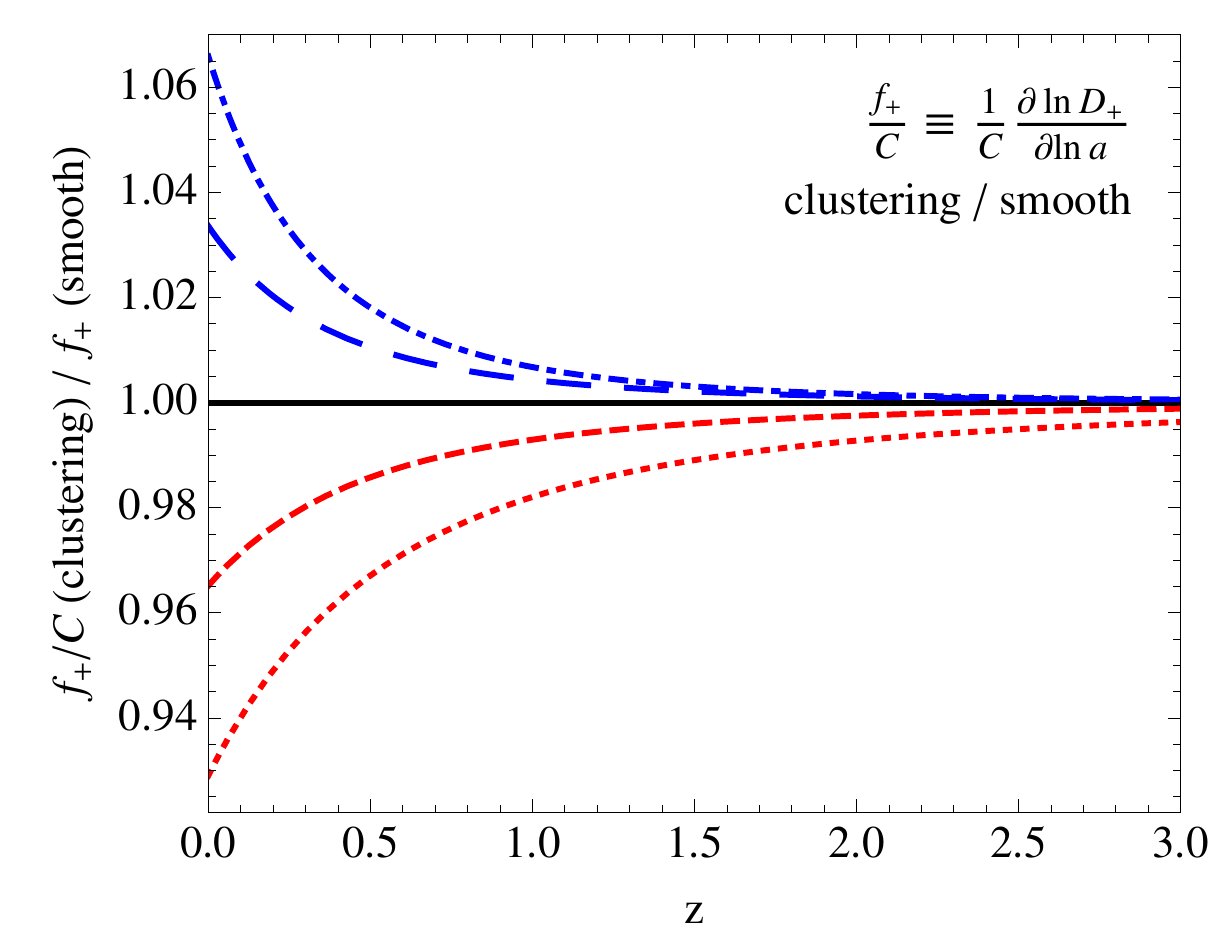}}
{\includegraphics[width=0.49\textwidth]{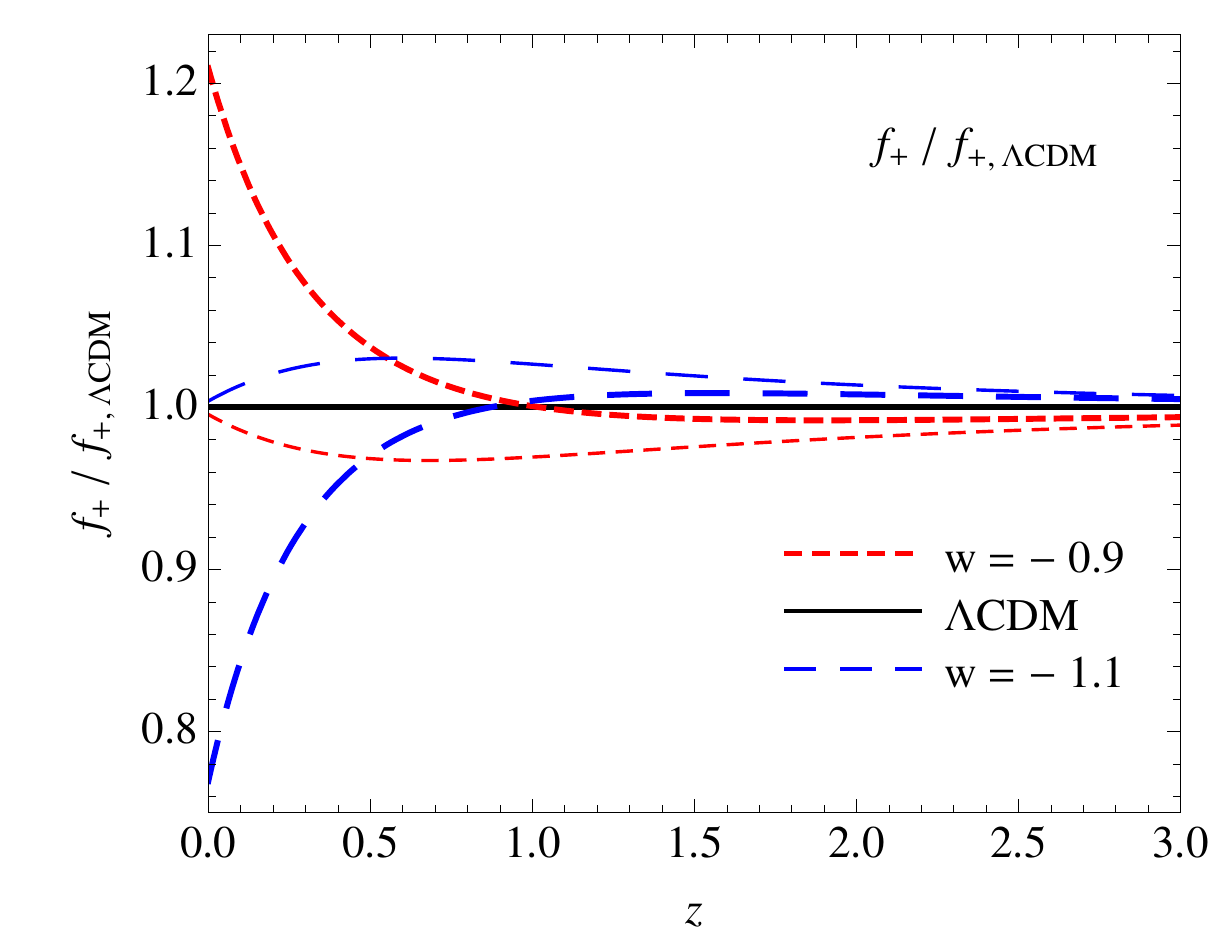}}
{\includegraphics[width=0.49\textwidth]{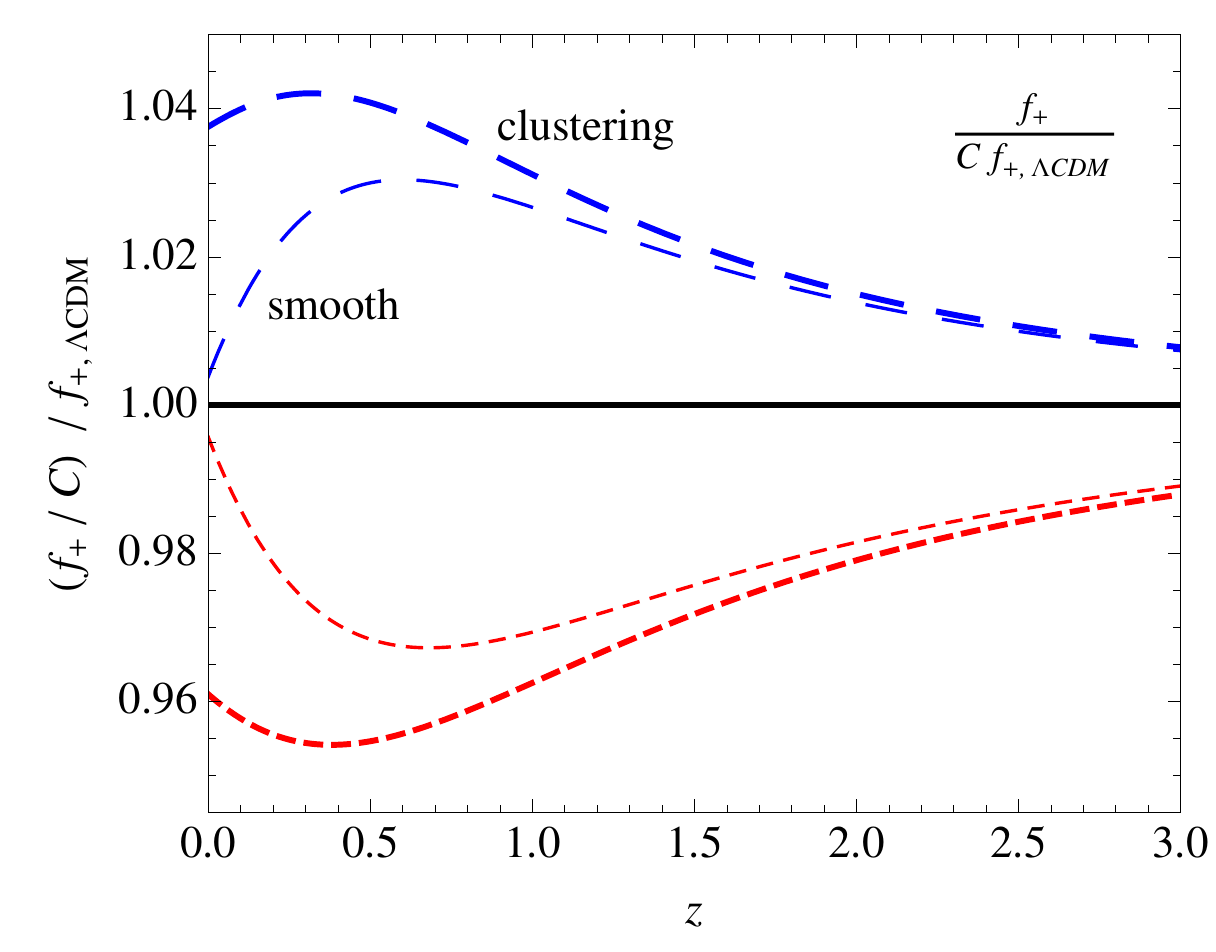}}
\caption{{\em Upper panels}: ratio of the growth rate $f_+$ ({\em left}) and of the quantity $f_+/C$ ({\em right}) for clustering quintessence to their counterparts in the smooth quintessence scenario (\ie, in both cases, $f_+$) as a function of redshift for several value of $w$. {\em Bottom panels}: ratio of the  $f_+$ ({\em left}) and of the quantity $f_+/C$ ({\em right}) for clustering ({\em thick lines}) and smooth quintessence ({\em thin lines}) to $f_+$ in the $\Lambda$CDM case.}
\label{fig:f}
\end{center}
\end{figure}
In Fig.~\ref{fig:f} the left panels show the effects of clustering quintessence on $f_+$. In particular, in the upper left panel we plot the ratio of the growth rate $f_+$ for clustering quintessence to the same quantity for a smooth quintessence component, for several values of $w$. In the lower left panel we plot the ratio of $f_+$ in the clustering and smooth cases to $f_+$ in a $\Lambda$CDM cosmology. At high redshift the background evolution dominates the growth, as already discussed. At lower redshift $f_+$ changes rapidly with time due to the clustering of quintessence. At $z=0$ the effect is quite large, of the order of $\sim 20\%$ for $|1+w| = 0.1$.

In the right panels of the same figure, we show the effects of clustering quintessence on the ratio $f_+/C$. As discussed in more details in appendix~\ref{app:zdist}, it is this quantity, and not $f_+$ alone, that describes the effect on redshift distortions at linear level.
We show the ratio between $f_+/C$ for clustering quintessence and $f_+$ for a smooth quintessence component ({\em upper right panel}) and for $\Lambda$CDM ({\em lower right panel}). As one can see, the growth of $C$ in the ratio $f_+/C$ is very important, compensating---and marginally dominating---the growth of $f_+$.  
Note that, from eq.~\eqref{f_D}, increasing the growth factor corresponds to decreasing the ratio $f_+/C$. Indeed, quintessence with zero sound speed reduces $f_+/C$ for $w>-1$, favoring the clustering, while the opposite happens when $w<-1$.

This qualitative behavior can be understood by introducing the {\em growth index}  $\gamma$, so that $f_+ \equiv \Omega_m^\gamma$, where $\gamma$ is generally a time-dependent quantity. From eq.~(\ref{f_evol}) and using 
\be
\frac{d \Omega_m}{d \ln a} = 3\, w\, \Omega_Q \Omega_m\;,
\ee
the evolution equation for $\gamma$ in terms of the matter abundance $\Omega_m$ is given in our case by
\be
\label{eq:alpha}
3\, w\, \Omega_Q \ln \Omega_m \frac{d \gamma}{d \ln \Omega_m} + \frac12
+ \Omega_m^\gamma +  3\, w\, \left(\gamma-\frac12\right) \Omega_Q + \frac{ 3 w (C-1)}{C} - \frac32 \Omega_m^{1-\gamma} C =0\;.
\ee

Close to matter dominance, one can solve this equation as an expansion in $\Omega_Q$. In the smooth case, derived by solving eq.~\eqref{eq:alpha} with $C=1$, one finds the growth index computed in \cite{Wang:1998gt},\footnote{Note that in the first line of eq.~\eqref{alpha_smooth} we have corrected a typo in the denominator of eq.~(B11) of \cite{Wang:1998gt}.} \ie
\be
\begin{split}
\gamma_{\rm smooth} & =  \frac{3(1-w)}{5-6w} + \frac{ 3  (1-w) (2-3w) }{2 (5 - 6 w)^2 (5 - 
   12 w)} \Omega_Q + {\cal O}(\Omega_Q^2) \\
   & \simeq  0.545 + 0.007 \, \Omega_Q  +  (1+w) \big(  0.025 + 0.005 \, \Omega_Q \big)   + \ldots   \;, \label{alpha_smooth}
 \end{split}
\ee   
where in the second line we have expanded as well with respect to $1+w$. This can be compared to the growth index in the clustering case computed by solving eq.~\eqref{eq:alpha} with $C$ given by eq.~\eqref{eq:C}, \ie
\be
\begin{split}
\gamma & = \frac{6 w^2}{5-6w} - \frac{ 3 w (   72 w^4- 48 w^3- 34 w^2+ 56 w -25) }{(5 - 6 w)^2 (5 - 
   12 w)} \Omega_Q + {\cal O}(\Omega_Q^2) \\ 
   & \simeq 0.545 +0.007 \, \Omega_Q   - (1+w) \big(  0.793 +  0.443 \, \Omega_Q \big)  + \ldots   \;.\label{alpha_clustering}
\end{split}
\ee
Let us stress that $\gamma$ refers to the growth index of the {\em total} density contrast, which is different from the growth index of the density contrast of dark matter only computed, for instance, in the clustering case in \cite{Ballesteros:2008qk}.

The values of $\gamma_{\rm smooth}$ and $\gamma$ given in eqs.~\eqref{alpha_smooth} and \eqref{alpha_clustering} are consistent with what shown in the upper left panel of Fig.~\ref{fig:f}, as one can check by approximating the behavior of the ratio between $f_+$ in the clustering and smooth cases with $\Omega_m^{\gamma-\gamma_{\rm smooth}}$, where $\gamma-\gamma_{\rm smooth} = -(1+w) (0.818+0.448 \; \Omega_Q)$. Note also that $\gamma-\gamma_{\rm smooth}$ strongly depends on $\Omega_Q$. Indeed, whereas in the smooth case the time dependence of the growth index is very weak, so that one can consistently approximate it as a constant also at low redshifts \cite{Linder:2005in}, in the clustering case this time dependence is more severe and such approximation is not viable.

An alternative description of the growth function could be then given in terms of a {\em reduced} growth index $\gamma_{\rm red}$, defined by $f_+/C\equiv\Omega_m^{\gamma_{\rm red}}$, which presents a weaker dependence on cosmology and redshift. Indeed, one can check that $\gamma_{\rm red}-\gamma_{\rm smooth} = (1+w) (0.181+0.053 \; \Omega_Q)$, which agrees with what shown in the upper right panel of Fig.~\ref{fig:f}. An analogous behavior for the growth rate is discussed in \cite{Di Porto:2007ym} in the context of modified gravity models.\footnote{Di Porto and Amendola study in \cite{Di Porto:2007ym} the growth of structures in a modified gravity model, parametrizing the growth rate $f_+ = \Omega_m^{\gamma_{\rm DA}} (1+\eta_{\rm DA}) $. Their case is similar to ours. Indeed, in our case $\gamma_{\rm red}$ and $(1+w)\Omega_Q/\Omega_m$ play the role of $\gamma_{\rm DA}$ and $\eta_{\rm DA}$, respectively. Taking $\eta_{\rm DA}$ to be constant, the authors of this reference constrain the values of $\gamma_{\rm DA}$ and $\eta_{\rm DA}$ from a set of galaxy and Lyman-$\alpha$ observations, mostly at high redshift ($z\gtrsim 2$). They find $\gamma_{\rm DA}=0.6^{+0.4}_{-0.3}$ and $\eta_{\rm DA}=0.0^{+0.3}_{-0.2}$ at 1-$\sigma$ CL. In our model for $|1+w|\simeq 0.1$, $(1+w)\Omega_Q/\Omega_m$  is very close to zero at $z\simeq 2$, while it grows to $\sim 0.2$ at $z=0$. Note, however, that due to the strong redshift dependence of $(1+w)\Omega_Q/\Omega_m$ it is difficult to extrapolate the validity of their results to our case.}

Previous works (for instance \cite{Ballesteros:2008qk,Sapone:2009mb,Sapone:2010uy}) have studied the clustering of quintessence concentrating on its effect on the {\em matter} perturbation $\delta_m$, instead of the total perturbation $\delta$. 
To connect to these works, we define the linear growth rate for {\em matter} perturbations as
\be
f_m \equiv \frac{d\ln D_m}{d\ln a}\;. \label{fm_def}
\ee
From eq.~\eqref{eq:Dm}, it is related to $f$ by
\be
D_m\,f_m=D\, \frac{f}{C}\, . \label{f_fm}
\ee


\subsection{Fitting functions}

As mentioned above, when quintessence dominates the Universe $\gamma$ becomes strongly time-dependent and the parameterization $f_+ = \Omega_m^\gamma$, with $\gamma$ given in eq.~\eqref{alpha_clustering}, becomes a poor approximation (see, for instance, \cite{Ferreira:2010sz}). We have found the following approximation to be accurate in the range $0.1\lesssim \Omega_m\le1$, 
\be
f_+  = C \left[ \Omega_m^{4/7} +\left( \frac1{70} - \frac{1+w}4 \right) \Omega_Q \left(1 + \frac{\Omega_m}{2} \right) \right]\;.
\ee 
As shown in the right panel of Fig.~\ref{fig:fits_Df}, for $\Omega_m \ge 0.2$ the accuracy is better than $1\%$ level for $ - 1.15 \le w \le - 0.85$. For $w=-1$ this formula reduces to the well-known fit given in \cite{Lahav:1991wc} for $\Lambda$CDM. Following this reference, inverting eq.~\eqref{f_D} and using in this equation the  above approximation for $f_+$, eq.~\eqref{f_def}, it is possible to find the fitting formula for the growth function
\be
\frac{D_+}{a} = \frac52 \Omega_m \left[\Omega_m^{4/7} + \frac32 \Omega_m +\left( \frac1{70} - \frac{1+w}4 \right) \Omega_Q \left(1 + \frac{\Omega_m}{2} \right) \right]^{-1}\;.
\ee
As shown in the left panel of Fig.~\ref{fig:fits_Df}, for $\Omega_m \ge 0.2$ and $ w\ge - 1.5$ the error is again less than $1 \%$. When $w=-1$  the commonly used fitting formula given in \cite{Carroll:1991mt} for $\Lambda$CDM is recovered. 

\begin{figure}[t]
\begin{center}
{\includegraphics[width=0.49\textwidth]{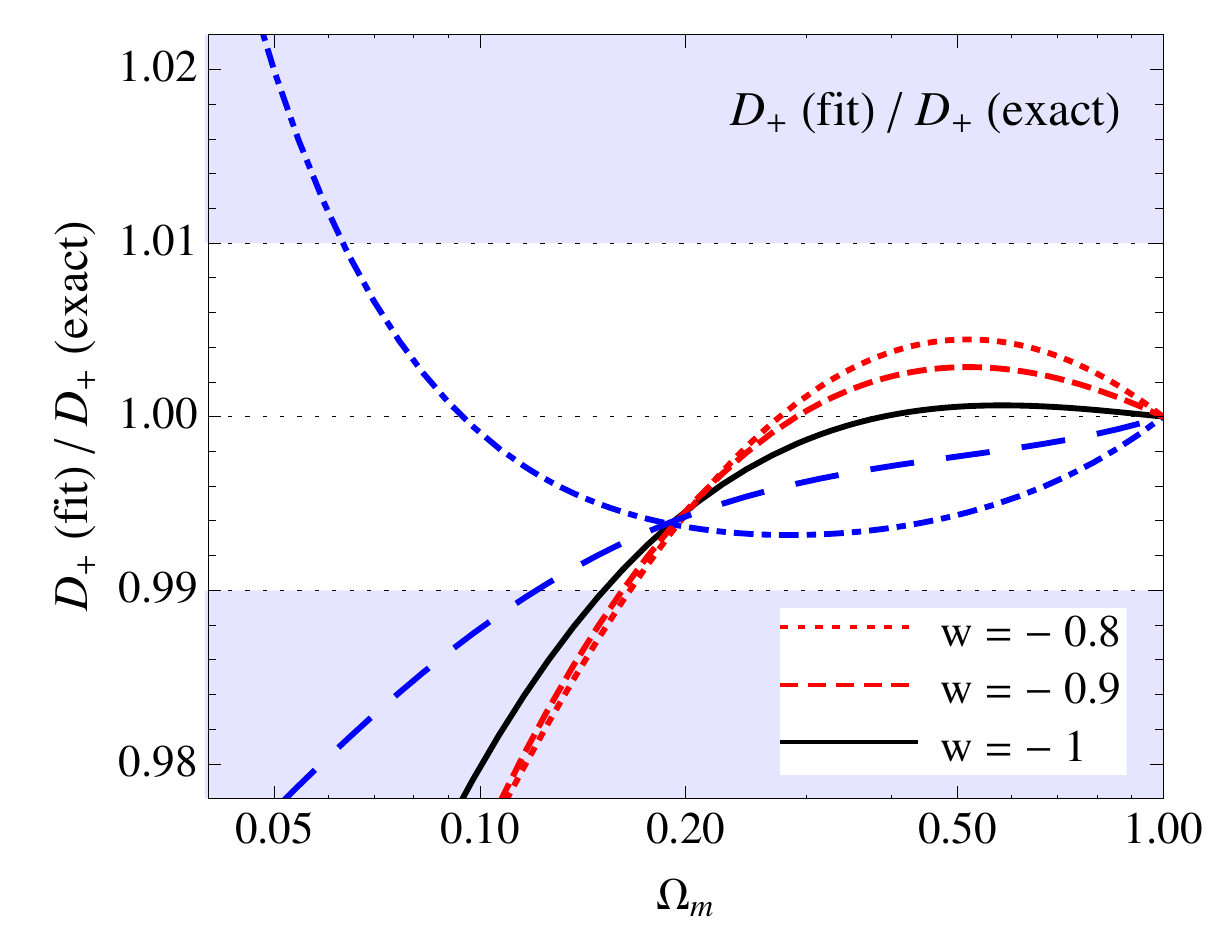}}
{\includegraphics[width=0.49\textwidth]{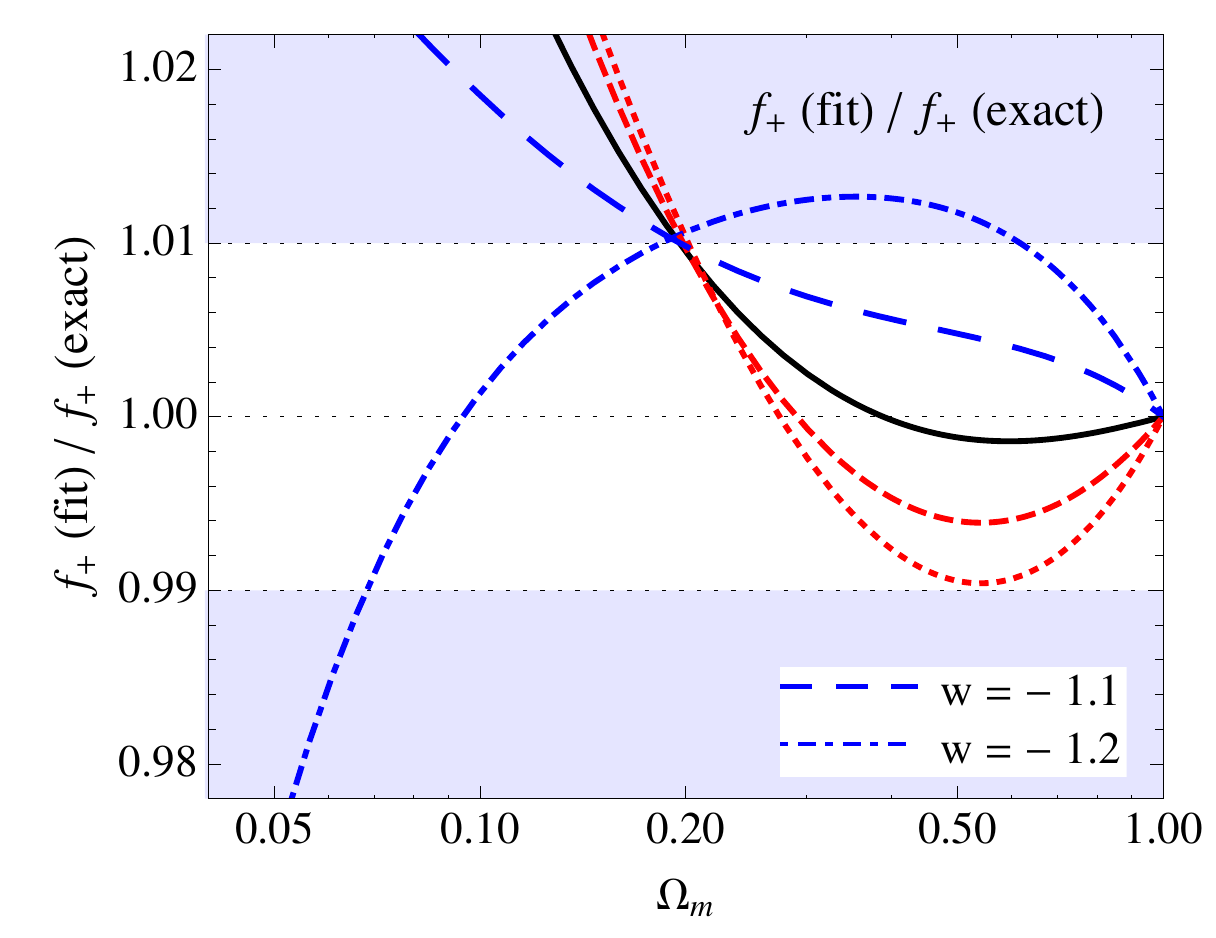}}
\caption{Ratio of the fitting functions for $D_+$ ({\em left panel}) and $f_+$ ({\em right panel}) to their exact values as a function of the value of $\Omega_m$.}
\label{fig:fits_Df}
\end{center}
\end{figure}


\section{Non-linear evolution}
\label{sec:nonlinear}

The effects of a dominant quintessence component at low redshift are clearly not entirely captured by the linear growth factor. In particular, as quintessence clusters on all observable scales, we expect it to significantly affect the {\em nonlinear} evolution of structures. Thus, a proper description of the density correlators in the mildly nonlinear regime is required. Indeed, current and future redshift and weak lensing surveys will target this range of scales with great accuracy.

As a first step in this direction, in this section we discuss the equations of motion~(\ref{continuity_tot}) and (\ref{euler_tot}) in the framework of EPT. In particular, we will derive their second-order solutions and use these to derive the tree-level expression for the bispectrum of the total density fluctuation $\delta$, a valid approximation at large-scales. We leave for future work the study of nonlinear corrections to the density power spectrum and the bispectrum.

\subsection{Perturbative expansion}


In perturbation theory, it is useful to rewrite the fluid equations~(\ref{continuity_tot}) and (\ref{euler_tot}) using the growing function $D_+$ as time. In particular, we define $\eta \equiv \ln D_+$.
Moreover, it is convenient to rescale $\theta$ by defining
\be
\tt \equiv  - \frac{C }{\HH f_+} \theta \;, \label{Theta_def}
\ee
such that, according to eq.~\eqref{v_delta}, at linear order $\tt=\delta$, \ie
\be
\tt^{\rm lin}_{\vec k} (\eta) =  \delta_{\vec k}^{\rm lin} (\eta)= D_+ (\eta) \delta^{\rm in}_{\vec k} \;.  \label{theta_lin}
\ee
Note, again, that  the definition of $\tt$, eq.~\eqref{Theta_def}, contains also a factor $C$ at the numerator, which is absent in the standard case.

In terms of $\tt$, the evolution equations \eqref{continuity_tot} and \eqref{euler_tot}  become
\begin{align}
&\frac{\partial \delta_{\vec k}}{\partial \eta} -   \tt_{\vec k} = \frac{\alpha(\vec q_1, \vec q_2)}C \tt_{\vec q_1} \delta_{\vec q_2} \label{continuity_tot_eta}\,,\\
&\frac{\partial \tt_{\vec k}}{\partial \eta} - \tt_{\vec k} - \frac{ f_- }{f_+^2} (\tt_{\vec k} - \delta_{\vec k}) =  \frac{\beta(\vec q_1, \vec q_2)}C \tt_{\vec q_1} \tt_{\vec q_2}\;, \label{euler_tot_eta}
\end{align}
where we have used eq.~\eqref{f-} to replace $-(3/2) \Omega_m C$ by $f_-$. Written in this form, the continuity and Euler equations are the same as those derived for a smooth quintessence or $\Lambda$CDM cosmology \cite{Bernardeau:2001qr} except that the kernels $\alpha$ and $\beta$ multiply the time-dependent function $C^{-1}$. Thus, the non-standard behavior of clustering quintessence is now encoded in the time dependence of the nonlinear couplings.

The solutions to these equations can be written in perturbative form as 
\be
\label{eq:PTdelta}
\delta =  \sum_{n=1}^{\infty} \delta^{(n)} \;, \qquad  \tt = \sum_{n=1}^{\infty} \tt^{(n)} \;,
\ee
where $\delta^{(1)} \equiv \delta^{\rm lin}$ and $\tt^{(1)} \equiv \tt^{\rm lin}$.
The $n$-th order solutions are proportional to $n$-th powers of the initial density $\d^{\rm in}$ and can be parameterized in terms of the kernels $F_n$ and $G_n$ and of powers of the linear growing mode solution $D_+$ as \cite{Bernardeau:2001qr}
\begin{align}
\delta^{(n)}_{\vec k}(\eta) &= F_n ( \vec q_1, \ldots ,\vec q_n; \eta) \, D_+^n(\eta) \, \delta_{\vec q_1}^{\rm in} \ldots \delta_{\vec q_n}^{\rm in}  \;, \label{F_n} \\
\tt^{(n)}_{\vec k}(\eta) &=  G_n ( \vec q_1, \ldots ,\vec q_n; \eta) \, D_+^n(\eta)   \,\delta_{\vec q_1}^{\rm in} \ldots \delta_{\vec q_n}^{\rm in} \;,\label{G_n}
\end{align}
where $F_n$ and $G_n$ are homogeneous functions of degree zero of the momenta $\vec q_1, \ldots, \vec q_n$, with $F_1=G_1=1$. In these expressions a multiple integration over the momenta $ \int d^3 q_1 \ldots d^3 q_n \delta_D(\vec k - \sum_{i=1}^n\vec q_i  )$ is implied on the right hand side. In the special case of matter dominance $D_+ = a$ and $F_n$ and $G_n$ become time-independent and can be constructed from algebraic recursion relations \cite{Goroff:1986ep}. In general, however, the kernels are time-dependent.

As already pointed-out in section~\ref{sec:eom}, since dark matter and quintessence are comoving, the system of equations involving $\delta$ and $\tt$ (or $\theta$) can be closed and we do not need to separately study the dark matter and quintessence to compute their evolution. 
However, it is interesting to compare the evolution of $\delta$ with that of the dark matter and quintessence density contrasts, respectively $\delta_m$ and $\delta_Q$. Using the definition of $\tt$,  eq.~\eqref{Theta_def}, eqs.~\eqref{continuity_m_FS} and \eqref{continuity_Q_FS} can be rewritten as 
\begin{align}
&\frac{\partial \delta_{m,\vec k}}{\partial \eta} -   \frac1C \tt_{\vec k} = \frac{\alpha(\vec q_1, \vec q_2)}C \tt_{\vec q_1} \delta_{m,\vec q_2} \label{continuity_m_eta}\,, \\
&\frac{\partial \delta_{Q,\vec k}}{\partial \eta}  -\frac{3 w}{f_+} \delta_{Q,\vec k} - \frac{1+w}{C}  \tt_{\vec k} = \frac{\alpha(\vec q_1, \vec q_2)}{C} \tt_{\vec q_1} \delta_{Q,\vec q_2} \label{continuity_Q_eta}\,.
\end{align}
One can then write the solutions to these equations perturbatively, similarly to what done above.


\subsection{Second-order solutions}

We now derive the second-order growing solutions in perturbation theory for the total density contrast $\delta$ and the velocity divergence $\tt$. Thus, we replace on the right hand side of eqs.~\eqref{continuity_tot_eta} and \eqref{euler_tot_eta} the linear growing solutions $\delta^{\rm lin}_{\vec k} (\eta) =  D_+ (\eta) \delta^{\rm in}_{\vec k}$ and $\tt^{\rm lin}_{\vec k} (\eta) =  D_+ (\eta) \delta^{\rm in}_{\vec k} $.
This yields, after symmetrization over the momenta,
\begin{align}
&\frac{\partial \d^{(2)}_{\vec k}}{\partial \eta} -   \tt^{(2)}_{\vec k} = \frac{D_+^2}C\, \alpha_s(\qv_1,\qv_2)\, \delta^{\rm in}_{\vec q_1}\, \delta^{\rm in}_{\vec q_2}  \label{continuity_tot_eta_ts}\,,\\
&\frac{\partial \tt^{(2)}_{\vec k}}{\partial \eta} - \tt^{(2)}_{\vec k} - \frac{ f_- }{ f_+^2}\, (\tt^{(2)}_{\vec k} - \d^{(2)}_{\vec k}) =  \frac{D_+^2}C\, \beta(\qv_1,\qv_2)\, \delta^{\rm in}_{\vec q_1}\, \delta^{\rm in}_{\vec q_2} \;, \label{euler_tot_eta_ts}
\end{align}
where $\alpha_s(\qv_1,\qv_2)\equiv[\alpha(\qv_1,\qv_2)+\alpha(\qv_2,\qv_1)]/2$ is the symmetrized projection of $\alpha(\qv_1,\qv_2)$.

The solutions to these equations are usually parameterized in terms of $F_2$ and $G_2$, defined by eqs.~\eqref{F_n} and \eqref{G_n} for $n=2$ as
\begin{align}
\d^{(2)}_{\vec k} (\eta) & = F_{2}(\vec q_1, \vec q_2; \eta) D_+^2(\eta) \delta^{\rm in}_{\vec q_1} \delta^{\rm in}_{\vec q_2}\;, \label{F2} \\
\tt^{(2)}_{\vec k} (\eta) & = G_{2}(\vec q_1, \vec q_2; \eta)D_+^2(\eta) \delta^{\rm in}_{\vec q_1} \delta^{\rm in}_{\vec q_2}\;,\label{G2}
\end{align}
so that the evolution equations \eqref{continuity_tot_eta_ts} and \eqref{euler_tot_eta_ts} become
\begin{align}
&\frac{\partial F_2}{\partial \eta} + 2 F_2 -   G_2  = \frac{\alpha_s}C  \label{continuity_tot_eta_ts_F}\,,\\
&\frac{\partial G_2}{\partial \eta} + G_2  - \frac{ f_-  }{f_+^2}\, (G_2 - F_2) =  \frac{\beta}C  \;. \label{euler_tot_eta_ts_G}
\end{align}

One can verify that the solutions to these equations can be formally written as
\begin{align}
F_2 &= \int_{-\infty}^\eta d \tilde \eta \frac{e^{\tilde \eta-\eta}}{ C(\tilde \eta)} \left[  \frac{3 \alpha_s + 2 \beta }{5} 
+  \frac{D_-(\eta)}{D_-(\tilde \eta)} e^{\tilde \eta- \eta} \frac{2 \alpha_s - 2 \beta }{5} \right]  \;, \label{sol_F2}\\
G_2 &= \int_{-\infty}^\eta d \tilde \eta \frac{e^{\tilde \eta-\eta}}{C(\tilde \eta)} \left[ \frac{3 \alpha_s + 2 \beta }{5}  
- \frac{3}{2} \frac{f_-(\eta) f_+(\tilde \eta) D_-(\eta)}{f_-(\tilde \eta) f_+(\eta) D_-(\tilde \eta)} e^{\tilde \eta- \eta} \frac{2 \alpha_s - 2 \beta }{5} \right]  \;,\label{sol_G2}
\end{align}
where we have used $D_+(\eta) = e^{\eta}$. Note that these solutions take the same form as in the smooth case, except for the factor $C(\tilde \eta)$ at the denominator inside the time integrals, which in the smooth case is absent. 

Before discussing the clustering case, let us consider more carefully the smooth case. For $C=1$ the terms that do not contain the linear decaying solution $D_-$ can be explicitly integrated to give a constant in time. Instead, the terms containing $D_-$ are generally time-dependent, so that each kernel---$F_2$ and $G_2$---is characterized by a function expressing their time-dependence. For instance, one can use the angular average of $F_2$ and $G_2$, defining the functions $\nu_2$ and $\mu_2$ respectively as $\nu_2 \equiv 2 \langle F_2 \rangle$ and $\mu_2 \equiv 2 \langle G_2 \rangle$ \cite{Bernardeau:1992zw}. These quantities also represent the second-order vertices of the nonlinear density contrast in the spherical collapse model (see appendix~\ref{app:vertices} for a generalization  to higher order in perturbation theory).
During matter dominance these functions become constant, \ie~$\nu_2 = 34/21$ and $\mu_2 = 26/21$. Indeed, in this case $f_+=1$, $f_-=-3/2$ so that $D_-(\eta) \propto e^{-3\eta/2 }$, and the integrals can be solved to give constant $F_2$ and $G_2$. As we will see, for a smooth quintessence or in a $\Lambda$CDM cosmology, $f_-$ and $f_+$ present a very weak dependence on time and the functions $\nu_2$ and $\mu_2$ are very well approximated by their constant Einstein-de Sitter values.

However, due to the presence of $C(\tilde \eta)$ in eqs.~\eqref{F2} and \eqref{G2}, in the case of clustering quintessence we need a third time-dependent function to fully characterize $F_2$ and $G_2$. We choose to parameterize this additional time dependence with the function $\epsilon$, defined as 
\be
\epsilon(\eta) \equiv  \int_{-\infty}^\eta d \tilde \eta e^{\tilde \eta-\eta} \left(1-\frac{1} {C(\tilde \eta)} \right) = 1- e^{-\eta} \int_{-\infty}^\eta d \tilde \eta  \frac{e^{\tilde \eta}} {C(\tilde \eta)}\;, \label{epsilon}
\ee
such that it does not involve the decaying mode $D_-$ and vanishes when $C=1$. With this definition $F_2$ and $G_2$ read
\begin{align}
F_2 (\vec q_1, \vec q_2; \eta) &= -\frac12 \left[1- \epsilon(\eta) - \frac{3\nu_2(\eta)}{2}  \right] \alpha_s(\vec q_1, \vec q_2) + \frac32 \left[1-\epsilon(\eta) - \frac{\nu_2(\eta)}{2} \right] \beta(\vec q_1, \vec q_2) \;, \label{F2_alpha} \\ 
G_2 (\vec q_1, \vec q_2; \eta) &= -\frac12 \left[1- \epsilon(\eta) - \frac{3  \mu_2(\eta)}{2} \right] \alpha_s(\vec q_1, \vec q_2) + \frac32 \left[1-\epsilon(\eta) - \frac{\mu_2(\eta)}{2} \right] \beta(\vec q_1, \vec q_2) \;.\label{G2_alpha}
\end{align}
The evolution equations for $\nu_2$ and $\mu_2$ can be found by averaging over the angles eqs.~\eqref{continuity_tot_eta_ts_F} and \eqref{euler_tot_eta_ts_G}, using that $\langle \alpha_s \rangle=1$ and $\langle \beta \rangle =1/3$. They read
\begin{align}
\frac{\partial \nu_2}{\partial\eta}+2\,\nu_2-\mu_2 & =  \frac{2}{C}\, , \label{nu_evol}\\
\frac{\partial \mu_2}{\partial\eta}+\,\mu_2 - \frac{f_-}{f_+^2}(\mu_2-\nu_2) & =  \frac{2}{3\,C} \label{mu_evol} \, ,
\end{align}
with the initial conditions set by their solutions in matter dominance. An integral form of these functions can be given by taking the angular average of eqs.~\eqref{sol_F2} and \eqref{sol_G2} and one can verify that they depend both on the linear growing {\em and} decaying solutions.

It is possible to relate $\epsilon$ to matter or quintessence fluctuations. Indeed, using $D_+ = e^\eta$ one can formally integrate eq.~\eqref{eq:Dm}. This yields
\be
D_{m,+} =  \int_{-\infty}^\eta d \tilde \eta \frac{e^{\tilde \eta}}{ C(\tilde \eta)}\;,
\ee
so that the second equality in eq.~\eqref{epsilon} can be rewritten as
\be
\epsilon =  1-\frac{D_{m,+}}{D_+}  =  \frac{\Omega_Q}{\Omega_m} \frac{D_{Q,+} }{D_+}\;. \label{eps_Dm}
\ee
Thus, $\epsilon$ is the ratio between the linear quintessence perturbations and the total perturbations, $\epsilon = {\delta \rho_Q^{\rm lin}}/{\delta \rho^{\rm lin}}$.
\begin{figure}[t]
\begin{center}
{\includegraphics[width=0.49\textwidth]{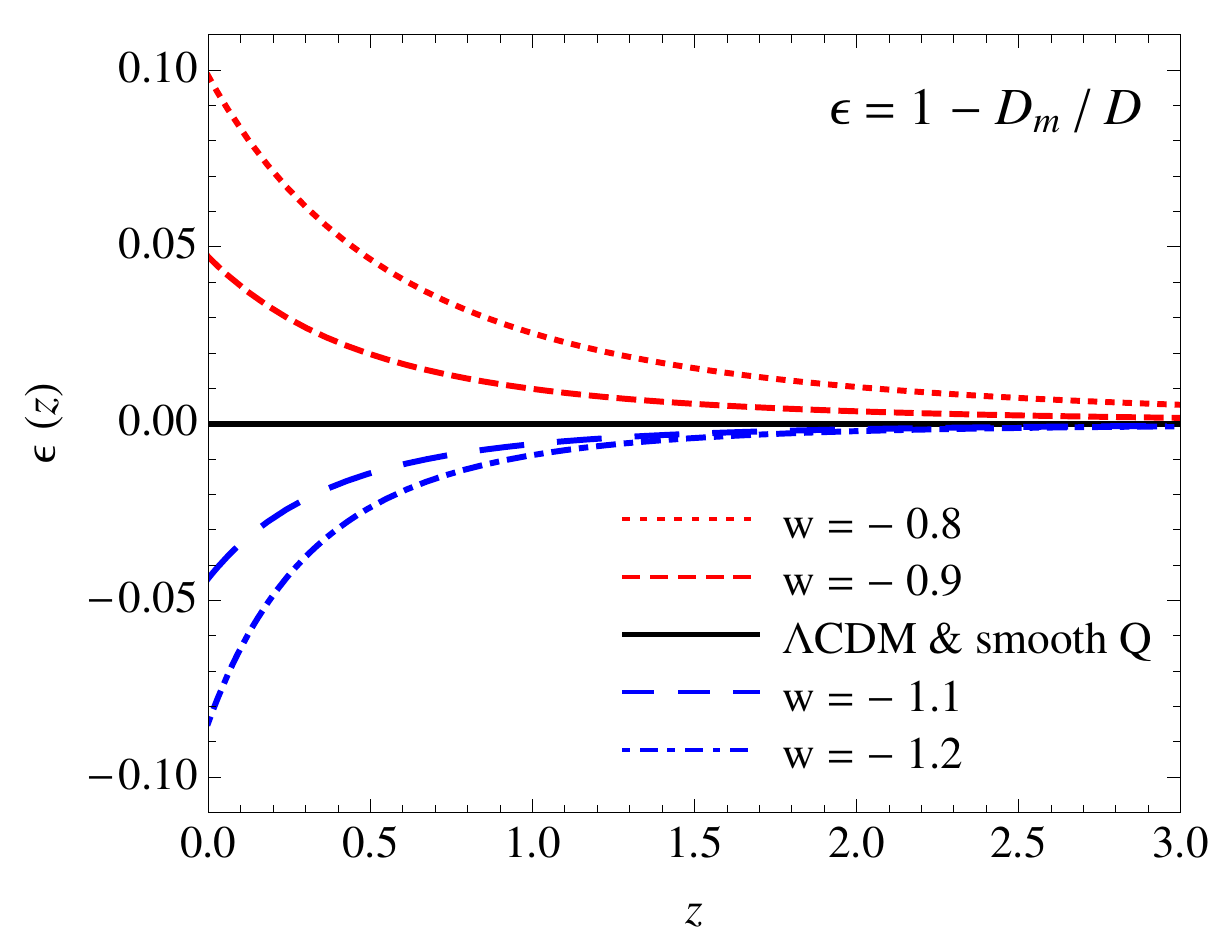}}
\caption{The function $\epsilon$ defined in eq.~\eqref{epsilon} as a function of redshift for several values of $w$. The black continuous line corresponds to the zero value of $\epsilon$ in $\Lambda$CDM and smooth quintessence cosmologies.}
\label{fig:epsilon}
\end{center}
\end{figure}
In Fig.~\ref{fig:epsilon} we show the function $\epsilon$ as a function of the redshift for several values of $w$. As quintessence becomes a non negligible component of the Universe, $\epsilon$ becomes larger.
Today it is of the order of $\sim 0.05$ for $|w+1| = 0.1$. As we will see, this translates into similar corrections to the tree-level expression of the total bispectrum. 

\begin{figure}[t]
\begin{center}
{\includegraphics[width=0.49\textwidth]{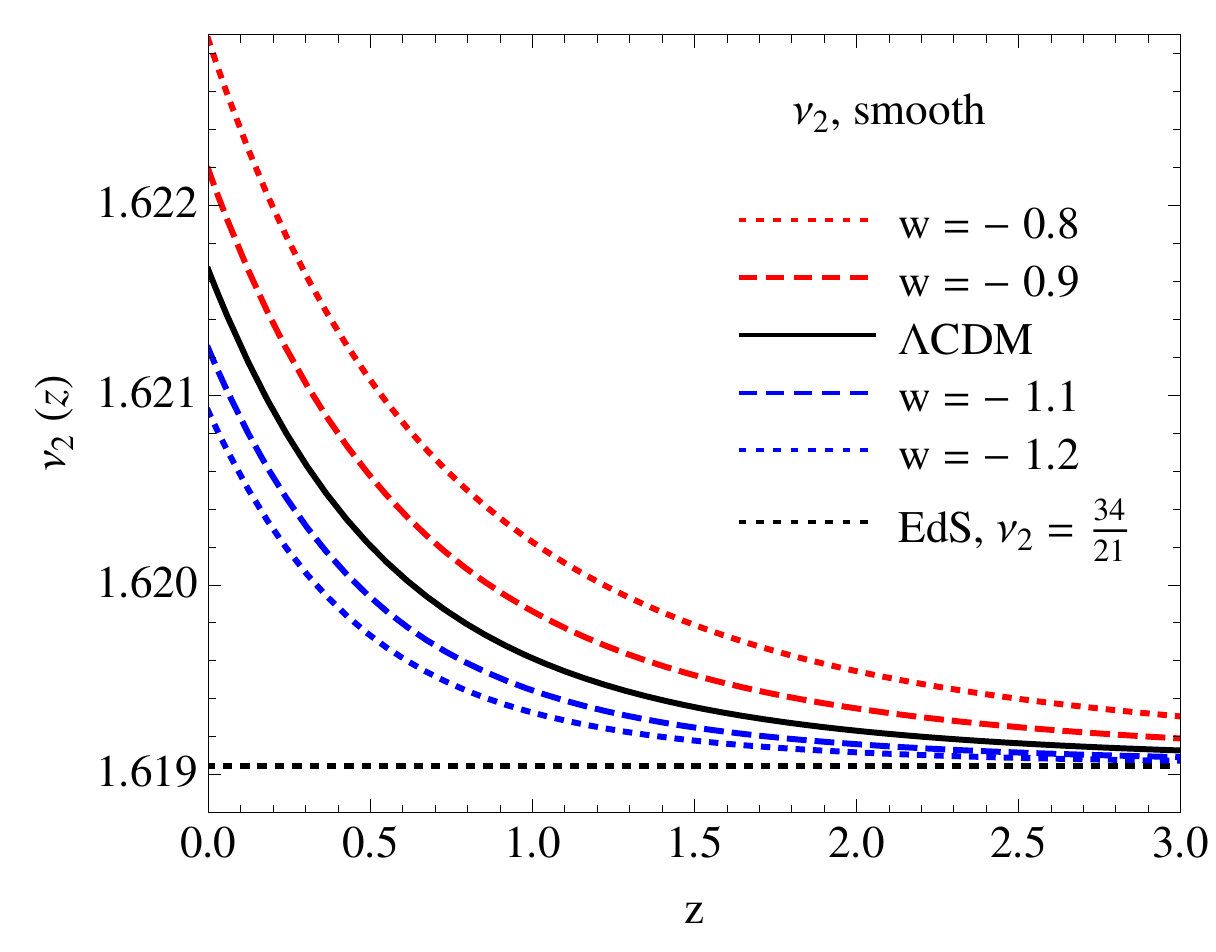}}
{\includegraphics[width=0.49\textwidth]{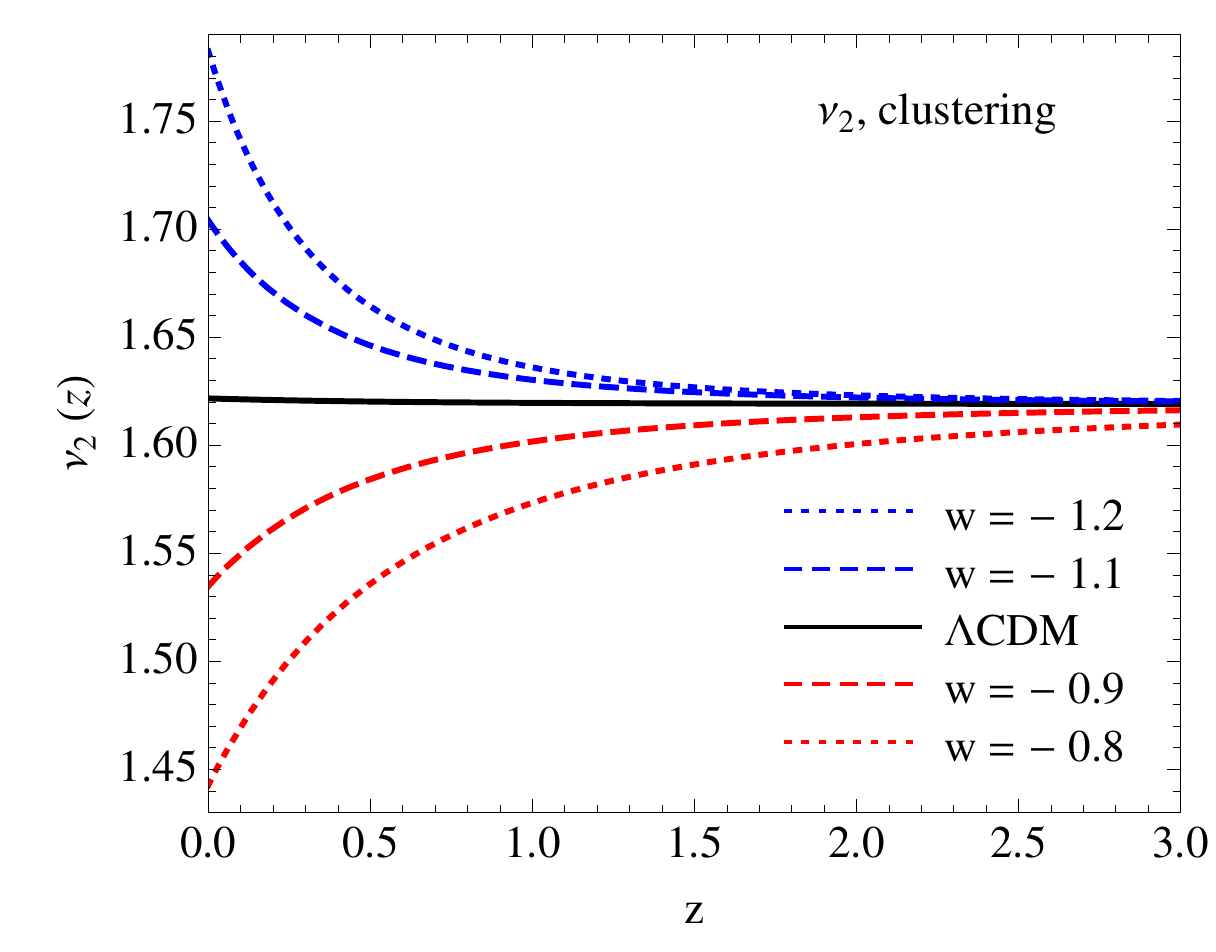}}\\
{\includegraphics[width=0.49\textwidth]{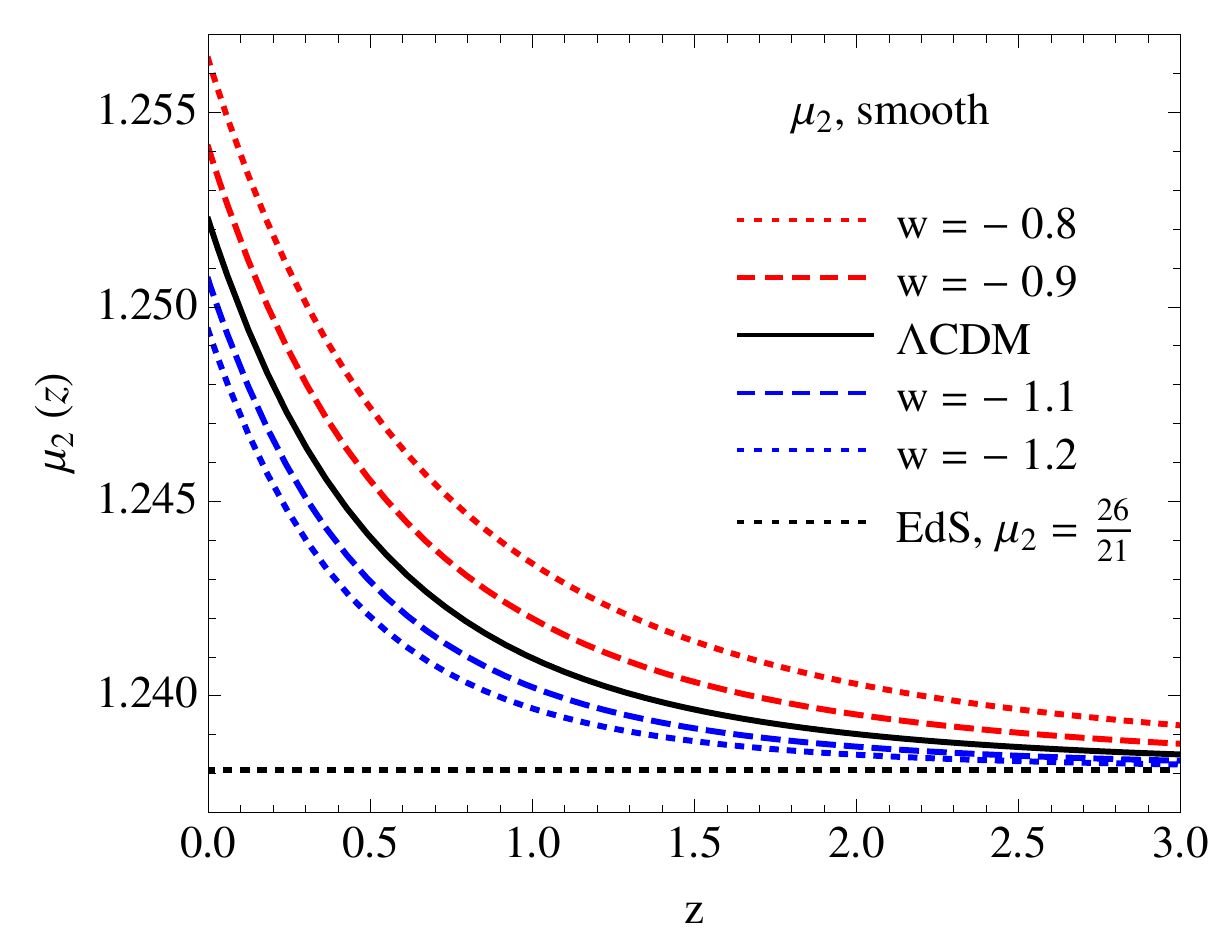}}
{\includegraphics[width=0.49\textwidth]{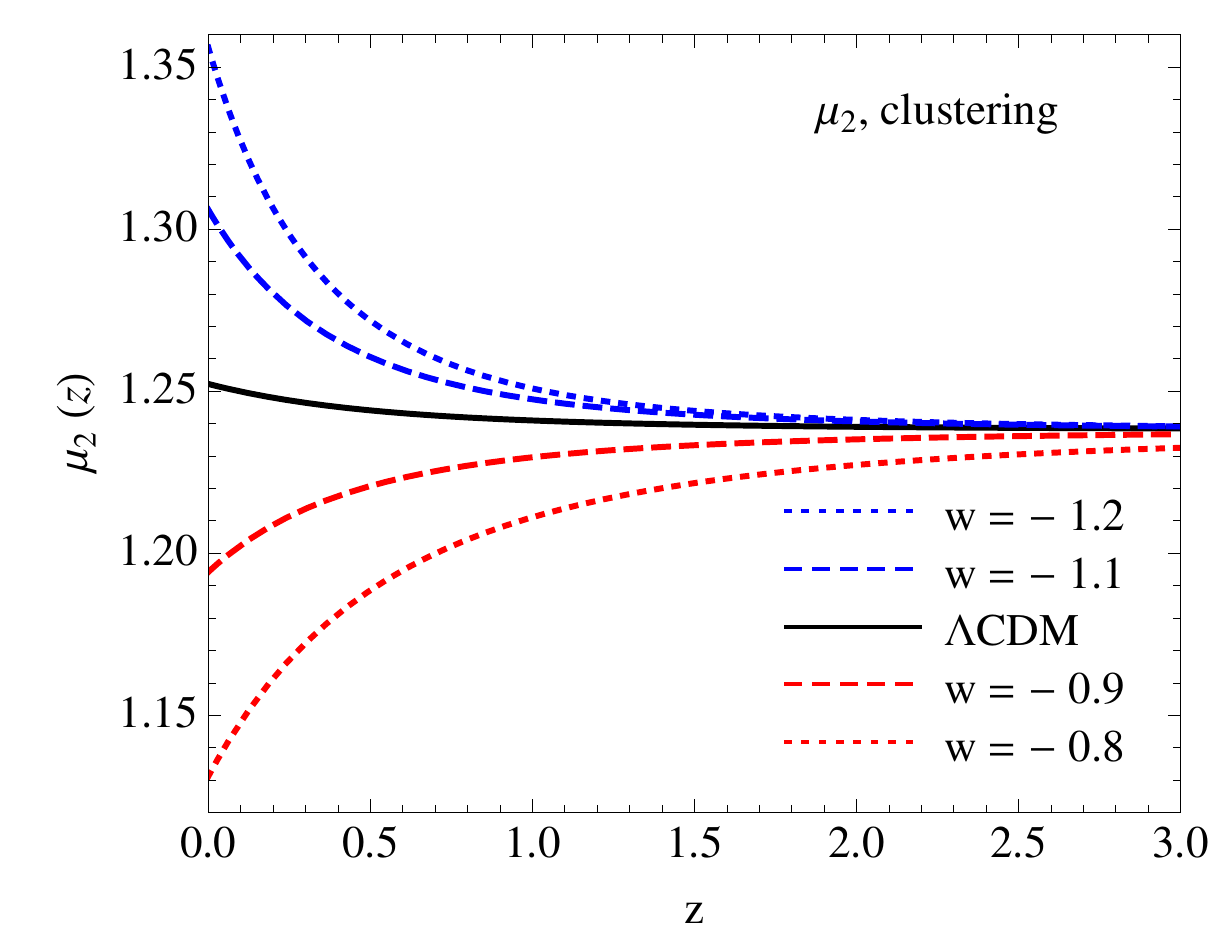}}
\caption{{\em Upper panels}: vertex $\nu_2$ for the pseudo-matter perturbations in the spherical collapse approximation in the smooth ({\em left}) and clustering ({\em right}) quintessence scenarios. {\em Lower panels}: same plots for the vertex $\mu_2$.}
\label{fig:nu2mu2}
\end{center}
\end{figure}
In Fig.~\ref{fig:nu2mu2} we show the effect of quintessence on the vertices $\nu_2$ ({\em upper panels}) and $\mu_2$ ({\em lower panels}) as a function of redshift and for different values of $w$. In particular, on the left panels we show the solutions in the smooth case, where we can notice the relatively small corrections induced by the different cosmological background to the Einstein-de-Sitter constant solutions. On the other hand, in the clustering case, shown in the right panels, the corrections are significantly larger, of the order of 5\% for $|1+w|\simeq 0.1$ for both $\nu_2$ and $\mu_2$.  These corrections can be estimated using eq.~\eqref{corrections}.

As $\epsilon$ explicitly characterizes the contribution of quintessence perturbations to the total one, it is useful to estimate the deviation of $\nu_2$ and $\mu_2$ from their values in the smooth case in terms of this quantity. The exact explicit dependence cannot be written down in a simple form. However, we find that a very good approximation is given by
\be
\nu_2 = \nu_{2,\rm smooth} - (1.669 - 0.205\, w)\,\epsilon\;, \qquad \mu_2=  \mu_{2,\rm smooth} - (1.3+0.033\, w) \,\epsilon\;, \label{corrections}
\ee
where $\nu_{2,\rm smooth}$ and $\mu_{2,\rm smooth}$ are computed in the smooth case. As shown in Fig.~\ref{fig:fits_numu}, for $\Omega_m \ge 0.2$, the error made using these approximations is less than about $0.1\%$ for $-1.2 \le w \le -0.8$.
\begin{figure}[t]
\begin{center}
{\includegraphics[width=0.49\textwidth]{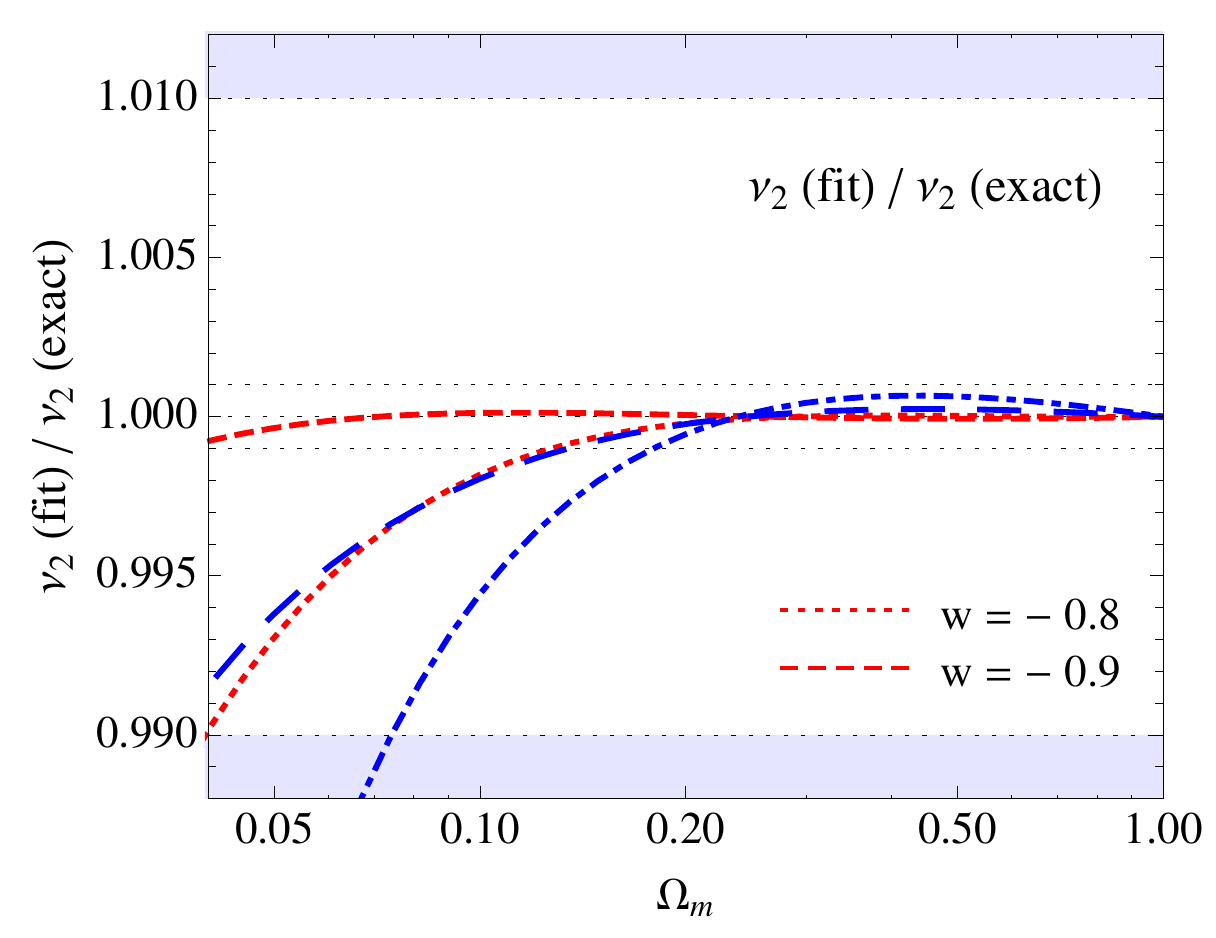}}
{\includegraphics[width=0.49\textwidth]{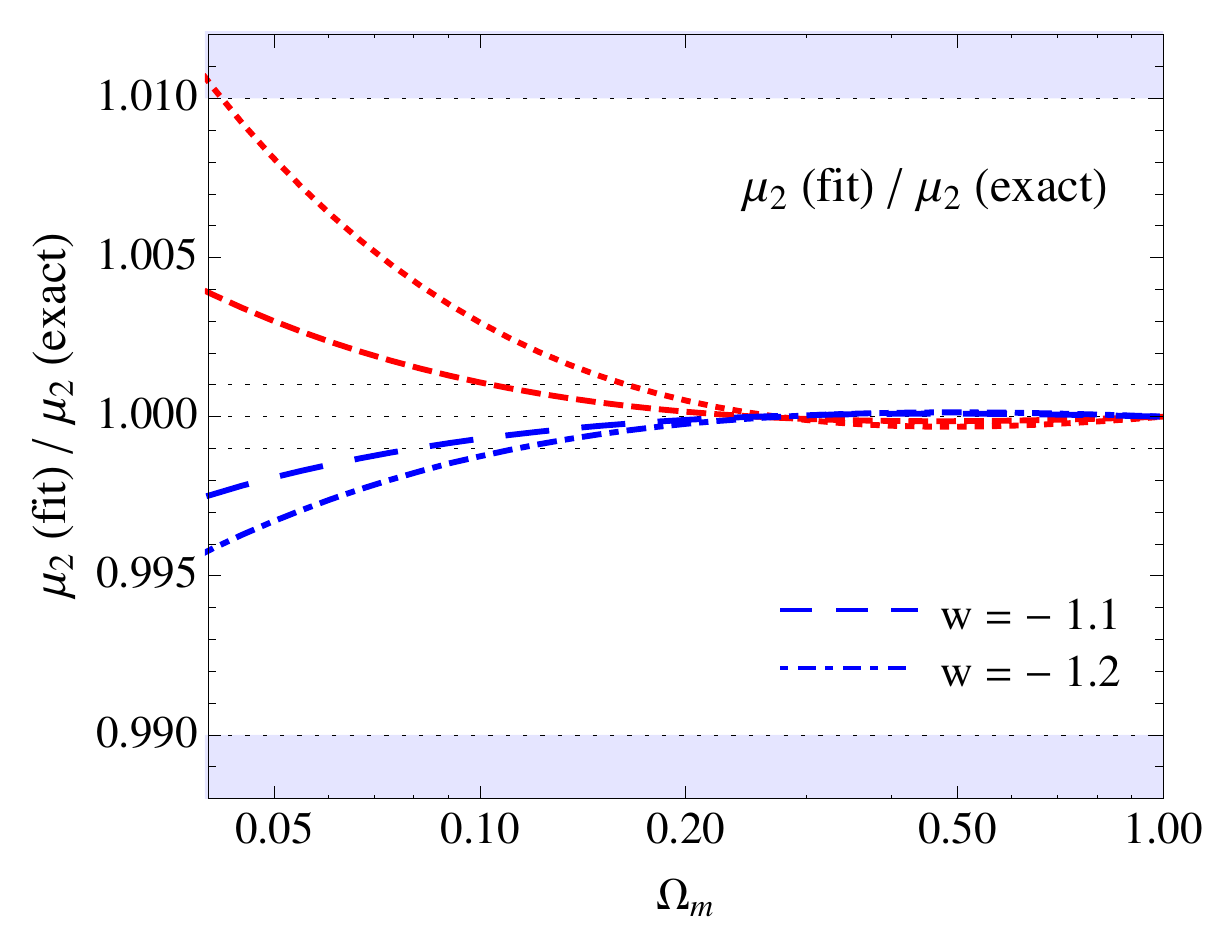}}
\caption{Ratio of the approximations for $\nu_2$ ({\em left panel}) and $\mu_2$ ({\em right panel}) in Eq.~(\ref{corrections}) to their exact values as a function of the value of $\Omega_m$.}
\label{fig:fits_numu}
\end{center}
\end{figure}

Let us rewrite $\alpha_s(\qv_1,\qv_2)$ and $\beta(\qv_1,\qv_2)$ by making more explicit the dependence on the scalar product $\hat q_1 \cdot \hat q_2$, where $\hat q_i \equiv \vec q_i/q_i$. From their definitions, eqs.~\eqref{alpha_def} and \eqref{beta_def}, one finds 
\begin{align}
\alpha_s (\vec q_1,\vec q_2) &= 1 + \frac{\hat q_1 \cdot \hat q_2}{2} \left(\frac{q_1}{q_2} + \frac{q_2}{q_1} \right)\;, \\
\beta (\vec q_1,\vec q_2) &=  \frac{\hat q_1 \cdot \hat q_2}{2} \left(\frac{q_1}{q_2} + \frac{q_2}{q_1} \right) + (\hat q_1 \cdot \hat q_2)^2\;.
\end{align}
Using these expressions we can reorganize the kernels $F_2$ and $G_2$ as multipolar expansions, in terms of their monopole, dipole and quadrupole contributions \cite{Bouchet:1992uh,Scoccimarro:2004tg}
\begin{align}
F_2 (\vec q_1, \vec q_2) &= \frac{\nu_2}2 + (1-\epsilon) \frac{\hat q_1 \cdot \hat q_2}{2} \left(\frac{q_1}{q_2} + \frac{q_2}{q_1} \right) - \frac12 \left(1-\epsilon - \frac{\nu_2}{2} \right)\left[1-3 (\hat q_1 \cdot \hat q_2)^2 \right]\;, \label{F2_multipole} \\ 
G_2 (\vec q_1, \vec q_2) &= \frac{\mu_2}2 + (1-\epsilon) \frac{\hat q_1 \cdot \hat q_2}{2} \left(\frac{q_1}{q_2} + \frac{q_2}{q_1} \right) - \frac12 \left(1-\epsilon - \frac{\mu_2}{2} \right)\left[1-3 (\hat q_1 \cdot \hat q_2)^2 \right] \label{G2_multipole}\;.
\end{align}
As already mentioned, the first term corresponds to the second-order evolution in the spherical collapse dynamics. The dipole of the middle term is due to the nonlinear transformation from following mass elements in the Lagrangian description to describing the dynamics of the fluids in the Eulerian formalism. In the standard $\Lambda$CDM and smooth cases, this term is time-independent. In the clustering case quintessence contributes to the mass of a clustered object. As this mass is not conserved---the energy density does not scale as the volume---this nonlinear transformation receives a time-dependent correction proportional to $\epsilon$. The last term is due to the tidal gravitational field. Indeed, the dipole can be rewritten as
\be
\left[1-3 (\hat q_1 \cdot \hat q_2)^2 \right] \delta_{\vec q_1}  \delta_{\vec q_2} = -3\, (\hat q_1^i \hat q_1^j - \frac13 \delta^{ij})  \delta_{\vec q_1}  ( \hat q_2^i \hat  q_2^j - \frac13 \delta^{ij}) \delta_{\vec q_2} \;,
\ee
where $(\hat k^i \hat k^j - \frac13 \delta^{ij})  \delta_{\vec k} $ is the Fourier representation of the tidal gravitational field $(\partial^i \partial^j - \frac13 \delta^{ij} \nabla^2 )  \Phi $.

Let us now consider the nonlinear correction to the density contrast of dark matter only, $\delta_m$. We can plug the linear growing solutions \eqref{theta_lin} and \eqref{deltam_lin} on the right hand side of eq.~\eqref{continuity_m_eta}. This yields the second-order continuity equation for matter perturbations,
\be
\frac{\partial \d^{(2)}_{m,\vec k}}{\partial \eta} -   \frac1C \, \tt^{(2)}_{\vec k} = \frac{D_+\,D_{m,+}}C \alpha_s(\qv_1,\qv_2)\,\delta^{\rm in}_{\vec q_1} \delta^{\rm in}_{\vec q_2}  \label{continuity_m_eta_ts} \;.
\ee
Making use of eq.~\eqref{eq:Dm} we can rewrite the partial derivative with respect to $\eta = \log D_+$ as a derivative with respect to $\log D_{m,+}$.
Then, after multiplying this equation by $D_m/D$, we can use eq.~\eqref{eps_Dm} to rewrite the continuity equation as 
\be
\frac{\partial \d^{(2)}_{m,\vec k}}{\partial \log D_{m,+}} -  (1-\epsilon) \, \tt^{(2)}_{\vec k} = D_{m,+}^2 \alpha_s(\qv_1,\qv_2)\,\delta^{\rm in}_{\vec q_1} \delta^{\rm in}_{\vec q_2}   \;. \label{cont_m_new}
\ee
The velocity divergence on the left hand side is  now weighted by a factor $1-\epsilon = D_m/D$. 
Note also that the $1/C$ time dependence in the vertex present in the evolution of the total perturbation (eq.~\eqref{continuity_tot_eta_ts}) is now absent. As a consequence, we can describe the full time dependence of $F_{m,2}$ in terms of its angular average, $\nu_{m,2} \equiv 2 \langle F_{m,2} \rangle$.

The solution to this equation can be parameterized as
\be
\d^{(2)}_{m,\,\vec k} (\eta)  = F_{m,2}(\vec q_1, \vec q_2; \eta) D_{m,+}^2(\eta) \delta^{\rm in}_{\vec q_1} \delta^{\rm in}_{\vec q_2}\;, \label{Fm2} 
\ee
with 
\be
F_{m,2}(\vec q_1, \vec q_2)  = -\frac12 \left(1- \frac32 \nu_{m,2} \right) \alpha_s(\vec q_1, \vec q_2) + \frac32 \left(1 - \frac{\nu_{m,2}}{2} \right) \beta(\vec q_1, \vec q_2)\;.\label{eq:F2m}
\ee
The evolution equations for  $\nu_{m,2}$, obtained by taking the average of eq.~\eqref{cont_m_new}, reads
\be
\frac{\partial \nu_{m,2}}{\partial\log D_{m,+}} + 2 \nu_{m,2}  - (1-\epsilon) \mu_
2 =  2    \, .
\ee
Note that if we rewrite $F_{m,2}$ in eq.~\eqref{eq:F2m} in terms of multipoles as in eq.~\eqref{F2_multipole}, the $\epsilon$ correction in the dipole term is absent. As already mentioned, this is a consequence of matter conservation and its relation to the transformation from Lagrangian to Eulerian space. More precisely, to the fact that the Jacobian of the coordinate transformation from Eulerian position $\xv$ to Lagrangian position $\qv$ is given by matter conservation, $\bar{\rho}_m(1+\d_m)d^3x=\bar{\rho}_m d^3q$, as $J = (1+ \d_m(\xv))^{-1}$.\\


\section{Power spectrum and bispectrum}
\label{sec:bispectrum}

In this section we consider the leading-order contributions in perturbation theory to the power spectrum and bispectrum of the total density contrast $\delta$, which provide good approximations on large scales. The power spectrum $P(k,\eta)$ is defined as
\be
\langle \delta_{\vec k} (\eta) \delta_{\vec k'} (\eta) \rangle \equiv \d_D(\vec k + \vec k') P(k,\eta)\;.
\ee
Assuming Gaussian initial conditions,  it is possible to derive a perturbative expression for the nonlinear power spectrum  in terms of the expansion in eq.~\eqref{eq:PTdelta}. We have
\be
\langle \delta_{\vec k}\,\delta_{\vec k'}\rangle = \langle \d_{\vec k}^{\rm lin}\,\d_{\vec k'}^{\rm lin}\rangle+\langle \d_{\vec k}^{(2)}\,\d_{\vec k'}^{(2)}\rangle+\langle \d_{\vec k}^{\rm lin}\,\d_{\kv'}^{(3)}\rangle+\dots\;,
\ee
where the first term on the right hand side defines the linear power spectrum, $\langle \d_{\kv}^{\rm lin}\,\d_{\kv'}^{\rm lin} \rangle \equiv \d_D(\vec k+\vec k') P_{\rm lin}(k,\eta)$, while the second and the third ones constitute {\em one-loop} corrections of the same order, \ie~fourth-order in the initial field $\d^{\rm in}$. 

In terms of the power spectrum we have
\be
P(k,\eta)=P_{\rm lin}(k,\eta)+P_{\rm 1-loop}(k,\eta)+\dots\;, \label{P_exp}
\ee
where the linear contribution can be written as a function of the {\em initial} power spectrum, $\langle \d_{\kv}^{\rm in}\,\d_{\kv'}^{\rm in} \rangle \equiv \d_D(\vec k+\vec k') P_{\rm in}(k)$, simply as
\be
P_{\rm lin}(k,\eta) \equiv D_+^2 (\eta) P_{\rm in}(k)\;, \label{P_delta} \\
\ee
and where $P_{\rm 1-loop}(k,\eta)$ includes the two one-loop contributions mentioned above. Thus, at leading order in perturbation theory the effect of clustering quintessence is encoded in the linear growth function $D_+$. On smaller scales one needs to consider nonlinear corrections. A consistent evaluation of the one-loop corrections for the power spectrum requires both the second- as well as the {\em third}-order solutions for the density field. We leave this for future work.

The second-order solution $\delta^{(2)}$, on the other hand, provides the tree-level expression for the density bispectrum, the leading contribution in EPT. The bispectrum is defined as
\be
\langle \delta_{\vec k_1}(\eta) \delta_{\vec k_2}(\eta) \delta_{\vec k_3}(\eta) \rangle \equiv \d_D(\kv_1 + \kv_2 + \kv_3) B (k_1,k_2,k_3;\eta)\;.
\ee
Using Wick theorem, the tree-level solution for the bispectrum is given by
\be
B(k_1,k_2,k_3;\eta)=2\,F_2(\kv_1,\kv_2;\eta) \,P_{\rm lin}(k_1,\eta)\,P_{\rm lin}(k_2,\eta) +2~ {\rm cyclic}\,. \label{B_delta}
\ee
The time evolution depends on the linear growth factor $D_+$ via the linear power spectrum \eqref{P_delta} and on the functions $\epsilon(\eta)$ and $\nu_2(\eta)$ in the kernel $F_2$, eq.~\eqref{F2_alpha}.
For instance, for equilateral configurations, {\em i.e.} $k_1=k_2=k_3=k$, 
\begin{align}
\label{eq:Beq}
B(k,k,k;\eta) & =  6\,F_2(\kv,\kv;\eta)\,P_{\rm lin}^2(k,\eta)\\
& =  3\left\{\frac{9}{8}\,\nu_2(\eta)-\frac{5}{4}\left[1-\epsilon(\eta)\right]\right\}\,D_+^4(\eta)\,P_{\rm in}^2(k) \label{eq:Beq2}\,, 
\end{align}
where in the second equality we have used eq.~\eqref{P_delta} to express the time dependence of $P_{\rm lin}(k,\eta)$.

To highlight the shape-dependence of the bispectrum it is customary to introduce a {\em reduced} bispectrum $\Q(k_1,k_2,k_3)$ defined as
\be
\Q(k_1,k_2,k_3;\eta)\equiv \frac{B(k_1,k_2,k_3;\eta)}{P(k_1,\eta)\,P(k_2,\eta)+2~ {\rm cyclic}}\,. \label{Q}
\ee
In the tree-level approximation, this is independent of the linear growth factor, as one can verify by using eqs.~\eqref{P_exp}, \eqref{P_delta} and \eqref{B_delta} in \eqref{Q}, which yields
\be
\Q(k_1,k_2,k_3;\eta) = \frac{2\,F_2(\kv_1,\kv_2;\eta)\,P_{\rm in}(k_1)P_{\rm in}(k_2)+2~{\rm cyclic}}{P_{\rm in}(k_1)\,P_{\rm in}(k_2)+2~ {\rm cyclic}}\,.
\ee

In an Einstein-de Sitter universe (matter dominance), as $F_2$ is time-independent the reduced bispectrum is independent of redshift. Even in the $\Lambda$CDM and in the smooth quintessence scenarios, it shows a very mild time evolution.
By contrast, as already explained, in the clustering case the time-evolution of the functions $\epsilon$ and $\nu_2$ is important and $\Q$ significantly departs from the standard value. Thus, the reduced bispectrum is sensitive to clustering quintessence. In particular, while the effect of clustering quintessence on the power spectrum and bispectrum is expected to be strongly degenerate with $\Omega_m$ and $\sigma_8$, this degeneracy disappears in the reduced bispectrum.
Note that in {\em redshift space} the bispectrum depends on $F_2$ but also on the kernel for the velocity divergence $G_2$ which is affected through $\mu_2$ by corrections of similar magnitude (see appendix~\ref{app:vertices}).

In the specific case of the equilateral configuration the reduced bispectrum becomes
\be
\Q(k,k,k; \eta)=2\,F_2(\kv,\kv;\eta)=\frac{9}{8}\,\nu_2(\eta)-\frac{5}{4}\left[1-\epsilon(\eta)\right]\,,
\ee
which reduces to the constant value of $4/7$ at early times, during matter domination. Note that the normalized skewness, $S_3 = \langle \delta^3 \rangle/\langle \delta^2 \rangle^2$, is simply given by $S_3 = 3 \nu_2$, so that the effect of clustering quintessence on the skewness is, using eq.~\eqref{corrections}, of the order $\sim - 5 \epsilon$.

\begin{figure}[t]
\begin{center}
{\includegraphics[width=0.49\textwidth]{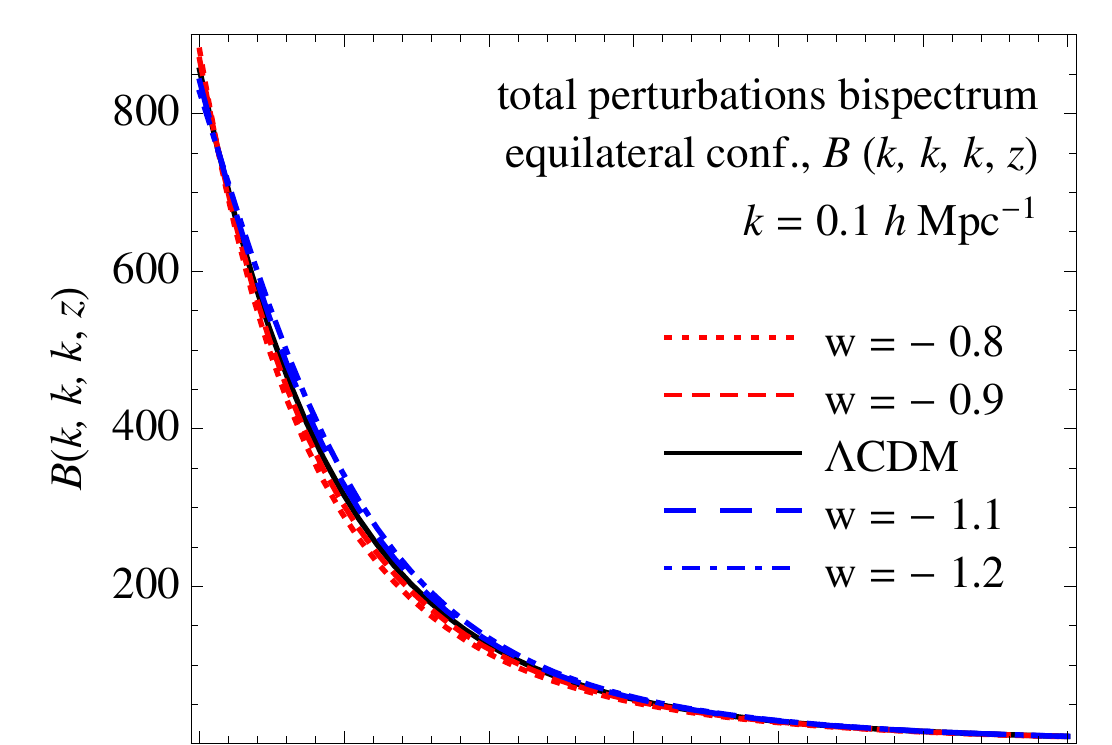}}
{\includegraphics[width=0.49\textwidth]{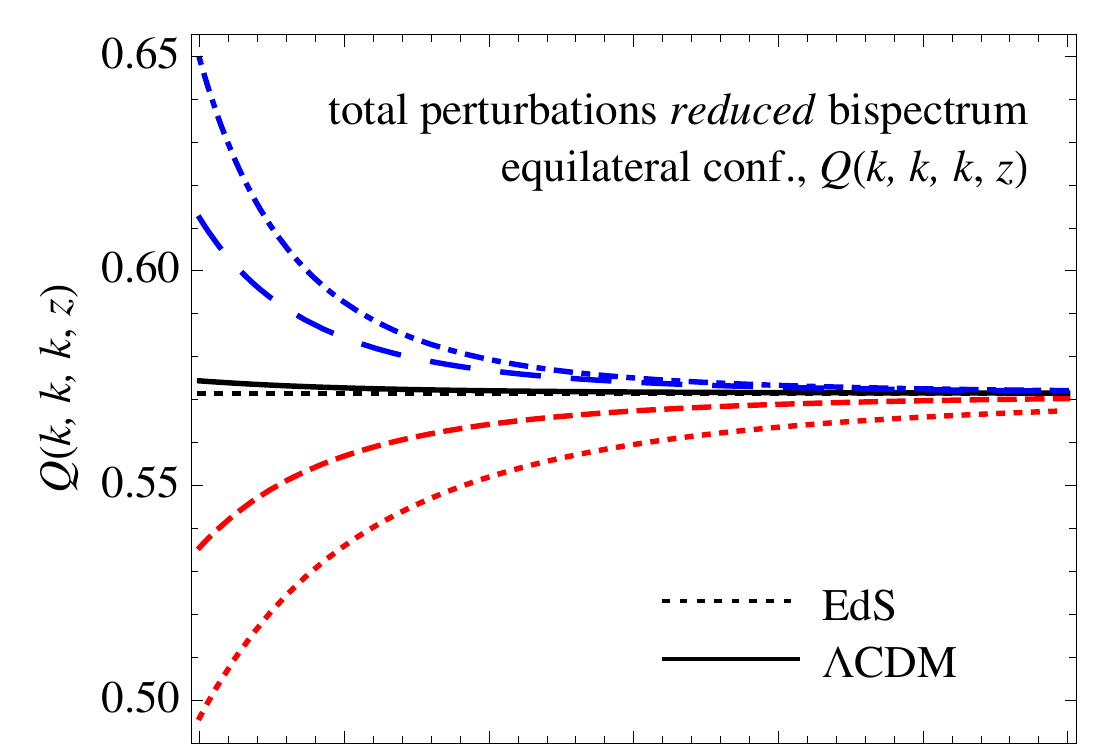}}\\
{\includegraphics[width=0.49\textwidth]{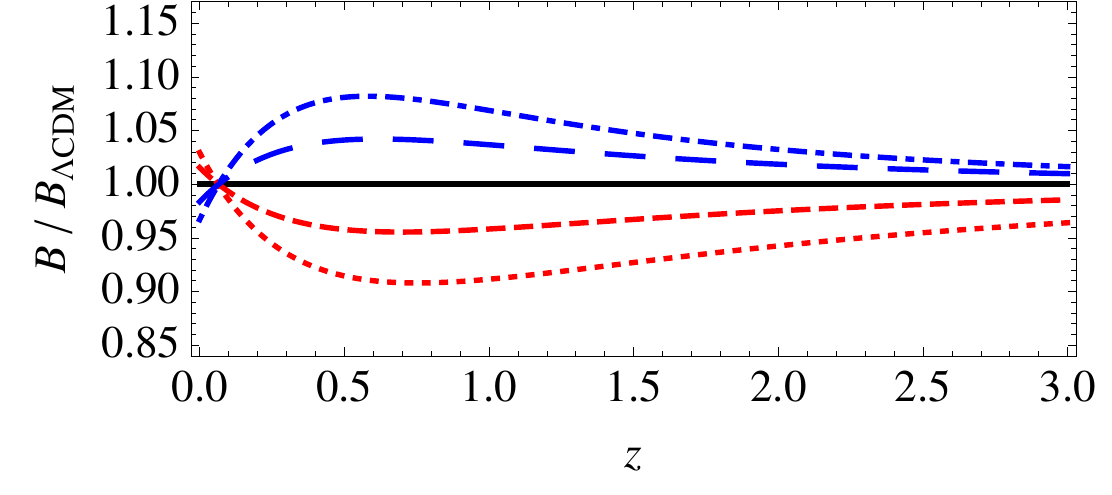}}
{\includegraphics[width=0.49\textwidth]{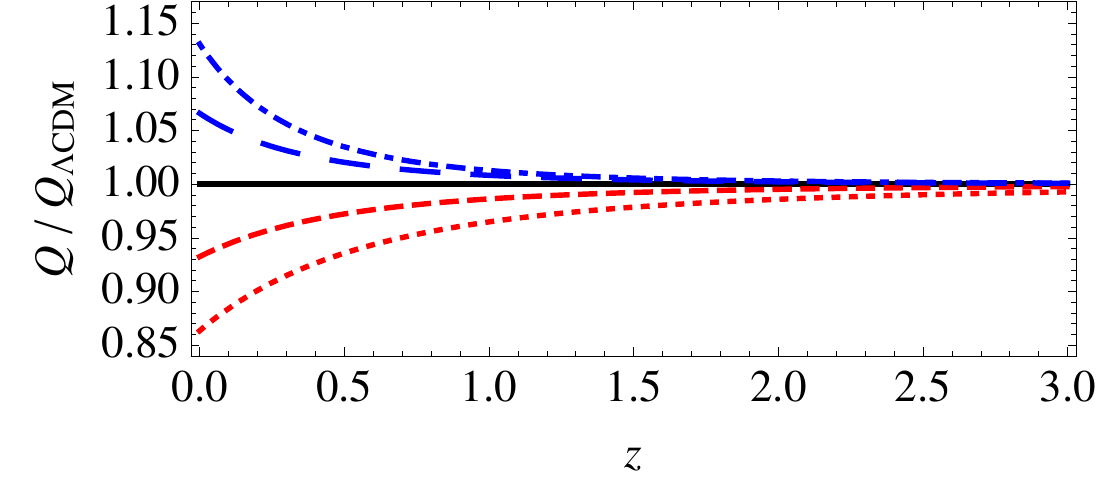}}
\caption{Effects of clustering quintessence on the tree-level bispectrum $B$ ({\em left panels}) and reduced bispectrum $\Q$ ({\em right panels}) for the total density contrast $\d$, in the equilateral configuration, as a function of redshift. At tree-level ${\cal Q}$ is $k$-independent, while $B$ depends on $k$, chosen here to be $k=0.1\kMpc$. In the upper right panel the constant value $4/7$ for an Einstein-de Sitter cosmology is also shown as a dotted black line. The lower panels show the ratio with the corresponding $B$ and $\Q$ for a $\Lambda$CDM cosmology.}
\label{fig:Qeqz}
\end{center}
\end{figure}
In Fig.~\ref{fig:Qeqz}, in the  left panels, we show the tree-level bispectrum in the equilateral configuration $B(k,k,k;z)$ for $k=0.1\kMpc$, as a function of redshift. In particular, the lower left panel shows the ratio between $B(k,k,k;z)$ in the clustering and $\Lambda$CDM cases, for various values of $w$.
Two effects are into play: the evolution of the linear growth function $D_+$ and that of the second-order kernel $F_2$, through the functions $\nu_2$ and $\epsilon$. The evolution is dominated by the fourth power of the linear growth function $D_+$, see eq.~\eqref{eq:Beq2}. Indeed, $D_+$ is responsible for the turnaround at low redshift, as one can check in Fig.~\ref{fig:D}. We do not include a similar plot for the ratio between $B(k,k,k;z)$ in the clustering and smooth cases, which is dominated by the linear evolution.  The right panels of Fig.~\ref{fig:Qeqz} shows instead the corresponding reduced bispectrum $\Q$. The evolution is entirely due to the second-order kernel $F_2$ and is more significant at low redshift, with corrections larger than 5\% at $z=0$ for $|1+w|=0.1$. Again, we do not include a similar plot for the ratio between $\Q(k,k,k;z)$ in the clustering and smooth cases, which shows a very similar behavior.

Other configurations of particular interest are the collinear ones, where the triangle formed by the three momenta is flattened, such that the three sides satisfy the relation $k_3=k_1+k_2$. Thus, the scalar product between $\vec k_1$ and $\vec k_2$ is $+1$ while the other two are $-1$. In this case the dependence of $F_2$ on  $\nu_2$ drops off and the reduced bispectrum becomes
\be
\Q(k_1,k_2,k_3;\eta) = [1-\epsilon(\eta)]\left[2+\frac{\left(k_1/k_2+k_2/k_1\right)\,P_{\rm in}(k_1)\,P_{\rm in}(k_2)-2~{\rm cyclic}}{P_{\rm in}(k_1)\,P_{\rm in}(k_2)+2~{\rm cyclic}}\right]\, ,
\ee
where the time-dependence is entirely factorized in the $(1-\epsilon)$ term. Thus, for collinear configurations the effects of clustering dark energy is described in terms of $\epsilon(\eta)$ alone, while no effect is expected in the smooth case. 

\begin{figure}[t]
\begin{center}
{\includegraphics[width=0.49\textwidth]{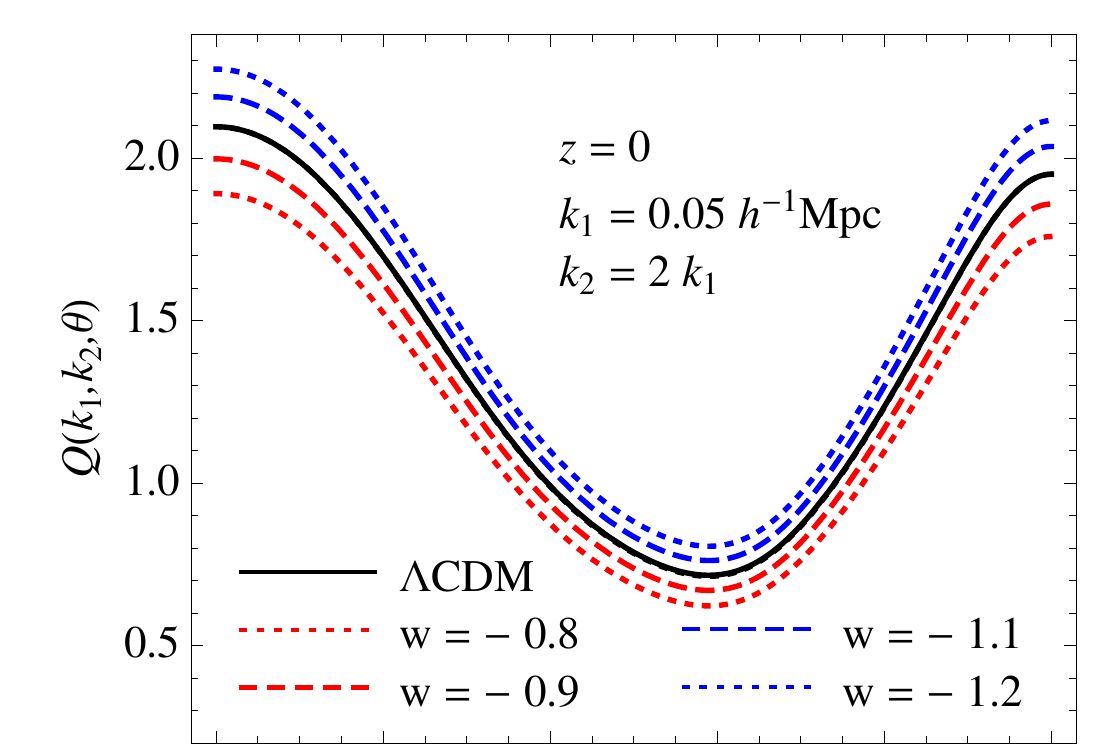}}
{\includegraphics[width=0.49\textwidth]{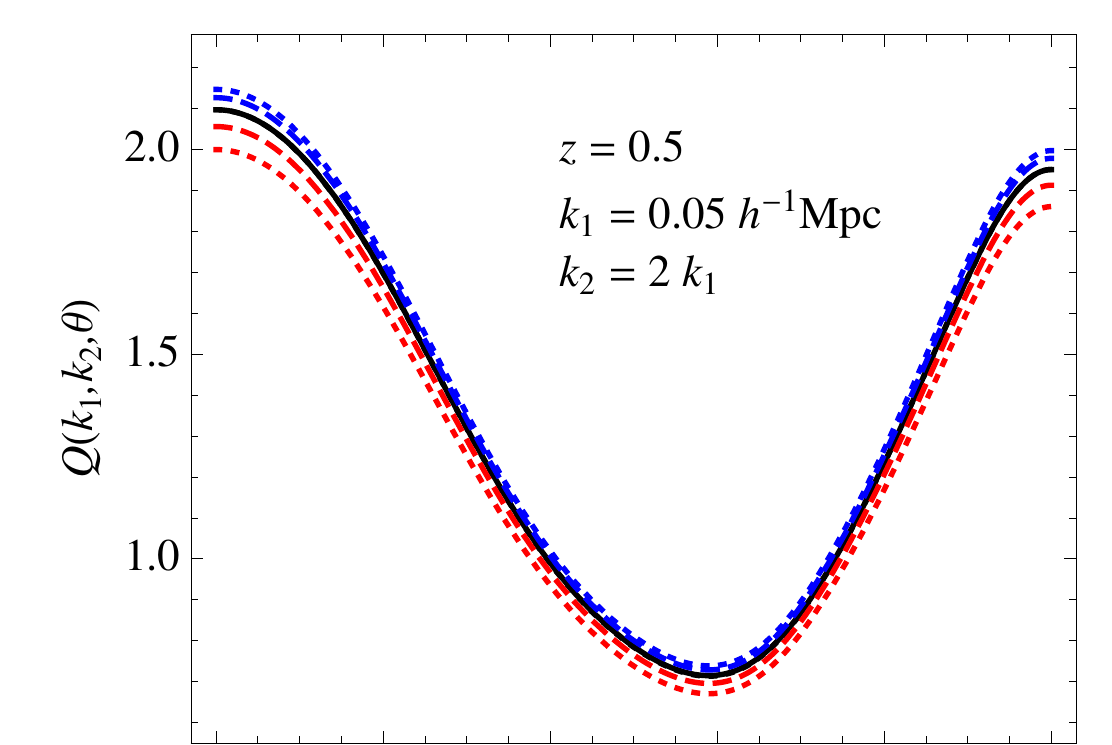}}\\
{\includegraphics[width=0.49\textwidth]{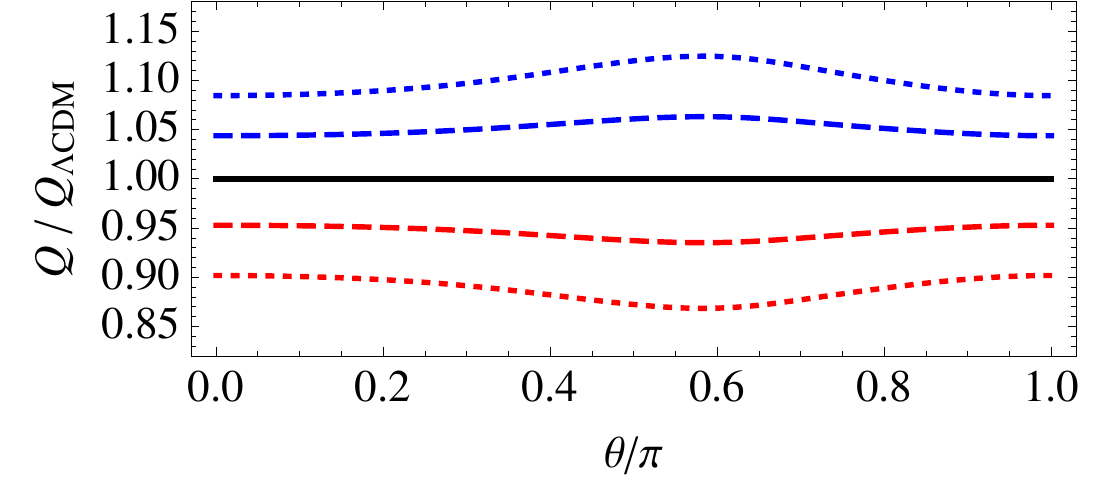}}
{\includegraphics[width=0.49\textwidth]{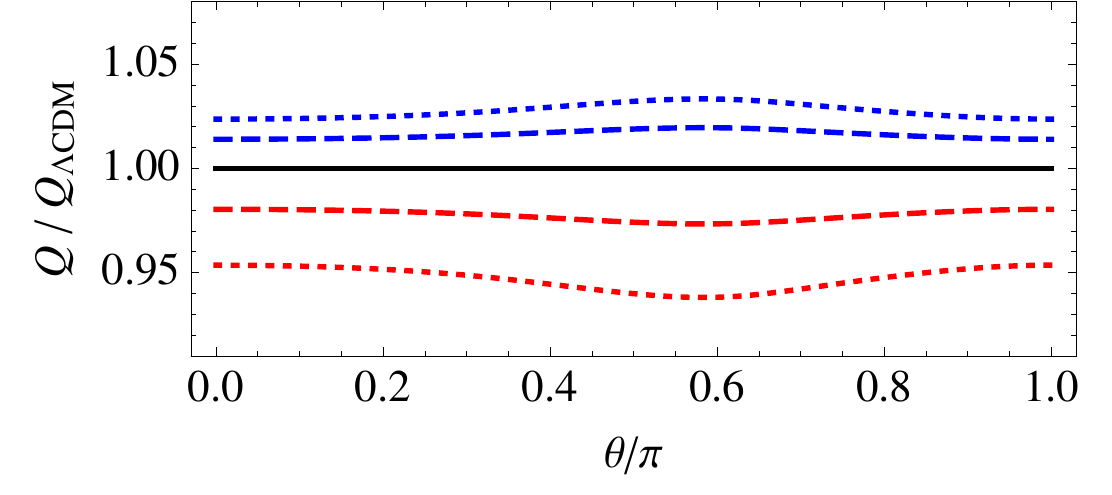}}\\
\caption{{\em Upper panels}: the total reduced bispectrum $Q(k_1,k_2,k_3,z)$ for $k_1=0.05\kMpc$ and $k_2=2k_1$ as a function of the angle $\theta$ between the two wavenumbers at $z=0$ ({\em left}) and $z=0.5$ ({\em right}). {\em Lower panels}: ratio with respect to the $\Lambda$CDM value.}
\label{fig:Q}
\end{center}
\end{figure}
Let us study the shape dependence of the corrections due to clustering quintessence. In Fig.~\ref{fig:Q} we plot the reduced bispectrum $\Q(k_1,k_2,k_3;z)$ with fixed $k_1=0.05\kMpc$ and $k_2=2 k_1$, as a function of the angle $\theta$ between $\vec k_1$ and $\vec k_2$. At $z=0$ ({\em left panels}) the effect is of the order of 5\% for $|1+w|=0.1$ roughly for all the configurations considered and degrades to a few percent at $z=0.5$ ({\em right panels}). 
One can notice that the effect of clustering quintessence is roughly independent of $\theta$. In general, in the tree-level approximation the effect of clustering quintessence is similar on different scales, \ie~for different values of $k_1$ and $k_2$. However, we expect higher-order corrections to become relevant on small scales and change this behavior. Notice that analogous corrections to the matter reduced bispectrum in the DGP model \cite{Dvali:2000hr} are instead strongly dependent on the triangle shape, see for instance Fig.~10 in \cite{Scoccimarro:2009eu} and Fig.~10 in \cite{Chan:2009ew} for a comparison with N-body simulations.


\section{Signal-to-noise}
\label{sec:StoN}

It is important to get an idea of the size of the corrections induced by clustering quintessence compared to cosmic variance. Since there is no way of distinguishing between matter and quintessence perturbations in any observation relying on gravitational effects, in the following we will only consider the power spectrum and bispectrum of the {\em total} perturbation $\delta$.

The effect of clustering quintessence on the power spectrum is encoded in the linear growth function $D_+$, see eq.~\eqref{P_delta}. As the power spectrum at fixed redshift is simply a function of one variable, \ie~the wavenumber $k$, the comparison of the corrections induced by quintessence with the expected statistical error is relatively straightforward. 
In the case of the bispectrum, however, a few plots of specific sets of triangular configurations can be misleading. In fact, the statistical significance of an observable, such as the galaxy bispectrum in redshift surveys, can only be assessed when {\em all measurable configurations} are accounted for. In this respect, it has been shown that the {\em cumulative signal-to-noise} expected for the galaxy bispectrum in surveys of the size of the SDSS main sample is comparable to the signal-to-noise for the power spectrum when all triangles down to mildly nonlinear scales are included in the analysis \cite{Sefusatti:2004xz}. This holds true also when the full nonlinear covariance of the correlators and the survey geometry is considered. Moreover, a joint analysis of power spectrum and bispectrum can improve the constraints on cosmological parameters, particularly for those parameters responsible for the amplitude of perturbations and when the dark energy equation of state parameter $w$ is allowed to vary \cite{Sefusatti:2006pa}.

Before presenting our conclusions, in this section we consider some simple estimates of the signal-to-noise ratio that we expect for the effect of clustering {\em and} smooth quintessence on the power spectrum and bispectrum in their leading-order, perturbative approximations. Such estimates do not relate to any specific large-scale structure observable. Thus, they should not be directly interpreted in terms of ``measurable'' departures from a $\Lambda$CDM cosmology. Rather, they provide a helpful comparison between the power spectrum and  the bispectrum as large-scale structure probes of quintessence. More broadly, they aim to motivate further research more closely related to current and future observations in weak lensing or redshift surveys. 

For this purpose we consider a cosmological volume of $1\cGpc$, significantly smaller than the volumes that will be probed by future surveys, such as Euclid, BOSS or LSST. We choose $z=0.5$ as an intermediate redshift. However, since the time-evolution of the effect of clustering quintessence is very important, see Fig.~\ref{fig:D}, the redshift dependence could be used to increase our ability to detect this effect. Furthermore, we limit our calculation to scales $k\le 0.2\kMpc$, where we expect the linear and tree-level approximations to be sufficiently accurate for our purposes. Note that excluding small scales and neglecting higher-order corrections in $P$ and $B$, which are relevant in the mildly nonlinear regime, is a conservative stance. Indeed, similarly to what we have discussed for $F_2$ and $G_2$, higher-order kernels are further affected by quintessence clustering. Thus, we expect higher-order corrections to $P$ and $B$ in the mildly nonlinear regime to be sensitive to clustering quintessence.

The cumulative signal-to-noise for the power spectrum is simply defined here as
\be
\left(\frac{S}{N}\right)^2_{k_{\rm max}}=\sum_{k=k_f}^{k_{\rm max}}\frac{\left[P_{Q}(k)-P_{\Lambda}(k)\right]^2}{\Delta P_\Lambda^2(k)}\,.
\ee
The sum is intended over the measurable wavenumbers $k$ from the fundamental frequency $k_f=2\pi/L$ (defined by the assumed volume $V=L^3$) to the smallest scale $k_{\rm max}$ in steps of $k_f$.  By $P_Q(k)$ we denote the power spectrum for a (smooth or clustering) quintessence cosmology while $P_\Lambda$ corresponds to the power spectrum in the $\Lambda$CDM case. The power spectrum variance $\Delta P^2(k)$ is approximated by its leading, Gaussian contribution,
\be
\Delta P^2(k)=\frac{k_f^2}{2\pi k^2}P^2(k)\,.
\ee

In the bispectrum case we define, {\em mutatis mutandis}, 
\be
\left(\frac{S}{N}\right)^2_{k_{\rm max}}=\sum_{k_f\le k_1,\,k_2,\,k_2}^{k_{\rm max}}\frac{\left[B_{Q}(k_1,k_2,k_3)-B_{\Lambda}(k_1,k_2,k_3)\right]^2}{\Delta B_\Lambda^2(k_1,k_2,k_3)}\,,
\ee
where the sum runs over {\em all} triangular configurations defined by the wavenumbers $k_1$, $k_2$ and $k_3$ smaller than or equal to $k_{\rm max}$. The bispectrum variance, again in the Gaussian approximation, is given by \cite{Scoccimarro:1997st, Scoccimarro:2003wn}
\be
\Delta B^2(k_1,k_2,k_3)=\frac{s_B}{8\pi^2 k_1 k_2 k_3}P(k_1)P(k_2)P(k_3)\,,
\ee
with $s_B=6$, $2$ or $1$ for equilateral, isosceles or scalene triangles respectively. 

Finally, we can define a cumulative signal-to-noise for the reduced bispectrum as
\be
\left(\frac{S}{N}\right)^2_{k_{\rm max}}=\sum_{k_f\le k_1,\,k_2,\,k_2}^{k_{\rm max}}\frac{\left[\Q_{Q}(k_1,k_2,k_3)-\Q_{\Lambda}(k_1,k_2,k_3)\right]^2}{\Delta \Q_\Lambda^2(k_1,k_2,k_3)}\,,
\ee
where we assume the reduced bispectrum variance to be dominated by the error on the bispectrum, so that $\Delta \Q(k_1,k_2,k_3)\simeq \Delta B(k_1,k_2,k_3) / [P(k_1)P(k_2)+2~{\rm cyclic}]$. 

Therefore, for these simple estimates we neglect the non-Gaussian contributions to the correlators variance, which we expect to be subdominant in the range of scales considered here, $k\le 0.2\kMpc$. Most importantly, we neglect the effect of covariance between different triangular configurations. As we have already stressed, these computations have purely illustrative purposes and do not assume any specific observable. Thus, we neglect any shot-noise contribution to the variances and any effect of a survey geometry, naturally expected, for example, in redshift surveys. 

\begin{figure}[th!]
\begin{center}
{\includegraphics[width=0.49\textwidth]{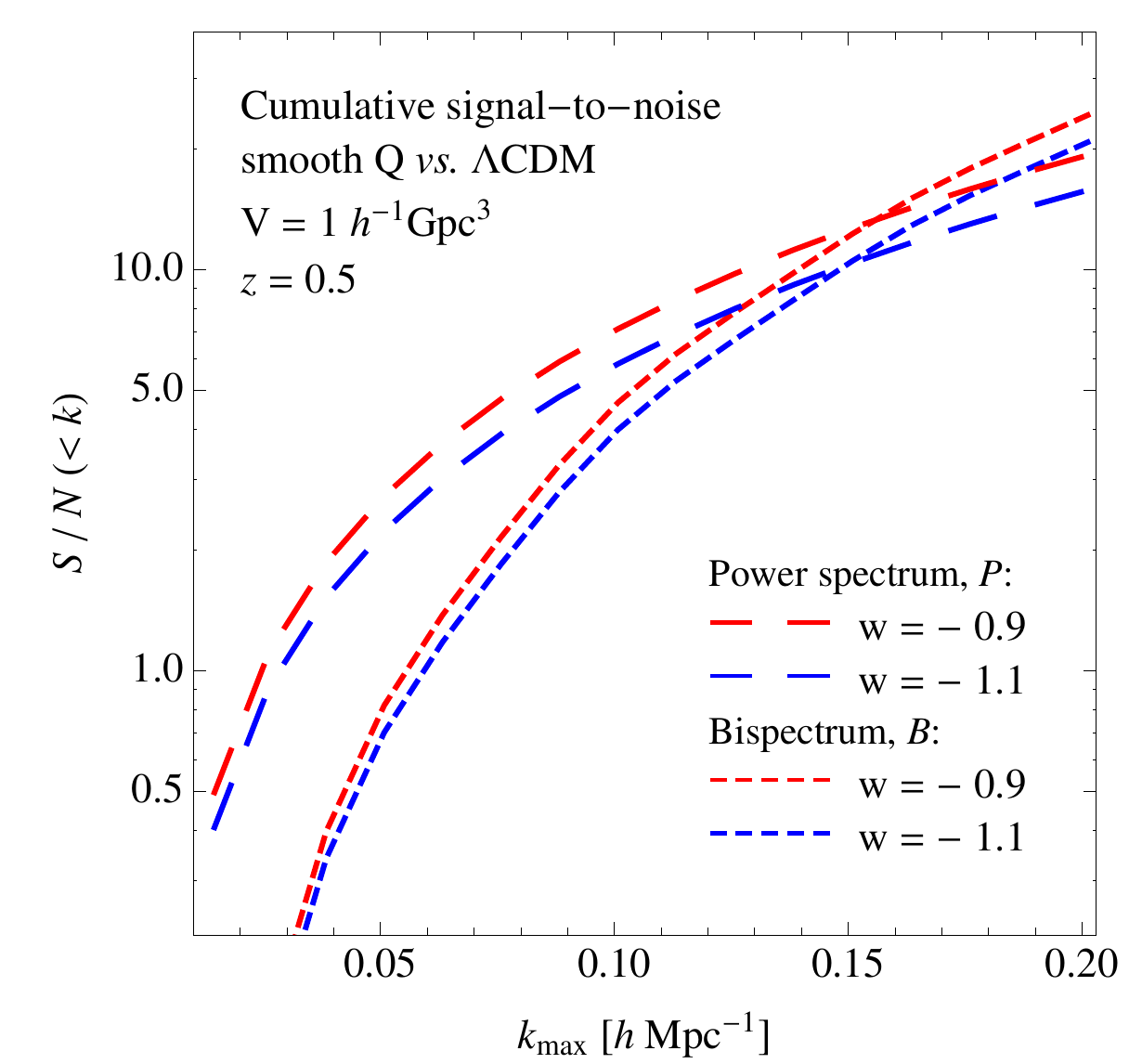}}
{\includegraphics[width=0.49\textwidth]{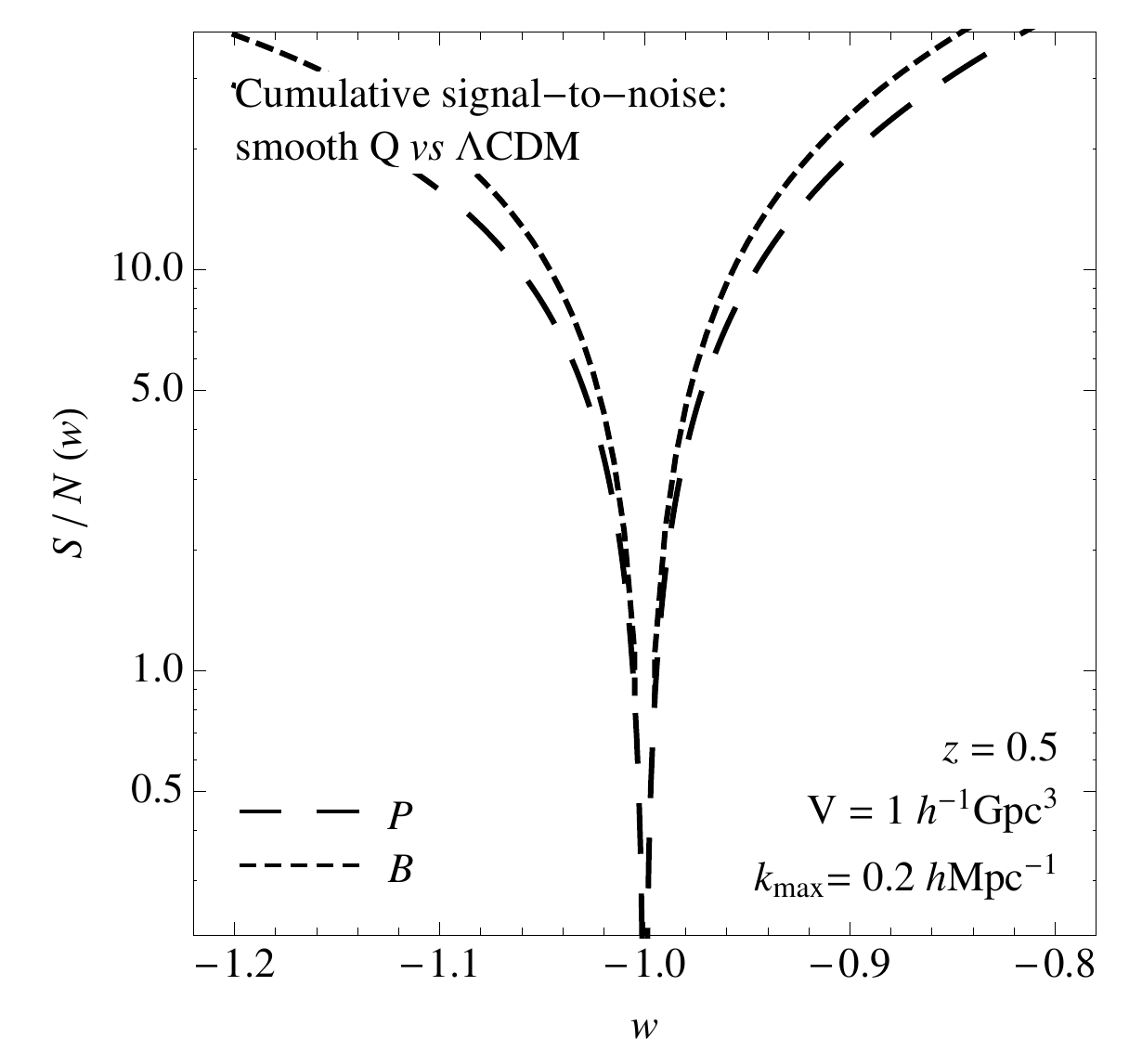}}
{\includegraphics[width=0.49\textwidth]{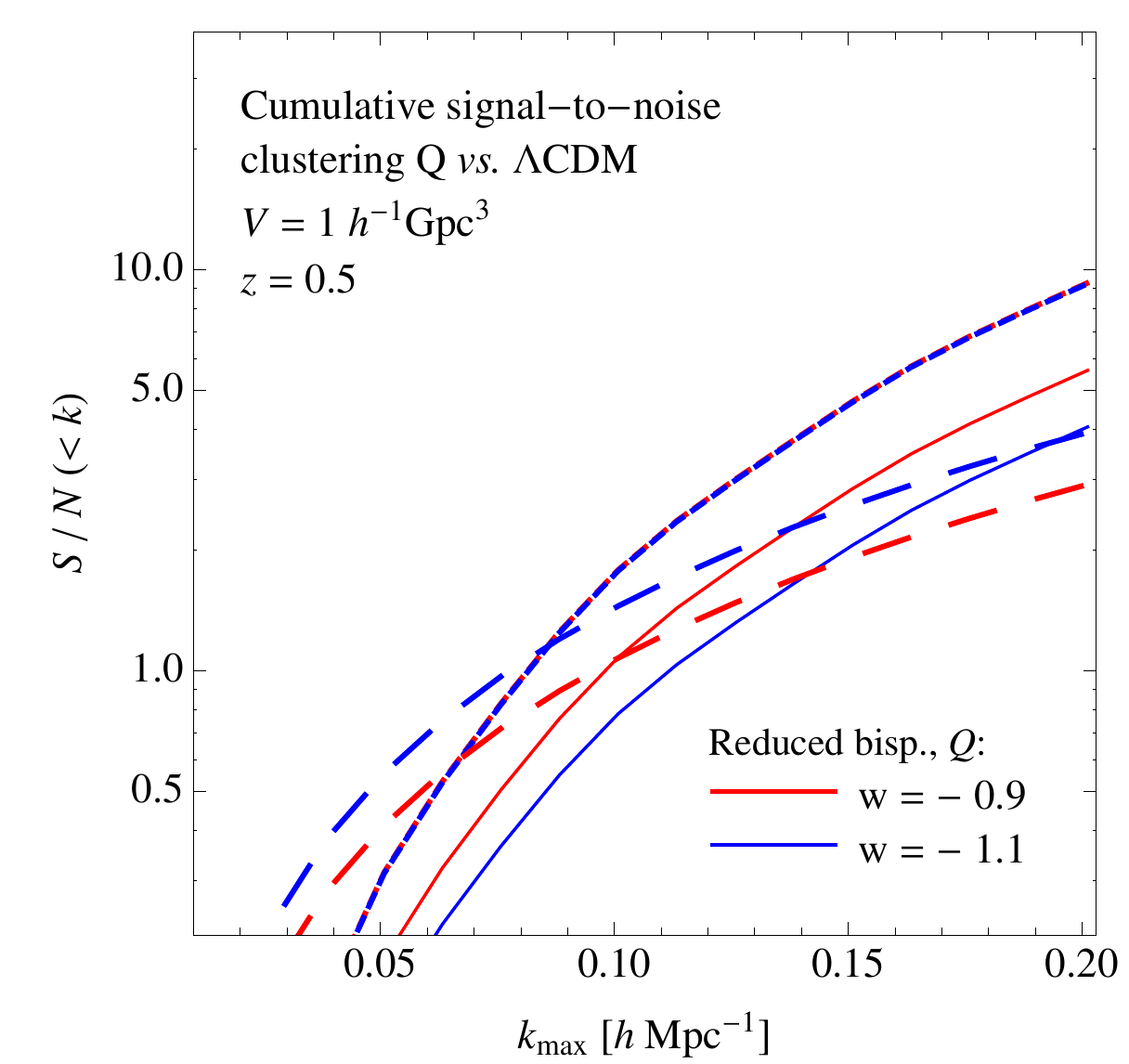}}
{\includegraphics[width=0.49\textwidth]{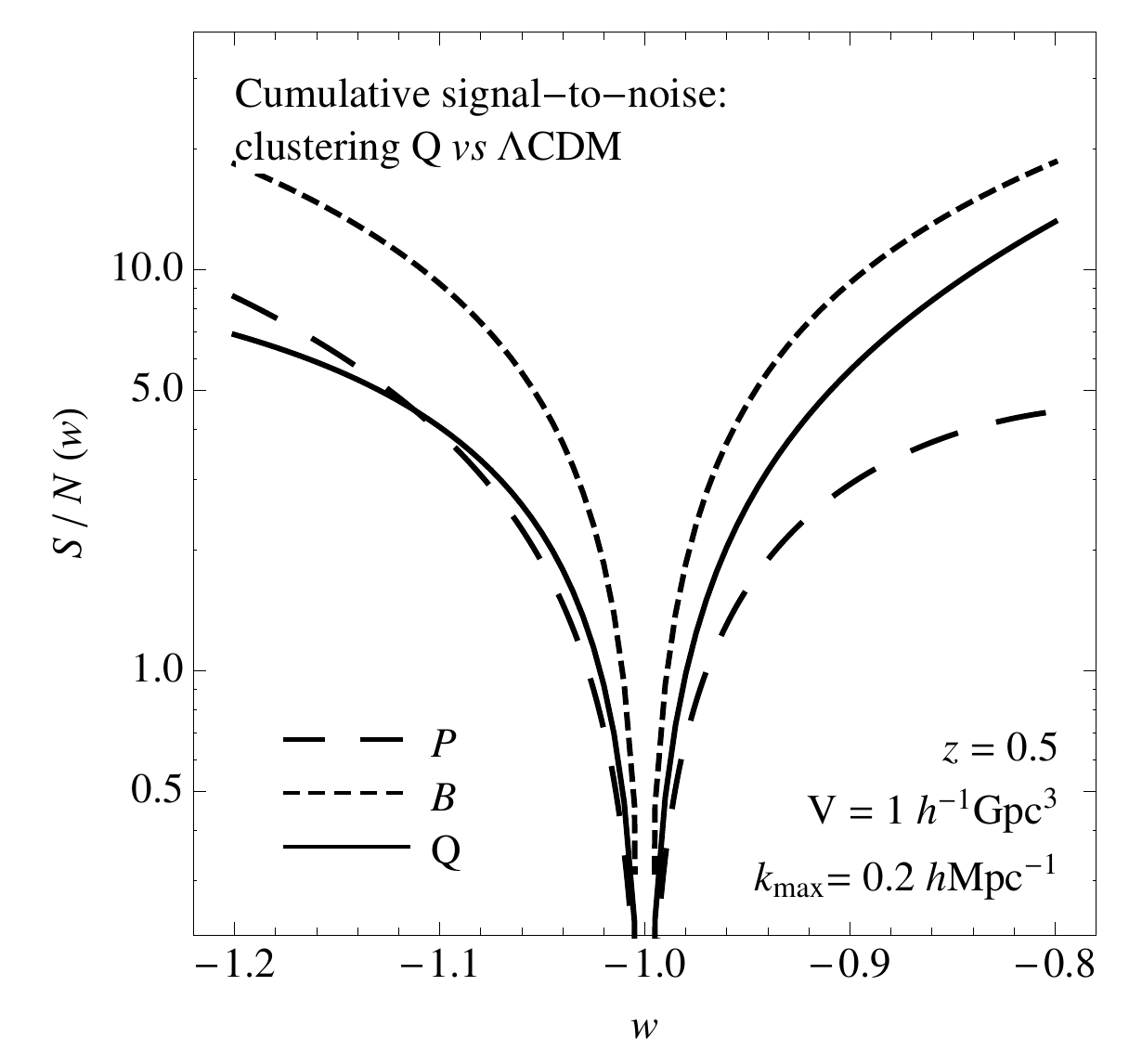}}
\caption{Cumulative signal-to-noise for the effect of smooth ({\em upper panels}) and clustering ({\em lower panels}) quintessence on the total power spectrum ({\em long-dashed lines}), bispectrum ({\em short-dashed lines}) and reduced bispectrum ({\em continuous lines}) for a volume of $1\cGpc$ at $z=0.5$. On the left panel the signal-to-noise is shown as  a function of the smallest scale considered, {\em i.e.} $k_{\rm max}$ for $w=-0.8$ ({\em thick, red lines}) and $w=-1.1$ ({\em thin, blue lines}). In the right panels the signal-to-noise for power spectrum, bispectrum and reduced bispectrum  is shown as a function of $w$, assuming $k_{\rm max}=0.2\kMpc$. }
\label{fig:StoNz0p5}
\end{center}
\end{figure}
In Fig.~\ref{fig:StoNz0p5} we show the cumulative signal-to-noise for the effect of smooth ({\em upper panels}) and clustering ({\em lower panels}) quintessence with respect to a $\Lambda$CDM cosmology, on the total power spectrum ({\em long-dashed lines}), bispectrum ({\em short-dashed lines}) and reduced bispectrum ({\em continuous lines}). Note that $w=-0.9$ and $w=-1.1$ lead to very close bispectrum signals with the two corresponding curves almost coinciding. The reduced bispectrum is considered only in the clustering case since in the smooth case the corrections are negligible. In particular, on the left panels the signal-to-noise is shown as  a function of the smallest scale considered, {\em i.e.}~$k_{\rm max}$ for $w=-0.9$ ({\em thick, red lines}) and $w=-1.1$ ({\em thin, blue lines}). In the right panels, instead, it is shown as a function of $w$, assuming $k_{\rm max}=0.2\kMpc$.

The dependence on $k_{\rm max}$ of the signal-to-noise for the power spectrum and the bispectrum is quite different. This has two reasons. On one hand the signal-to-noise for an {\em individual} triangular configuration of the bispectrum is  smaller than that of the power spectrum for a given wavenumber of the same order. Thus, when only large-scale modes are considered the cumulative signal-to-noise in the bispectrum is smaller because only a few triangles can be measured on those scales. On the other hand,  as we include smaller scales the number of available triangles grows more rapidly than the number of available wavenumbers. This is the case for the signal associated to the measurement of the correlators \cite{Sefusatti:2004xz} as for the signal from the corrections induced by clustering dark energy, as studied here. One can check that the scale at which the two statistics display comparable signals depends on the volume, as expected.

An interesting point, illustrated by the lines for the power spectrum and bispectrum on the right panels of Fig.~\ref{fig:StoNz0p5}, is that the dependence of  the signal-to-noise on the equation of state $w$ strongly depend on whether quintessence is clustering or not. For instance, in the particular example at hand, the ideal constraints on $w$ for clustering quintessence can be a factor a few larger than in the smooth case. Indeed, the power spectrum and bispectrum are both mainly sensitive to the equation of state through the linear growth function $D_+$, and Fig.~\ref{fig:D} shows that, at $z=0.5$, $D_+$ for clustering quintessence is very close to the one in $\Lambda$CDM, contrarily to $D_+$ in the smooth case. For the reduced bispectrum, which does not depend on $D_+$, we find the opposite. Indeed, in the smooth case the dependence of $\Q$ on $w$ is very mild while in the clustering case it becomes stronger, due to the modifications discussed in the previous section.

\begin{figure}[t]
\begin{center}
{\includegraphics[width=0.49\textwidth]{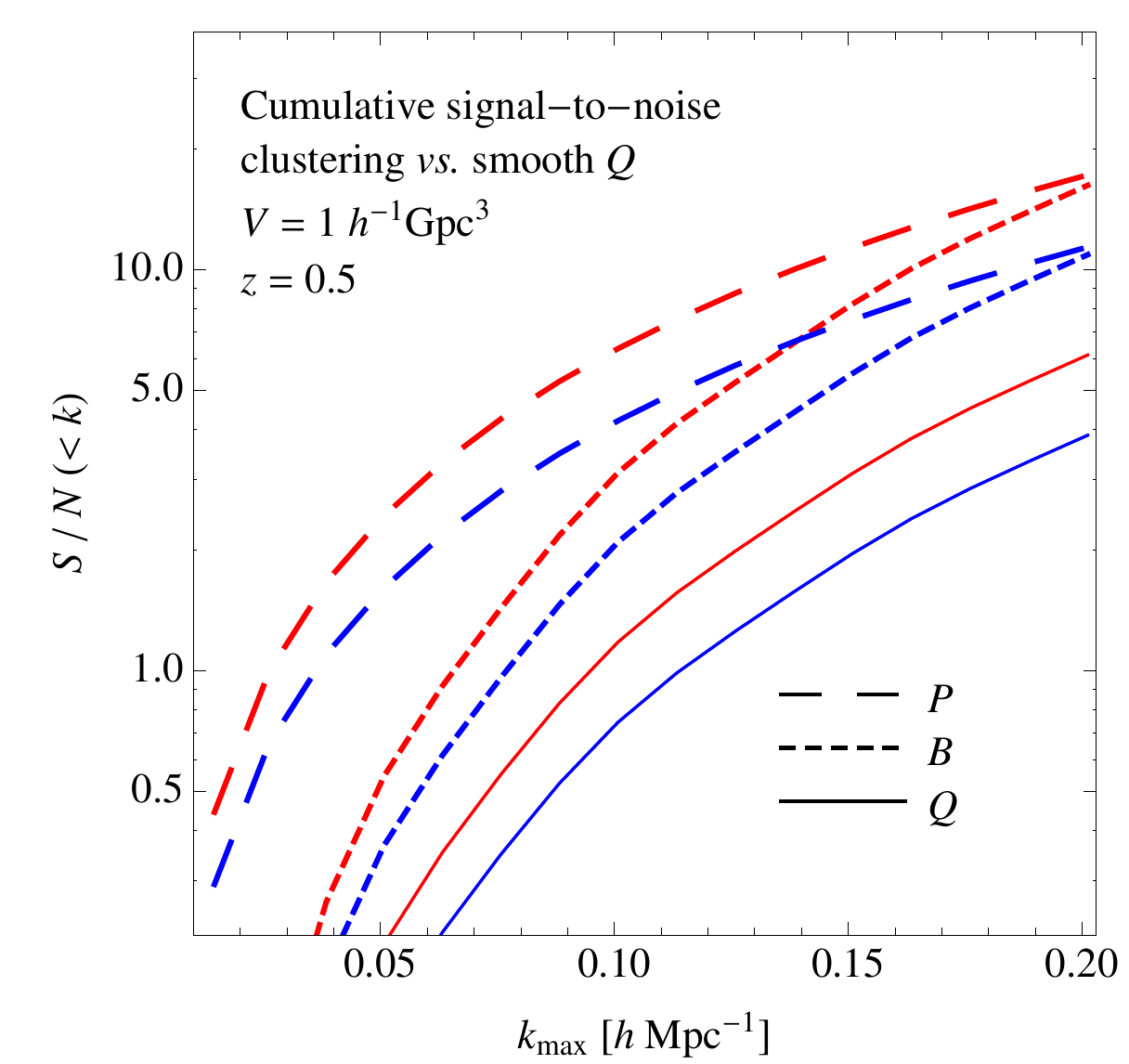}}
{\includegraphics[width=0.49\textwidth]{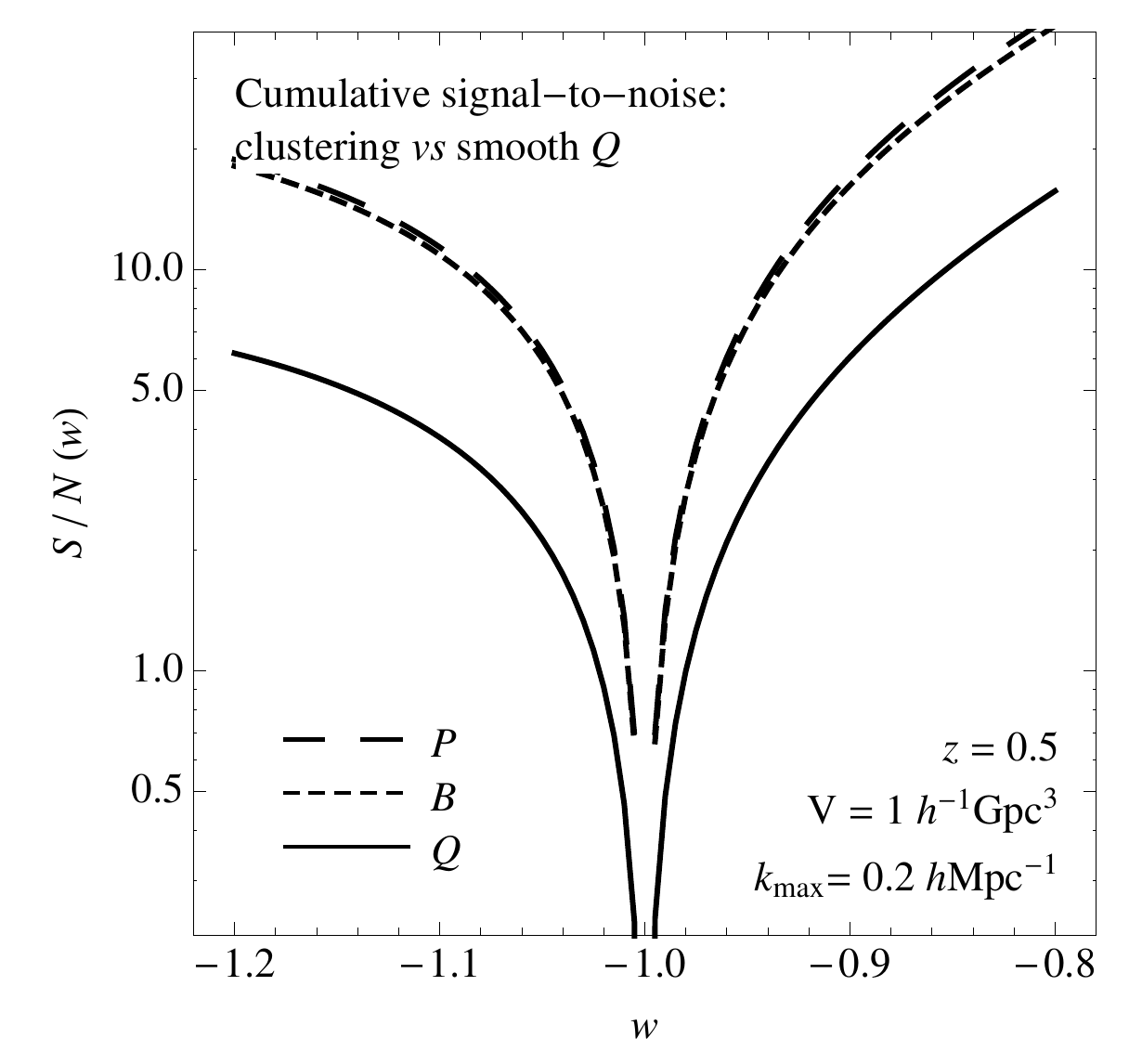}}
\caption{Cumulative signal-to-noise for the effect of clustering quintessence on the total power spectrum ({\em long-dashed lines}), bispectrum ({\em short-dashed lines}) and reduced bispectrum ({\em continuous lines}) with respect to the smooth quintessence scenario assuming a volume of $1\cGpc$ at $z=0.5$. On the left panel the signal-to-noise is shown as a function of the smallest scale considered, {\em i.e.} $k_{max}$ for $w=-0.9$ ({\em thick, red lines}) and $w=-1.1$ ({\em thin, blue lines}). In the right panel the signal-to-noise for power spectrum, bispectrum and reduced bispectrum is shown as a function of $w$, assuming $k_{\rm max}=0.2\kMpc$.}
\label{fig:StoN_c_vs_s}
\end{center}
\end{figure}
In Fig.~\ref{fig:StoN_c_vs_s} we show the signal to noise for the corrections induced by clustering dark energy with respect to the smooth case. For the power spectrum the expression is given by
\be
\left(\frac{S}{N}\right)^2_{k_{\rm max}}=\sum_{k=k_f}^{k_{\rm max}}\frac{\left[P_{Qc}(k)-P_{Qs}(k)\right]^2}{\Delta P_{Qs}^2(k)}\,,
\ee
where $P_{Qc}$ and $P_{Qs}$ represent the power spectrum for clustering and smooth quintessence, respectively. Analogous expressions can be written for the bispectrum and the reduced bispectrum. 
Note that the fact that the signal-to-noise for the power spectrum and the bispectrum in the right panel of Fig.~\ref{fig:StoN_c_vs_s} are so close is a coincidence due to the choice of $k_{\rm max}$, as can be seen from the left panel.

The results of Figs.~\ref{fig:StoNz0p5} and \ref{fig:StoN_c_vs_s}, particularly the relative size of the signal-to-noise of the power spectrum versus that of the bispectrum, are strongly dependent on the volume. For instance, larger volumes favor bispectrum measurements as the number of configuration increases significantly. 
Furthermore, these results strongly depend also on the redshift through $D_+$ and $\epsilon$ (see Fig.~\ref{fig:D} and \ref{fig:epsilon}). Thus, one should be particularly careful in extrapolating them to other redshifts. For instance,  as can be seen in Fig.~\ref{fig:Qeqz}, the choice of $z=0.5$ corresponds to the specific case where the effect of clustering quintessence is maximum on the bispectrum, while it gets significantly larger at smaller $z$ for the reduced bispectrum.

For these reasons, what shown here does not  provide an exhaustive comparison between different statistics or a complete picture of the signal expected. However, this analysis indicates that both correlators should in principle be equally sensitive to corrections induced by quintessence, both in the clustering and in the smooth case.

The relatively high signal-to-noise expected in the ideal set-up assumed here motivates further studies. Indeed, an actual detection of such features would require the ability of distinguishing them from other nonlinear effects or, in the case of redshift surveys, from galaxy bias and redshift distortions. The detailed study of the detectability of possible departures from a $\Lambda$CDM cosmology in actual observation is thus left to future work.


\section{Conclusions}
\label{sec:conclusions}

In this work we considered the case of a quintessence characterized by a vanishing speed of sound. In this case quintessence perturbations grow on all observables scales, inducing relevant effects on the evolution of structures when the dark energy component comes to dominate the energy density of the Universe.

At late time, both dark matter and quintessence  perturbations act as a source for the gravitational potential and they are practically indistinguishable by gravitational observations. Based on this, we introduce a {\em total density perturbation} as a weighted sum of matter and quintessence perturbations. Since quintessence is {\em comoving} with dark matter, the evolution of such a quantity is determined by a closed set of equations: the continuity equation for the total density field, the Euler equation for the common velocity field of the two components, and the Poisson equation relating the gravitational potential to the total density perturbation. This allows us to study the evolution of the total density fluctuation in Eulerian Perturbation Theory, in complete analogy with the usual treatment of matter fluctuations in a $\Lambda$CDM or smooth quintessence cosmology.

The equations of motion for the total perturbations are equivalent to those for the matter perturbation alone, with a simple correction: the linear term in the velocity divergence of the continuity equation is proportional to the function $C= 1 + (1+w) \Omega_Q/\Omega_m$. At early times, when quintessence is negligible, $C=1$ and we recover the standard evolution. At late times clustering quintessence increases the growth rate of fluctuations.

At linear order it is possible to obtain an exact integral expression for the growth of the total perturbation, eq.~\eqref{int_sol}. This solution relies on the fact that for a quintessence with vanishing speed of sound comoving regions behave as independent FRW universes. This integral expression allowed us to derive simple fitting functions for both the growth function and the growth rate of the total fluctuations. 

Beyond linear order, our set-up allows to straightforwardly apply to the clustering quintessence scenario standard EPT, but also more efficient resummation techniques such as Renormalized Perturbation Theory \cite{Crocce:2005xy, Crocce:2005xz, Crocce:2007dt, Bernardeau:2008fa,Bernardeau:2010md} and the Renormalization Group approach \cite{Matarrese:2007wc, Matarrese:2007aj,Pietroni:2008jx, Anselmi:2010fs, Saracco:2009df}. 

We showed that linear theory does not fully describe the rich phenomenology of a quintessence with zero speed of sound. Indeed, we found significant effects on the late-time evolution of higher-order perturbations. These can affect the total power spectrum over a wide range of observable scales, where the evolution of perturbations becomes nonlinear.

Since they directly depend on nonlinear corrections over the linear density field, also higher-order correlation functions, such as the bispectrum,  are affected by the clustering of quintessence at low redshift. In particular, we studied second-order solutions in EPT for the density contrast and velocity fields. In terms of these solutions we derived the leading-order (or tree-level) contribution in EPT to the total bispectrum. On large scales, this is expected to be a good approximation to the fully nonlinear bispectrum. 

In particular, we showed that the {\em reduced} bispectrum, which is normalized in such a way as to be independent of the linear evolution, receives significant corrections {\em only} in the clustering case. These corrections are of the order of $\delta \rho_Q/(\delta \rho_m + \delta \rho_Q)$, \ie~the ratio between the quintessence and total density perturbations, which at $z=0$ amounts to $5\%$ for $|1+w| =0.1$. These signatures offer a practical way of distinguishing the clustering scenario from the smooth one with the next generation of redshift and weak lensing surveys such as BOSS or Euclid. Corrections of the same magnitude to the reduced matter bispectrum are expected as well from non-Gaussian initial conditions. However, notice that, at least on large scales, such corrections present a different (in fact, opposite) redshift evolution as well as different dependences on scales and shapes (see, for instance, \cite{Liguori:2010hx}).

In Section~\ref{sec:StoN} we provided a simple estimate of the signal-to-noise ratio expected for the {\em effect} of quintessence on the power spectrum and bispectrum of the total density field. We limited our analysis to an ideal box of $1\cGpc$ at fixed redshift $z=0.5$. In particular, for the linear power spectrum we considered the signal expected for the {\em difference} between the predictions of smooth quintessence and $\Lambda$CDM, $P_{Q_s} - P_\Lambda$, of clustering quintessence and $\Lambda$CDM, $P_{Q_c} - P_\Lambda$, and of clustering and smooth quintessence, $P_{Q_c} - P_{Q_s}$. We performed the same analysis for the tree-level bispectrum and the reduced bispectrum. In order to provide a fair comparison between power spectrum and bispectrum, we included all measurable triangular configurations down to a given $k_{\max}$. Below a given scale, the signal-to-noise ratio of the bispectrum becomes more important than the one of the power spectrum. Interestingly, this takes place at a larger scale for the clustering case.

This preliminary analysis is clearly very limited. We discussed simply the density perturbations without considering a particular observable. Moreover, we did not include higher-order nonlinear corrections. These are expected to be relevant  for wavenumbers close to the maximum value considered here, \ie~$k_{\rm max} = 0.2\kMpc$.  Nevertheless, our results emphasize the importance of a joint analysis of the power spectrum {\em and} bispectrum in future redshift and weak lensing surveys, possibly extending over a large redshift range to take full advantage of the rich time-dependence in the clustering quintessence scenario. Furthermore, they suggest that smaller scales, where the evolution of perturbations is nonlinear, are likely to be affected significantly by quintessence clustering. Thus, extending this analysis to the nonlinear regime---somehow a necessary task---can significantly improve the constraints on dark energy with respect to those forecasted assuming only linear theory predictions.

Given the significantly different phenomenology between the smooth and clustering cases, it is crucial to develop accurate theoretical predictions to compare with cosmological observations. As shown, the inclusion of higher-order perturbations, either as corrections to the power spectrum or in higher-order statistics, can be very important in this process. In this work we presented a first step in this direction using perturbative techniques based on Eulerian Perturbation Theory. We leave to future study the extension to more efficient resummation schemes. \\

\bigskip
{\bf Acknowledgments}\\

We are grateful to Francis Bernardeau, Pierstefano Corasaniti, Paolo Creminelli, Guido D'Amico, Enrique Gazta\~ naga, Christian Marinoni, Rom\'an Scoccimarro and Atsushi Taruya for useful discussions. ES acknowledges support by the European Commission under the Marie Curie Inter European Fellowship and he is grateful to the Center for Cosmology and Particle Physics of New York University for kind hospitality during the completion of this project. 

\appendix

\section{Appendices}

\subsection{Scalar field and fluid equations with pressure}
\label{app:scalar}

In this section we show that a scalar field with zero speed of sound of fluctuations satisfies eqs.~\eqref{continuity}, \eqref{euler} and \eqref{poisson}.

Let us consider the action for $k$-essence, \ie~\cite{ArmendarizPicon:1999rj,ArmendarizPicon:2000dh}
\be
S = \int d^4 x \sqrt{-g} \; P(\phi,X) \;,
\qquad X = - g^{\mu \nu} \partial_\mu \phi \partial_\nu \phi \;.
 \label{kaction}
\ee
The evolution equation of $\phi$ derived from this action is
\be
 \nabla_\mu ( \; 2 P_{,X} \partial^\mu \phi ) + P_{,\phi} =0\;,
\label{phi_evolution}
\ee
where $P_{,f} \equiv \partial P/\partial f$ and $\nabla_\mu$ is the covariant derivative with respect to the metric $g_{\mu \nu}$.
The energy-momentum tensor of this field can be derived using 
\be
T_{\mu \nu} = - \frac{2}{\sqrt{-g}} \frac{\delta S}{\delta g^{\mu \nu}} \;,
\ee
and can be written in the perfect fluid form as \cite{Garriga:1999vw}
\be
T_{\mu \nu} = (\rho + p) u_\mu u_\nu + p g_{\mu \nu}\;,
\ee
once we identify the rest-frame energy density and pressure and the 4-velocity of the fluid, respectively as
\be 
\rho=2 P_{,X} X - P \;, \qquad p=P\;, \qquad u_\mu = - \frac{\partial_\mu \phi}{\sqrt{X}}\;.
\label{rho_P_u}
\ee

Now we will show that the scalar field satisfies relativistic fluid equations. 
The relativistic continuity equation follows from the evolution equation \eqref{phi_evolution}. Using the third equality in \eqref{rho_P_u} to replace $\partial^\mu \phi$ by the 4-velocity $u^\mu$, and multiplying it by $\sqrt{X}$ this equation reads
\be
- u^\mu \partial_\mu ( \; 2 P_{,X} \sqrt{X} ) \sqrt{X} + P_{,\phi} \sqrt{X} - 2 P_{,X} X \nabla_\mu u^\mu =0 \;.
\ee
By employing that $\sqrt{X}= u^\mu\partial_\mu \phi $ this equation can be rewritten as
\be
 u^\mu \left[ \partial_\mu ( \; 2 P_{,X} X ) -  P_{,X} \partial_\mu X -  P_{,\phi} \partial_\mu \phi \right] + 2 P_{,X} X \nabla_\mu u^\mu =0 \;, \label{pass_1}
\ee
where, using the first equality in eq.~\eqref{rho_P_u}, we recognize in the quantity in brackets the gradient of $\rho$, \ie~$ \partial_\mu \rho$. Thus, eq.~\eqref{pass_1} can be rewritten as a continuity equation of the energy,
\be
u^\mu \partial_\mu \rho + (\rho +p) \nabla_\mu u^\mu =0\;, \label{continuity_rel}
\ee
which, alternatively, can be derived from projecting the conservation equation of the energy-momentum tensor along the fluid 4-velocity,  $u^\nu \nabla_\mu T^\mu_{\ \ \nu} =0$ \cite{Wald:1984rg}.

As shown in \cite{Creminelli:2009mu}, the relativistic Euler equation is an identity for the scalar field $\phi$. Taking the derivative of the definition of $X$ in eq.~\eqref{kaction},
\be
\partial^\nu(\partial^\mu\phi \partial_\mu\phi) = - \partial^\nu X \;,
\ee 
and rewriting it in terms of the 4-velocity $u^\mu = - \partial^\mu\phi/\sqrt{X}$ we have 
\be
2 u^\mu \sqrt{X} \nabla_\mu (\sqrt{X} u^\nu) = - \partial^\nu X \;,
\ee
and thus
\be
2 X u^\mu \nabla_\mu u^\nu = -  (g^{\nu \mu} + u^\nu u^\mu) \partial_\mu X \;.
\ee
In the parenthesis we recognize the projector on hypersurfaces orthogonal to $u^\mu$. 
Then we can use $\partial_\mu P = P_{,\phi} \partial_\mu \phi + P_{, X} \partial_\mu X$ to replace $\partial_\mu X$ in this equation, noting that the term proportional to $\partial_\mu \phi$ vanishes when multiplied by the projector orthogonal to $u^\mu$. Thus, we obtain
\be
(\rho+p) u^\mu \nabla_\mu u^\nu = -  (g^{\nu \mu} + u^\nu u^\mu) \partial_\mu p\;, \label{Euler_rel}
\ee
which, alternatively, can be derived from the conservation equation $(g^{ \nu \rho} + u^\nu u^\rho ) \nabla_\mu T^\mu_{\ \ \rho} =0$ \cite{Wald:1984rg}. In conclusion, as expected the scalar field satisfies the relativistic continuity and Euler equations, and thus it is dynamically equivalent to a perfect fluid. As the speed of sound of this fluid is defined as 
$c_s^2 \equiv p_{,X}/\rho_{,X}$ \cite{Garriga:1999vw}, the right hand side of the Euler equation \eqref{Euler_rel} can be rewritten as
\be
(\rho+p) u^\mu \nabla_\mu u^\nu = -  c_s^2 (g^{\nu \mu} + u^\nu u^\mu) \partial_\mu \rho\;, \label{Euler_rel_cs}
\ee
which shows that in the limit $c_s = 0$ the fluid satisfies geodesic motion $u^\mu \nabla_\mu u^\nu=0$ \cite{Creminelli:2009mu}.

Let us neglect for a moment metric perturbations, which will be reintroduced later on. Consider a coordinate system $(t,x^i)$ where the 4-velocity can be written as 
\be
u^\mu \equiv \gamma (1, v^i)\;, \label{u_v}
\ee 
with $v^i \equiv dx^i /dt$ being the 3-velocity of the fluid and $\gamma \equiv 1/\sqrt{1-v^2}$  the relativistic factor. Note that  the 3-velocity is related to the field by $\vec v = \vec \nabla \phi/\dot \phi$.
Multiplying eq.~\eqref{continuity_rel} by $\gamma$ and using in this equation the definition \eqref{u_v}, summing it to the $\nu=0$ component of eq.~\eqref{Euler_rel}, one finds 
\be
\partial_t \left[\gamma^2 (\rho +  v^2 p)  \right] + \vec \nabla \cdot \left[ \gamma^2 (\rho+p) \vec v \right]=0\;, \label{continuity_coords}
\ee
where we recognize in $\gamma^2 (\rho +  v^2 p)$ the energy density in the rest frame defined by $(t,x^i)$. 
Alternatively, this equation can be derived more directly from the $\nu=0$ component of the conservation equation of the energy-momentum, $ \nabla_\mu T^\mu_{\ \ 0} =0$. 

Multiplying the $\nu=0$ component of eq.~\eqref{Euler_rel} by $\vec v$ and subtracting it from the spatial component of the same equation one finds 
\be
\gamma^2(\rho+p) \left[ \partial_t \vec v + (\vec v \cdot \vec \nabla) \vec v \right] = - \vec \nabla p - \vec v \; \partial_t p \;, \label{euler_coords}
\ee
which can be also simply derived from $\nabla_\mu T^{\mu i} = 0$, after using eq.~\eqref{continuity_coords}. 

Let us consider fluctuations around an equilibrium state of the fluid, characterized by time dependent energy density and pressure $\bar \rho(t)$ and $\bar p(t) = w \bar \rho(t)$ and velocity $\vec v=0$.
Equations~\eqref{continuity_coords} and \eqref{euler_coords}  simplify in the limit of small velocity, \ie~for $v \ll c=1$. In this case they can be written as
\begin{align}
 \partial_t \rho  + \vec \nabla \cdot \left[ (\rho+p) \vec v \right]&=0\;, \label{continuity_nr} \\
 (\rho+p) \left[ \partial_t \vec v + (\vec v \cdot \vec \nabla) \vec v \right] &= - \vec \nabla p  - \vec v \; \partial_t p\;. \label{euler_nr}
\end{align}

We can now reintroduce gravity in these equations by assuming a perturbed flat Friedmann metric,
\be
ds^2 = - dt^2 (1+2 \Phi) + a^2(t) (1-2 \Psi) d\vec x^2\;,
\ee
with $\Phi,\Psi \ll1$. Gravity enters through the covariant derivatives of the 4-velocity on the left hand sides of eqs.~(\ref{continuity_rel}) and (\ref{Euler_rel}). In the continuity equation it yields the term $(\rho +p) u^\mu \nabla_\mu \ln \sqrt{-g} \simeq (\rho+p) 3 H$ and in the Euler equation  it introduces the term $(\rho + p) u^\mu \Gamma^i_{\mu \rho} u^\rho \simeq (\rho+p) (2 H u^i + \partial^i \Phi)$, where we have neglected subleading terms in $v/c$ and $\Phi,\Psi$. Thus, with these new terms and using comoving coordinates with $v^i \simeq a u^i$, eqs.~\eqref{continuity_nr} and \eqref{euler_nr} become
\begin{align}
& \partial_t \rho  + 3H(\rho+p) + \frac1a \vec \nabla \cdot \left[ (\rho+p) \vec v \right]=0\;, \label{continuity_nr_grav} \\
&  \partial_t \vec v + H \vec v+ \frac1a (\vec v \cdot \vec \nabla) \vec v + \frac1a \vec \nabla \Phi = - \frac1{\rho+p}\left( \frac1a \vec \nabla p  + \vec v \; \partial_t p \right)\;, \label{euler_nr_grav}
\end{align}
which are equivalent to eqs.~\eqref{continuity} and \eqref{euler} after introducing the conformal time $\tau$ related to $t$ by $d t = a d \tau$.

The gravitational potential $\Phi$ can be related to the energy density and pressure perturbations by the Einstein equation. Neglecting time-variations of $\Phi$ and $\Psi$ of order $H$, which are small in the sub-Hubble scale dynamics \cite{Creminelli:2009mu}, the $00$ component of the Einstein equation yields
\be
\nabla^2 \Psi = 4 \pi G a^2 \delta \rho\;,
\ee
while the traceless part of the $ij$ component yields
\be
\nabla^2 (\Phi-\Psi) = 12 \pi G a^2 \delta p\;.
\ee
By combining these equations one obtains the Poisson equation \eqref{poisson},
\be
\nabla^2 \Phi = 4 \pi G a^2 (\delta \rho + 3 \delta p)\;.
\ee

Note that in deriving eqs.~\eqref{continuity_nr_grav} and \eqref{euler_nr_grav} we did not assume that energy and pressure perturbations are small with respect to their background value. In this sense, these equations are nonlinear in the energy density and pressure. However, for large pressure gradients one expects that the velocity, sourced by the right hand side of eq.~\eqref{euler_nr_grav}, becomes close to relativistic values invalidating the assumption $v\ll c$. This does not happen when pressure gradients are suppressed by the smallness of the speed of sound, as for dust or clustering quintessence. Indeed, in this case the right hand side of eq.~\eqref{euler_nr_grav} vanishes, and these equations consistently describe the nonlinear regime.


\subsection{Vertices in the spherical collapse approximation}
\label{app:vertices}

In the spherical collapse model one assumes spherical symmetry around $\vec x=0$. As a consequence, the linear density field $\delta_{\rm lin}(\vec k)$ depends only on the norm of $\vec k$, $k = |\vec k|$ and the nonlinear solutions for the density contrast $\delta$ and velocity divergence $\tt$ can be written as  
\cite{Bernardeau:1992zw},
\begin{align}
\d_{\rm sc}(\tau) & =  \sum_{n=1}^{\infty}\frac{\nu_n(\tau)}{n!}D^n(\tau) \varepsilon^n\, , \\
\tt_{\rm sc}(\tau)& =  \sum_{n=1}^{\infty}\frac{\mu_n(\tau)}{n!}D^n(\tau)\varepsilon^n\, ,
\end{align}
where $\nu_n$ and $\mu_n$ are the angular averages of the kernels,
\begin{align}
\nu_n&= n! \langle F_n(\vec q_1, \ldots, \vec q_n) \rangle \equiv n! \int \frac{d \Omega_1}{4 \pi} \ldots \frac{d \Omega_n}{4 \pi} F_n(\vec q_1, \ldots, \vec q_n)\;, \\
\mu_n&= n! \langle G_n(\vec q_1, \ldots, \vec q_n) \rangle \equiv n! \int \frac{d \Omega_1}{4 \pi} \ldots \frac{d \Omega_n}{4 \pi} G_n(\vec q_1, \ldots, \vec q_n)\;, 
\end{align}
$\varepsilon\equiv \int d^3k\, \d_{\rm in}(k)$ and $\nu_1=\mu_1=1$. 

From the equations of motion (\ref{continuity_tot_eta}) and (\ref{euler_tot_eta}) we find the recursive equations for the evolution of $\nu_n$ and $\mu_n$,
\begin{align}
\frac{\partial \nu_n}{\partial\eta}+n\,\nu_n-\mu_n & =  \frac{1}{C}\sum_{m=1}^{n-1}\binom{n}{m}\mu_m\nu_{n-m}\, ,\\
\frac{\partial \mu_n}{\partial\eta}+(n-1)\,\mu_n+\frac{3}{2}\frac{\Omega_m C}{f_+^2}(\mu_n-\nu_n) & =  \frac{1}{C}\sum_{m=1}^{n-1}\binom{n}{m}\mu_m\mu_{n-m}\, .
\end{align}
In particular, for $n=2$ we have
\begin{align}
\frac{\partial \nu_2}{\partial\eta}+2\,\nu_2-\mu_2 & =  \frac{2}{C}\, ,\\
\frac{\partial \mu_2}{\partial\eta}+\,\mu_2+\frac{3}{2}\frac{\Omega_m C}{f_+^2}(\mu_2-\nu_2) & =  \frac{2}{3\,C}\, .
\end{align}


\subsection{Redshift distorsions}
\label{app:zdist}

The derivation of the Kaiser formula \cite{Kaiser:1987qv} can be straightforwardly extended to the case of clustering quintessence. We follow the presentation of \cite{Bernardeau:2001qr} but we focus our considerations to the fluctuations in the number density of {\em galaxies} $\d_g$, rather than the matter overdensity $\d_m$. As both these densities are conserved, there is in fact no essential difference. In this way we directly refer to a generic biased population like the galaxy distribution. For simplicity, we work in the parallel-plane approximation. 

The mapping between redshift and position space, in comoving coordinates, is given by 
\be
\sv=\xv+\frac{v_z}{\HH}\hat{z}\,,
\ee
where $\hat{z}$ is the direction along the line-of-sight. From the conservation of the number of galaxies we can find the relation
\be
(1+\d_{g,s})\,d^3s=(1+\d_g)\,d^3x
\ee
between the galaxy overdensity in redshift space $\d_{g,s}$ and the same quantity in position space $\d_g$. As the volume elements are related by $d^3s = J(\vec x)d^3 x$, where $J$ is the Jacobian of the coordinate transformation from $\vec s$ to $\vec x$, one finds
\be
\d_{g,s}(\sv)=\frac{\d_g(\xv)+1-J(\xv)}{J(\xv)}\,.
\ee

The Jacobian is explicitly given by $J=|1+\nabla_z v_z/\HH|$. In Fourier space, under the assumption of $\nabla_zv_z/{\cal H}\ll 1$,  this yields
\be
\d_{g,s}(\kv)=\int\frac{d^3x}{(2\pi)^3}e^{-i\kv\cdot\xv-ik_z\,v_z/\HH}\left[\d_g(\xv)-\nabla_z v_z(\xv)/\HH\right]\,.
\ee
This equation can be written in terms of $\tt = - C/({\cal H} f) \vec \nabla \cdot \vec v $ defined in eq.~(\ref{Theta_def}) (here for simplicity we use the convention $f=f_+$) as 
\be
\begin{split}
\label{eq:zdist_exp}
\d_{g,s}(\kv) = & \sum_{n=1}^{\infty}\frac{1}{(n-1)!}\left(\frac{f\mu k}C\right)^{n-1}\int d^3 q_1\cdots d^3 q_n\d_D(\kv-\sum_{i=1}^n\qv_{i}) \\ & \times
\left[\d_g(\qv_1)+\frac{f\mu_1^2}{C}\tt({\qv_1})\right]\frac{\mu_2}{q_2}\tt({\qv_2})\cdots\frac{\mu_n}{q_n}\tt({\qv_n})
\,,
\end{split}
\ee
where $\mu\equiv k_z/k$ and $\mu_i \equiv q_{z,i}/q_i$. For $C=1$ we recover the standard expression of \cite{Scoccimarro:1999ed}.
At linear order this expression reduces to 
\be
\d_{g,s}^{\rm lin}(\kv)=\d_g^{\rm lin}(\kv)+\frac{f}{C}\mu^2\, \tt^{\rm lin}({\kv})\,.
\ee

Assuming a linear bias relation between the galaxy density and the total linear perturbations, \ie~$\d_g^{\rm lin}=b_1\,\d^{\rm lin}$, and replacing $\tt^{\rm lin}$ using eq.~(\ref{theta_lin}), we can write, in the linear approximation
\be
\begin{split}
\d_{g,s}^{\rm lin}(\kv) & =  b_1 \,\d^{\rm lin}(\kv)+\frac{f}{C}\,\mu^2\,\d^{\rm lin}(\kv) = \left(b_1 +\frac{f}{C}\mu^2 \right)\d^{\rm lin}(\kv)\,.
\end{split}
\ee
This is the analog of the Kaiser formula, except that $f$ is here  replace by $f/C$.
Using this expression, the linear galaxy power spectrum in redshift space is thus given by
\be\label{eq:powzdist}
P_{g,s}(\vec k)=P_g(k)\left(1+\beta\,\mu^2\right)^2\,,
\ee
where we have introduced the parameter $\beta$, which in the case of clustering quintessence is given by 
\be
\beta = \frac{f}{C b_1}\;. \label{beta}
\ee 
Thus, at linear order the corrections to the redshift distortions formula due to clustering quin\-tes\-sence enter only through the function $C$ in the parameter $\beta$. 
However, we can also express $\beta$ in terms of the linear growth rate $f_m$ defined in eq.~\eqref{fm_def}, and the linear bias parameter $b_{m,1}$, assuming a linear bias relation between the galaxy density and the matter linear perturbations, \ie~$\d_g^{\rm lin}=b_{m,1}\,\d_m^{\rm lin}$. Using eq.~\eqref{f_fm}, we can rewrite $\beta$ in eq.~\eqref{beta} as
\be
\beta=\frac{f_m\,D_m}{D\,b_1}=\frac{f_m}{b_{m,1}}\,, \label{beta_m}
\ee
recovering the relation employed in \cite{Sapone:2010uy}.

From eq.~(\ref{eq:zdist_exp}) one can also derive an expression for the galaxy bispectrum in redshift space. Assuming a {\em nonlinear} local relation between the galaxy overdensity $\d_g$ and the total density contrast $\d$ and Taylor expanding it in terms of powers of $\d$ one obtains  \cite{Fry:1992vr}
\be
\d_g(\xv)=\sum_{n=1}^{\infty}\frac{b_n}{n!}\d^n(\xv)\,. \label{Taylor}
\ee
Furthermore, the left hand side of eq.~\eqref{eq:zdist_exp} can be expanded in terms of redshift-space kernels $Z_n$ as \cite{Verde:1998zr, Scoccimarro:1999ed}
\be
\d_{g,s}(\kv,\eta)=\sum_{n=1}^{\infty}D_{+}^n(\eta)\int d^3 q_1\cdots d^3 q_n\d_D(\kv-\sum_{i=1}^n\qv_{i})Z_n(\qv_1,\dots,\qv_n;\eta)\d^{\rm in}_{\qv_1}\cdots\d^{\rm in}_{\qv_n}\,. \label{Zn}
\ee
Fourier transforming eq.~\eqref{Taylor} and using it to replace $\d_g(\vec k)$ on the right hand side of eq.~\eqref{eq:zdist_exp} one obtains, up to second order, 
\be
Z_1(\kv, \eta)=b_1(1+\beta \mu^2)\,,
\ee
and
\be
\begin{split}
Z_2(\kv_1,\kv_2;\eta)  = &  \, b_1 \left[ F_2(\kv_1,\kv_2;\eta)+\beta \mu^2\,G_2(\kv_1,\kv_2;\eta) \right]\\
&  +
 b_1^2 \beta \frac{\mu k}{2}  \left[\frac{\mu_1}{k_1}  \left(1+ \beta \mu_2^2\right)+\frac{\mu_2}{k_2}\left(1 + \beta \mu_1^2\right)\right]+\frac{b_2}2\,,
\end{split}
\ee
with $F_2$ and $G_2$ defined in eqs.~\eqref{F2} and \eqref{G2}.
Thus, as before these expressions for $Z_1$ and $Z_2$ can be quickly recovered from the standard ones after replacing $f$ by $f/C$. 

From eq.~\eqref{Zn}, the redshift-space, tree-level, galaxy bispectrum is given by 
\be
B_{g,s}(\kv_1,\kv_2,\kv_3;\eta)=2 Z_2(\kv_1,\kv_2;\eta)Z_1(\kv_1,\eta)Z_1(\kv_2,\eta) P_{\rm lin}(k_1,\eta)P_{\rm lin}(k_2,\eta)+2~{\rm cyclic}\,.
\ee
The effects of clustering dark energy enters in the ratio $f/C$ but also in the corrections to the second-order kernels $F_2$ and $G_2$, which are both of the same order. In addition, one should carefully consider the linear and quadratic bias parameters $b_1$ and $b_2$ defined in terms of the total density. One can expect for instance that for an ideal population of galaxies characterized by a conserved comoving number density, the corresponding linear bias would present a significant dependence on redshift at late times. The implications of the clustering quintessence scenario for galaxy bias are beyond the scope of this work and will be considered elsewhere.

\end{document}